\documentclass[twocolumn,nofootinbib,superscriptaddress,preprintnumbers,floatfix]{revtex4-1}

\pdfoutput=1

\usepackage{amsmath}
\usepackage{amsfonts}
\usepackage{graphicx}
\usepackage[small]{subfigure}
\usepackage[hyperfootnotes=false]{hyperref}
\usepackage{braket}
\usepackage{cancel}
\usepackage{xspace}
\usepackage{fmtcount}

\usepackage{color}
\usepackage{multirow}

\newcommand{\eq}[1]{Eq.~\eqref{eq:#1}}
\newcommand{\eqs}[2]{Eqs.~\eqref{eq:#1} and \eqref{eq:#2}}

\newcommand{\tab}[1]{Tab.~\ref{tab:#1}}
\newcommand{\msbar}{\overline{\textrm{MS}}}

\newcommand{\ord}[1]{\mathcal{O}(#1)}

\newcommand{\df}{\mathrm{d}}

\newcommand{\tr}{\mathrm{tr}}

\newcommand{\nn}{\nonumber}

\allowdisplaybreaks[4]

\DeclareRobustCommand{\Sec}[1]{Sec.~\ref{#1}}
\DeclareRobustCommand{\Secs}[2]{Secs.~\ref{#1} and \ref{#2}}
\DeclareRobustCommand{\App}[1]{App.~\ref{#1}}

\DeclareRobustCommand{\Fig}[1]{Fig.~\ref{#1}}

\DeclareRobustCommand{\Eq}[1]{Eq.~(\ref{#1})}
\DeclareRobustCommand{\Eqs}[2]{Eqs.~(\ref{#1}) and (\ref{#2})}

\newcommand{\be}{\begin{equation}}
\newcommand{\ee}{\end{equation}}

\def\tr{{\rm tr}}

\begin{document}

\preprint{\vbox{\hbox{UWTHPH 2015-1}\hbox{MIT--CTP 4630}\hbox{LPN14-128}
}}

\title{A Precise Determination of $\alpha_s$ from the C-parameter Distribution}

\author{Andr\'e H.~Hoang}
\affiliation{University of Vienna, Faculty of Physics, Boltzmanngasse 5, A-1090 Wien, Austria}
\affiliation{Erwin Schr\"odinger International Institute for Mathematical Physics,
University of Vienna, Boltzmanngasse 9, A-1090 Vienna, Austria}
\author{Daniel W.~Kolodrubetz}
\affiliation{Center for Theoretical Physics, Massachusetts Institute of
  Technology, Cambridge, MA 02139, USA}
\author{Vicent Mateu}
\affiliation{University of Vienna, Faculty of Physics, Boltzmanngasse 5, A-1090 Wien, Austria}
\author{Iain W.~Stewart}
\affiliation{Center for Theoretical Physics, Massachusetts Institute of
  Technology, Cambridge, MA 02139, USA}

\begin{abstract}
We present a global fit for $\alpha_s(m_Z)$, analyzing the available C-parameter data measured at center-of-mass energies between $Q=35$ and $207$\,GeV. The experimental data is compared to a \mbox{N$^3$LL$^\prime$ + $\mathcal{O}(\alpha_s^3)$ + $\Omega_1$} theoretical prediction (up to the missing four-loop cusp anomalous dimension), which includes power corrections coming from a field theoretical nonperturbative soft function. The dominant hadronic parameter is its first moment $\Omega_1$, which is defined in a scheme which eliminates the $\ord{\Lambda_{\rm QCD}}$ renormalon ambiguity. The resummation region plays a dominant role in the C-parameter spectrum, and in this region a fit for $\alpha_s(m_Z)$ and $\Omega_1$ is sufficient. We find $\alpha_s(m_Z)=0.1123\pm 0.0015$ and $\Omega_1=0.421\pm 0.063\,{\rm GeV}$ with $\chi^2/\rm{dof}=0.988$ for $404$ bins of data. These results agree with the prediction of  universality for $\Omega_1$ between thrust and C-parameter within 1-$\sigma$.
\end{abstract}

\maketitle

\section{Introduction}
\label{sec:intro}

In order to study Quantum Chromodynamics (QCD) accurately in the high-energy regime, it is useful to exploit the wealth of data from previous $e^+\, e^-$ colliders such as LEP. Here the final states coming from the underlying partons created in the collisions appear as boosted and collimated groups of hadrons known as jets. Event shapes have proven to be very successful to study these collisions quantitatively. They combine the energy and momenta of all of the measured hadrons into an infrared- and collinear-safe parameter which describes the geometric properties of the whole event by a single variable distribution. Due to their global nature event shapes have nice theoretical properties, making it possible to obtain very accurate theoretical predictions using QCD.
Most $e^+e^-$ event shape variables quantify how well the event resembles the situation of two narrow back-to-back jets, called dijets, by vanishing in this limit.  Because the dijet limit involves restrictions that only allow collinear and soft degrees of freedom for the final-state radiation, such QCD predictions involve a number of theoretical aspects that go beyond the calculation of higher-order perturbative loop corrections. These include factorization, to systematically account for perturbative and nonperturbative contributions, and the resummation of large logarithmic corrections by renormalization group evolution. Comparisons of predictions for event shapes with experimental data thus provide non-trivial tests of the dynamics of QCD.

Due to the high sensitivity of event shapes to jets induced by gluon radiation they are an excellent tool to measure the strong coupling  $\alpha_s$. For more inclusive hadronic cross sections (like $e^+e^-\to $ hadrons) the $\alpha_s$ dependence is subleading because it only occurs in corrections to a leading order term, while for event shapes the $\alpha_s$ dependence is a leading-order effect. For this reason, the study of event shapes for determining $\alpha_s$ has a long history in the literature (see the review~\cite{Kluth:2006bw} and the workshop proceedings~\cite{Bethke:2012jm}), including recent analyses which include higher-order resummation and corrections up to ${\cal O}(\alpha_s^3)$~\cite{GehrmannDeRidder:2007bj,Davison:2008vx,Becher:2008cf,Dissertori:2009qa,GehrmannDeRidder:2009dp,Chien:2010kc,Abbate:2010xh,Abbate:2012jh,Gehrmann:2012sc,Hoang:2014wka}.

Several previous high-precision studies which determine $\alpha_s(m_Z)$~\cite{Davison:2008vx,Becher:2008cf,Abbate:2010xh,
Abbate:2012jh,Gehrmann:2012sc} focus on the event shape called thrust~\cite{Farhi:1977sg},
\begin{equation}
 \tau \,=\, 1-T \,=\,
 \min_{\vec{n}}\! \left( 1 -\, \frac{ \sum_i|\vec{n} \cdot \vec{p}_i|}{\sum_j |\vec{p}_j|} \right),
\end{equation}
where $\vec{n}$ is called the thrust axis and it follows from the above equation that $0 \le \tau \le 1/2$. Another event shape, known as C-parameter~\cite{Parisi:1978eg,Donoghue:1979vi}, can be written as:
\begin{equation} \label{eq:Cdef}
C=\frac{3}{2} \, \frac{\sum_{i,j} | \vec{p}_i | | \vec{p}_j | \sin^2 \theta_{ij}}
{\left( \sum_i | \vec{p}_i | \right)^2}\,,
\end{equation}
where $\theta_{ij}$ gives the angle between particles $i$ and $j$. It is straightforward to show that $0\le C\le 1$.
In a previous paper~\cite{Hoang:2014wka} we computed the C-parameter distribution with a resummation of large logarithms at N$^3$LL$^\prime$ accuracy, including fixed-order terms up to ${\cal O}(\alpha_s^3)$ and hadronization effects using a field-theoretic nonperturbative soft function. These results were achieved by using the Soft Collinear Effective Theory (SCET)~\cite{Bauer:2000ew, Bauer:2000yr, Bauer:2001ct, Bauer:2001yt, Bauer:2002nz}. Our results for $C$ are valid in all three of the peak, tail, and far-tail regions of the distribution, and are the most accurate predictions available in the literature, having a perturbative uncertainty of $\simeq 3\%$ at $Q=m_Z$ for the region relevant for $\alpha_s(m_Z)$ and $\Omega_1$ fits. The same accuracy was previously achieved for thrust, where the remaining perturbative uncertainty in the $\tau$ distribution is $\simeq 2\%$ in this region~\cite{Abbate:2010xh}.  In this paper we make use of these new C-parameter theoretical results~\cite{Hoang:2014wka} to carry out a global fit to all available data, comparing the results with the analogous global fit for thrust~\cite{Abbate:2010xh} where appropriate. 

Since both $\tau$ and $C$ vanish in the dijet limit, it is worthwhile to contrast them in order to anticipate differences that will appear in the analysis.  Differences between $C$ and $\tau$ include the following:
\begin{enumerate}
\item [a)] calculating $\tau$ requires identifying the thrust axis with a minimization procedure, while $C$ does not involve a minimization; 
\item [b)] $\tau$ has a single sum over particles while $C$ has a double sum; 
\item [c)] the size of the nonperturbative region, where the entire shape function is important, is larger for $C$ compared to $\tau$ due to an enhancement by a factor of $3\pi/2$; 
\item [d)] the resummation region for $C$ is larger than that for thrust since the logarithms appear as $\ln(C/6)$ compared to $\ln(\tau)$, which increases the range of $C$ values that are useful for $\alpha_s$ fits; 
\item [e)] fixed-order predictions for the thrust cross section are smooth across the threshold where non-planar events first contribute, $\tau=1/3$, while the fixed-order C-parameter cross section has an integrable singularity at this threshold, $C_\text{shoulder}=3/4$.  The singularity for $C$ comes from the fact that the leading-order distribution is not continuous at $C=C_\text{shoulder}$. (The $C$-distribution can be made smooth here using LL resummation~\cite{Catani:1997xc}.) 
\end{enumerate}
A key similarity is that in the dijet limit ($C,\,\tau \ll 1$) the partonic cross sections for thrust and C-parameter are related up to NLL by using \mbox{$\tau_{\text{NLL}}=C_{\text{NLL}}/6$}~\cite{Catani:1998sf}. This relation quantifies several qualitative relations between $C$ and $\tau$ in the dijet region.

Recent higher-order event-shape analyses~\cite{Davison:2008vx,Abbate:2010xh,Gehrmann:2012sc,Abbate:2012jh,Gehrmann:2012sc} have found values of $\alpha_s(m_Z)$ significantly lower than the world average of $\alpha_s(m_Z)=0.1185\pm 0.0006$~\cite{Agashe:2014kda} which is dominated by the lattice QCD determination~\cite{Chakraborty:2014aca}. This includes the determination carried out for thrust at N$^3$LL$^\prime\,$+$\,{\cal O}(\alpha_s^3)$ in Ref.\,\cite{Abbate:2010xh}\footnote{Note that results at N$^3$LL require the currently unknown QCD \mbox{four-loop} cusp anomalous, but conservative estimates show that this has a negligible impact on the perturbative uncertainties. Results at N$^3$LL$^\prime$ also technically require the unknown 3-loop non-logarithmic constants for the jet and soft functions which are also varied when determining our uncertainties, but these parameters turn out to only impact the peak region which is outside the range of our $\alpha_s(m_Z)$ fits in the resummation region.}, which is also consistent with analyses at N$^2$LL\,+\,${\cal O}(\alpha_s^3)$~\cite{Gehrmann:2009eh,Gehrmann:2012sc} which consider the resummation of logs at one lower order. In Ref.~\cite{Banfi:2014sua} a framework for a numerical code with N$^2$LL precision for many $e^+e^-$ event shapes was found, which could also be utilized for $\alpha_s$ fits in the future. In Ref.~\cite{Abbate:2010xh} it was pointed out that including a proper fit to power corrections for thrust causes a significant negative shift to the value obtained for $\alpha_s(m_Z)$, and this was also confirmed by subsequent analyses~\cite{Gehrmann:2012sc}. Recent results for $\alpha_s(m_Z)$ from $\tau$ decays~\cite{Boito:2014sta}, DIS data~\cite{Alekhin:2012ig}, the static potential for quarks~\cite{Bazavov:2014soa}, as well as global PDF fits~\cite{Ball:2011us,Martin:2009bu} also find values below the world average. With the new analysis we present here, we provide another event-shape determination of $\alpha_s(m_Z)$ with a high level of precision. We will also simultaneously examine the leading power correction to the distribution, which should be universal between thrust and \mbox{C-parameter}.

This paper is organized as follows. In Sec.~\ref{sec:theory} we review the theoretical
calculation of the C-parameter cross section, presented in more detail in Ref.~\cite{Hoang:2014wka}.
The details on the experimental data and fit procedure used in our analysis are given in Secs.~\ref{sec:data}
and~\ref{sec:fitprocedure}.
In Sec.~\ref{sec:results} we present the results of our fit for $\alpha_s(m_Z)$
and the first moment of the nonperturbative soft function, $\Omega_1^C$. Fits which include hadron-mass effects are discussed in Sec.~\ref{sec:hadmassresum}. In Sec.~\ref{sec:peak-tail} we make
predictions for the peak and far-tail regions of the distribution, which are not used in our fit,
and compare those regions to experimental data. The universality of $\Omega_1$ is
discussed in Sec~\ref{sec:universality}, where we compare our results with the previous fit done using thrust
in Ref.~\cite{Abbate:2010xh}. Finally, Sec.~\ref{sec:conclusions} contains our conclusions. We also include three appendices. The first, Appendix~\ref{ap:profile}, contains the formulae needed to calculate the profile functions,
the second Appendix~\ref{ap:subtractions} contains results that support our choice for the definition of the renormalon free $\Omega_1$ parameter to use for $C$; and the third compares fit results for thrust with our profiles and those from Ref.~\cite{Abbate:2010xh}.

\section{Theory Review}
\label{sec:theory}
Until a few years ago, the theoretical uncertainties related to perturbative contributions as well as hadronic power corrections were still larger than the experimental uncertainties. The situation on the theory side has dramatically changed with the calculation of ${\cal O}(\alpha_s^3)$ corrections~\cite{GehrmannDeRidder:2007bj, GehrmannDeRidder:2007hr, GehrmannDeRidder:2009dp,Ridder:2014wza, Weinzierl:2008iv,Weinzierl:2009ms}, and the pioneering use of SCET to obtain higher-order perturbative corrections in Ref.~\cite{Becher:2008cf}, and to obtain accurate predictions for the full spectrum and a precision description of power corrections in Ref.~\cite{Abbate:2010xh}. In this section we review the theoretical work behind our calculation of $\df \sigma/\df C$, presented in Ref.\,\cite{Hoang:2014wka}.

SCET separates the physics occurring at the different energy and momentum scales relevant to the underlying jets whose properties are characterized by an event shape. For the C-parameter distribution the relevant scales are:  (i) the hard scale $\mu_H$, which is related to the production of partons at short distances and is of the order of the center-of-mass energy $Q$; (ii) the jet scale $\mu_J\sim Q\sqrt{C/6}$, which governs the formation and evolution of the jets; and (iii) the soft scale $\mu_S\sim QC/6$, which is the scale of large-angle soft radiation. All three scales are widely separated in the dijet region $C\ll 1$, where most of the events occur and where the distribution is maximal. There is one function associated to each one of these scales in the factorization theorem that describes the dominant contribution of the C-parameter distribution in the dijet limit: (i) the hard function $H$ (the modulus squared of the QCD-to-SCET matching coefficient), which is common to all dijet event-shape factorization theorems; (ii) the jet function $J_\tau$ (given by matrix elements of quark fields with collinear Wilson lines), which is common for C-parameter~\cite{Hoang:2014wka}, thrust~\cite{Abbate:2010xh} and Heavy-Jet-Mass ($\rho$)~\cite{Chien:2010kc}; and (iii) the soft function $S_{\widetilde C}$ (defined by a vacuum matrix element of purely soft Wilson lines), which in general depends on the specific form of the event shape. Whereas the former two are perturbative ($\mu_H, \mu_J \gg \Lambda_{\rm QCD}$), permitting the calculation of the hard and jet functions as an expansion in powers of $\alpha_s$, the soft function has perturbative corrections ($\mu_S\gg \Lambda_{\rm QCD}$) as well as nonperturbative contributions that need to be accounted for ($\mu_S\gtrsim \Lambda_{\rm QCD}$). Renormalization evolution between the three scales $\mu_H$, $\mu_J$, and $\mu_S$ sums up large logarithms to all orders in perturbation theory. It turns out that the soft function anomalous dimension for $C$ and $\tau$ are identical~\cite{Hoang:2014wka}, providing a connection between these two event shapes at every order of perturbation theory. Only corrections related to non-logarithmic terms in their soft functions, and the associated towers of logarithms, differ between $C$ and $\tau$.

The soft function can be further factorized into a partonic soft function $\hat S_{\!\widetilde C}$, calculable in perturbation
theory, and a nonperturbative shape function $F_{\!C}$, which has to be obtained from fits to data.  In the strict $\overline {\rm MS}$ scheme this factorization was achieved in Refs.~\cite{Hoang:2007vb,Ligeti:2008ac}.  (Analytic power corrections for the C-parameter distributions have also been studied in other schemes and frameworks, see e.g.\
Refs.~\cite{Gardi:2003iv,Korchemsky:2000kp,Dokshitzer:1995zt}.) 

The treatment of hadronic power corrections greatly simplifies in the tail of the distribution, defined by \mbox{$QC\gg 3\pi \Lambda_{\rm QCD}$}, where the
shape function can be expanded in an OPE. Here the leading power correction is parametrized by $\Omega_1^C$, the first moment of the shape function. The main effect of this leading power correction is a shift of the cross section, \mbox{${\df\hat\sigma} (C) \to {\df\hat\sigma}(C \,-\, \Omega_1^C/Q)$}.  Interestingly, this matrix element is related to the corresponding one in thrust by
\begin{align} \label{eq:O1univ}
 \frac{\Omega_1^\tau}{2} = \frac{\Omega_1^C}{3\pi}\equiv \Omega_1 \,.
\end{align} 
This relation was first derived using dispersive models with a single soft-gluon approximation in Ref.~\cite{Dokshitzer:1995zt}. The equality can actually be derived to all orders in QCD just using quantum field theory~\cite{Lee:2006nr}, but ignoring hadron-mass effects~\cite{Salam:2001bd}. These hadron-mass effects can also be formulated purely with quantum field theory operators~\cite{Mateu:2012nk}. Although they may in general give large corrections, hadron-mass effects turn out to violate \Eq{eq:O1univ} at only the $2\%$ level, which is well below the 15\% level determination of $\Omega_1$ that we will achieve here. When presenting the results of our fits, we parametrize the power correction using $\Omega_1$ defined in \Eq{eq:O1univ} to ease comparison with our previous analysis which determined this $\Omega_1$ based on thrust~\cite{Abbate:2010xh,Abbate:2012jh}.
\subsection{Factorized Cross Section Formula}
\label{subsec:crosssection}
In order to understand the perturbative components of the C-parameter cross section we make use of the \mbox{C-parameter} factorization formula. To make
the connection to thrust simpler we will often use functions defined with the variable $\widetilde C = C/6$. For the perturbative
cross section we find~\cite{Hoang:2014wka}:
\begin{align}\label{eq:factorization-partonic-singular}
\!\!\!\frac{1}{\sigma_0}\frac{\df \hat\sigma_{\rm s}}{\df C}=\frac{Q}{6} H(Q,\mu)\!
\int\! \df s\, J_\tau(s,\mu) S_{\widetilde C}\bigg(\frac{Q C}{6}- \frac{s}{Q},\mu\bigg).
\end{align}
Here $J_\tau$ is the thrust jet function, obtained by the convolution of the two hemisphere jet
functions, $J_\tau = J \otimes J$. $J_\tau$ describes the collinear radiation in the direction of the two jets.
Its definition and expression up to $\mathcal{O}(\alpha_s^3)$\,\cite{Lunghi:2002ju, Bauer:2003pi,Becher:2006qw}
can be found in Refs.~\cite{Becher:2008cf,Abbate:2010xh}. The three-loop non-logarithmic coefficient of this jet function, $j_3$, is not known, 
and we vary it in our scans. The anomalous dimension of $J_\tau$ is known at
three loops, and can be obtained from Ref.~\cite{Moch:2004pa}.

The hard factor $H$ contains short-distance QCD effects and is obtained from the Wilson
coefficient of the QCD-SCET matching of the vector and axial-vector currents. The hard function is the same
for all event shapes for massless quarks, and its expression up to $\mathcal{O}(\alpha_s^3)$
\cite{Matsuura:1987wt,Matsuura:1988sm,Gehrmann:2005pd,Moch:2005id,Lee:2010cg,Baikov:2009bg,Lee:2010cg,Gehrmann:2010ue},
can be found in Ref.~\cite{Abbate:2010xh}. The full anomalous dimension of $H$ is known at three loops,
${\cal O}(\alpha_s^3)$~\cite{vanNeerven:1985xr,Matsuura:1988sm,Moch:2005id}.

The soft function $S_{\widetilde C}$ describes wide-angle soft radiation between the two jets. It is defined as
\begin{align}
 S_{\widetilde C}(\ell,\mu) = \frac{1}{N_C}\langle \,  0\,| \tr\, \overline{Y}_{\bar n}^T Y_n
    \delta\bigg(\!\ell- \frac{Q\widehat C}{6}\bigg) Y_n^\dagger \overline{Y}^*_{\bar n}\, |\, 0\,\rangle\,,
\end{align}
where $\widehat C$ is an operator whose eigenvalues on physical states correspond to the value of \mbox{C-parameter} for that state:
$\widehat C |X\rangle = C(X) |X\rangle$. Since the hard and jet functions are the same as for thrust, the anomalous dimension of the soft function has to coincide with the anomalous dimension of the thrust soft function. Hence one only needs to determine the non-logarithmic terms of the \mbox{C-parameter} soft function. In Ref.~\cite{Hoang:2014wka} we computed it analytically at
one loop, $s_1^{\widetilde C} = -\,\pi^2 C_F/6$, and used EVENT2 to numerically determine the two-loop non-logarithmic coefficient $s_2^{\widetilde C}$, with the result
\begin{align}
s_2^{\widetilde C}=-\,43.2\,\pm\,1.0 \,.
\end{align}
The three-loop non-logarithmic coefficient of the \mbox{C-parameter} soft function, $s_3^{\widetilde C}$, is currently not known, and we estimate it with a Pad\'e approximation, assigning a very conservative uncertainty. We vary this constant in our scan analysis. The precise definitions of $j_3$ and $s_2^{\widetilde C}$ as well as $s_3^{\widetilde C}$ can be found in Eqs.~(A10) and (A12) of Ref.~\cite{Hoang:2014wka}.

In Eq.~(\ref{eq:factorization-partonic-singular}) the hard, jet and
soft functions are evaluated at a common scale $\mu$. If fixed-order expressions are used for these functions, then there is no scale
choice that simultaneously minimizes the logarithms for these three
functions. One can instead renormalization-group evolve from $\mu$ to
the respective scales $\mu_H\sim Q$, $\mu_J\sim Q \sqrt{C/6}$ and
$\mu_S\sim Q C/6$ at which the logs in each of $H$, $J_\tau$, and $S_{\widetilde C}$ are minimized, and only use fixed-order expressions for these functions at these scales. In this way, large
logs of ratios of the scales are summed up in the renormalization group
evolution kernels $U_H$, $U_J^\tau$, and $U_S^\tau$:
\begin{align}\label{eq:singular-resummation}
\frac{1}{\sigma_0}\frac{\df \hat\sigma_{\rm s}}{\df C} & \,=\,
\frac{Q}{6} H(Q,\mu_H)\,U_H(Q,\mu_H,\mu)\,\times \\\nonumber
&\int\! \df s\, \df s^\prime\,
J_\tau(s,\mu_J)\,U_J^\tau(s-s^\prime,\mu,\mu_J) \\\nonumber
&\int\! \df k\,U_S^\tau(k,\mu,\mu_S)\, S_{\widetilde C}\bigg(\!\frac{Q C}{6}- \frac{s}{Q}-k,\mu_S\!\bigg).
\end{align}
\label{eq:power}

The evolution kernels $U_H$, $U_J^\tau$ and $U_S^\tau$ resum the large logarithms, $\ln(C/6)$, and explicit expressions can be found in Ref.~\cite{Hoang:2014wka}. The only unknown piece in our resummation of logarithms at N$^3$LL order is the small contribution from the four-loop cusp anomalous dimension, $\Gamma_{3}^\text{cusp}$, which we estimate using a Pad\'e approximation and conservatively vary in our analysis.

While \Eq{eq:singular-resummation} gives the part of the cross section that is singular and non-integrable as $C \rightarrow 0$, we also need to include the integrable or nonsingular contribution. This can be written as
\begin{align} \label{eq:nonsingular-expansion}
\frac{1}{\sigma_0} \frac{\df \hat{\sigma}_{\rm ns}}{\df C} & \,=\, \frac{\alpha_s(\mu_{\rm ns})}{2\pi}\, f_1(C)\,  \\
&+\left(\frac{\alpha_s(\mu_{\rm ns})}{2\pi}\right)^{\!\!2}\! \bigg[f_2(C) + \beta_0 \ln\!\Big(\frac{\mu_{\rm ns}}{Q}\Big) f_1(C)\bigg]\nn\\
&+\,\left(\frac{\alpha_s(\mu_{\rm ns})}{2\pi}\right)^{\!\!3}\! \bigg\{f_3(C) + 2\, \beta_0  \ln\!\Big(\frac{\mu_{\rm ns}}{Q}\Big)   f_2(C) \nn \\
& \quad 
+ \bigg[\frac{\beta_1}{2}  \ln\!\Big(\frac{\mu_{\rm ns}}{Q}\Big) + \beta_0^2 \ln^2\!\Big(\frac{\mu_{\rm ns}}{Q}\Big)\bigg] f_1(C) \bigg\}
\nn\\
&+\, {\cal O}(\alpha_s^4)
\,.\nonumber
\end{align}
The functions $f_1(C)$, $f_2(C)$, and $f_3(C)$ were determined  in Ref.~\cite{Hoang:2014wka} using the fixed-order results at ${\cal O}(\alpha_s^{1,2,3})$~\cite{Ellis:1980nc,Ellis:1980wv,Catani:1996jh,Catani:1996vz,GehrmannDeRidder:2007bj, GehrmannDeRidder:2007hr, GehrmannDeRidder:2009dp,Ridder:2014wza, Weinzierl:2008iv,Weinzierl:2009ms}.
The nonsingular cross section $\df\hat{\sigma}_{\rm ns}/\df C$ is independent of the renormalization scale $\mu$ order by order, and therefore we evaluate these pieces at the nonsingular scale $\mu_{\rm ns}$, and vary this scale to estimate higher-order perturbative nonsingular corrections. The scale variation of $\mu_{\rm ns}$ will be discussed further in Section \ref{subsec:profiles}.

The full partonic cross section is then given by
\begin{align} \label{eq:full-partonic}
\frac{1}{\sigma_0} \frac{\df \hat{\sigma}}{\df C} &= \frac{1}{\sigma_0} \frac{\df \hat{\sigma}_{\rm s}}{\df C} + \frac{1}{\sigma_0} \frac{\df \hat{\sigma}_{\rm ns}}{\df C}\,.
\end{align}
Nonperturbative effects are included by convolving Eq.~(\ref{eq:full-partonic}) with a shape function:
\begin{align}\label{eq:singular-nonperturbative}
\frac{1}{\sigma_0}\frac{\df \sigma}{\df C} = \!\int \!\df p \,\frac{1}{\sigma_0}\frac{\df \hat\sigma}{\df C}\Big(C-\frac{p}{Q}\Big)F_C(p)\,.
\end{align}
One important property of this shape function is that its first moment encodes the leading power correction to our cross section. In the $\overline{\rm MS}$ scheme this moment is given by
\begin{align}\label{eq:Omega_1-definition}
\overline \Omega_1^{\,C} &\equiv \!\int \df k\, k\, F_C(k)\,.
\end{align}
Up to the normalization factors shown in Eq.~(\ref{eq:O1univ}) we expect approximate universality between the $\Omega_1$ for \mbox{C-parameter} and thrust. For the calculation of the cross section, the shape function is expanded in a complete basis of functions obtained by an appropriate infinite-range mapping of the Legendre polynomials~\cite{Ligeti:2008ac}, with the coefficients chosen to maintain the first moment. For further details on the implementation of the shape function for C-parameter see Ref.~\cite{Hoang:2014wka}. We remove an $\ord{\Lambda_\text{QCD}}$ renormalon by using the Rgap scheme \cite{Hoang:2008fs,Hoang:2009yr}, which introduces a subtraction scale $R$ into our formula, as well as the gap parameter $\bar{\Delta}$ and the perturbative scheme-change gap parameter $\delta(R,\mu_S)$. Here, $\delta(R,\mu_S)$ is given by a perturbative series in $\alpha_s(\mu_S)$ whose mass dimension is set by an overall factor of $R$, and which also contains $\ln(\mu_S/R)$ factors. The convolution with the shape function now becomes,
\begin{align}\label{eq:singular-nonperturbative-gap}
\frac{\df \sigma}{\df C} = \!\int \!\df p \,e^{-3 \pi \frac{\delta(R,\mu_{s})}{Q}\frac{\partial}{\partial C}}\frac{\df \hat\sigma}{\df C}\Big(C-\frac{p}{Q}\Big)  &\\
 \times F_C\big(p - 3 \pi \bar{\Delta} (R,\mu_S)\big)&\,.\nn
\end{align}

The final component of our cross section is properly accounting for hadron-mass effects following Ref.~\cite{Mateu:2012nk}. Hadron-mass effects induce an 
additional series of large perturbative logarithms which start at NLL, $\alpha_s^k \ln^k(\mu_S/\Lambda_{\rm QCD})$, and also
break the exact universality between $\Omega_1^C$ and $\Omega_1^\tau$ given in Eq.~(\ref{eq:O1univ}). These effects are accounted for by including dependence on the transverse velocity, $r \equiv \frac{p_\perp}{\sqrt{p_\perp^2 + m_H^2}}$, in the nonperturbative matrix elements (here, $m_H$ gives the non-zero hadron mass). In particular, in the Rgap scheme the first moment of the shape function is actually given by
\begin{align} \label{eq:firstmomentwithhadron}
\int\! \df k \, k\, F_C\big(k-3 \pi \bar{\Delta}(R,\mu_S)\big)\, = \,\Omega_1^C(R,\mu_S) &\\
=3 \pi \!\int_0^1 \! \df r\, g_C(r) \, \Omega_1(R,\mu_S,r)&\,.\nn
\end{align}
In the $\overline{\rm MS}$ scheme the definition accounting for hadron-mass effects is the same as \Eq{eq:firstmomentwithhadron}, but one sets \mbox{$\bar\Delta=0$} and removes $R$ as an argument for these parameters.
Accounting for both the the $\overline{\rm MS}$ running due to hadron masses and the R-evolution running in the Rgap scheme, the evolution of the integrand on the right-hand side of \Eq{eq:firstmomentwithhadron} is given by,
\begin{align} \label{eq:hadron-masses}
& g_C(r)\,\Omega_1(R,\mu,r) = g_C(r)\!\left[ \frac{\alpha_s(\mu)}{\alpha_s(\mu_\Delta)} \right]^{\hat{\gamma}_1(r)}\!
\Omega_1(R_\Delta, \mu_\Delta,r) \nn\\
&+\Delta^{\rm diff}(R_\Delta,R,\mu_\Delta,\mu,r)\,,
\end{align}
where $R_\Delta$ and $\mu_\Delta$ give the initial scales where the function $\Omega_1(R_\Delta,\mu_\Delta,r)$ is defined. The perturbative evolution kernel $\Delta^{\rm diff}$ gives the $R$ and $\mu$ running for each value of $r$. The function $g_C$ encodes the event-shape dependence of the hadron-mass effects and $\hat{\gamma}_1$ gives the solution to the one-loop RGE for $\Omega_1$ with hadron masses derived in Ref.~\cite{Mateu:2012nk}. Since the two- and three-loop $r$-dependent anomalous dimensions are unknown, we do not treat the logs generated by hadron-mass effects to the same level of precision. When hadron-mass effects are accounted for we always sum the associated logarithms at NLL.  An analogous formula to \Eq{eq:hadron-masses} also holds for the thrust parameter $\Omega_1^\tau$.

Combining all of these elements gives the complete cross section. Note that we can resum to any order up to N${}^3$LL$^\prime$ and can choose to include or leave out the shape function, renormalon subtraction and hadron-mass effects. This flexibility allows us to see how the analysis changes when we take into account each of these additional physical considerations and enables us to test how robust the fits are to various changes in the theoretical treatment.

\subsection{Profile Functions}
\label{subsec:profiles}
In order to smoothly transition between the nonperturbative, resummation, and fixed-order regions we make use of profile functions $\mu_i(C)$ for the renormalization scales $\mu_H$, $\mu_J(C)$, $\mu_S(C)$, $R(C)$, and $\mu_{\rm ns}(C)$. In the three $C$ regions, the requirements on the scales which properly deal with large logarithms, nonperturbative effects, and the cancellations between singular and nonsingular contributions in the fixed-order region are
\begin{align} \label{eq:profile-constraints}
& \text{1) nonperturbative:~} C \lesssim 3\pi\,\Lambda_\text{QCD}\nn \\
&  \qquad  \mu_H  \sim Q,\  \mu_J \sim \sqrt{\Lambda_\text{QCD} Q},\  \mu_S \!\sim\! R \!\sim\!  \Lambda_\text{QCD} 
  \,, \nn \\[5pt]
& \text{2) resummation:~} 3\pi\,\Lambda_\text{QCD} \ll C < 0.75
 \\ %[-5pt]
& \qquad  \mu_H \sim Q,\  \mu_J \sim Q \sqrt{\frac{C}{6}},\   \mu_S \! \sim \! R \!\sim\! \frac{QC}{6}  \gg \Lambda_{\rm QCD}
  \,, \nn \\
& \text{3) fixed-order:~} C > 0.75
   \nn \\
 & \qquad \mu_{\rm ns} = \mu_H   = \mu_J
    = \mu_S = R \sim Q\gg \Lambda_{\rm QCD}\nn
  \,.\end{align}
In addition we take the fixed-order nonsingular scale $\mu_{\rm ns}\sim \mu_H$ in the nonperturbative and resummation regions. Our profile functions $\mu_i(C)$ satisfy these constraints, and provide continuous and smooth transitions between these $C$ regions.  The resummed perturbative cross section is independent of ${\cal O}(1)$ variations in all renormalization scales order by order in the logarithmic resummation. Therefore the dependence on parameters appearing in the profile functions gets systematically smaller as we go to higher orders, and their variation provides us with a method of assessing perturbative uncertainties.

For the hard renormalization scale we use the \mbox{$C$-independent} formula
\begin{align}
\mu_H \,=\, e_H \, Q\,,
\end{align}
where $e_H$ is a parameter that we vary from $0.5$ to $2$ in order to account for theory uncertainties.

The soft scale has different functional dependence in the three regions of \Eq{eq:profile-constraints}, and hence depends on the following parameters:
\begin{align} \label{eq:muSprofile2}
  \mu_S = \mu_S(C, \mu_0,r_s,\mu_H,t_0,t_1,t_2,t_s).
\end{align}
Here, $\mu_0$ controls the intercept of the soft scale at $C=0$, $t_0$ is near the boundary of the purely nonperturbative region and $t_1$ controls the end of this transition, where the resummation region begins. The transition from nonperturbative to perturbative is $Q$ dependent, so we use the $Q$-independent parameters $n_0 \equiv t_0\, Q/(1$\,{\rm GeV}) and $n_1 \equiv t_1\,Q/(1$\,{\rm GeV}). In the resummation region the parameter $r_s$ determines the slope of the soft scale relative to the canonical resummation region scaling, with $\mu_S = r_s  \mu_H C/ 6 $. The parameter $t_2$ controls where the transition occurs between the resummation and fixed-order regions and $t_s$ sets the value of $C$ where the renormalization
scales all become equal as required in the fixed-order region. For the jet scale we have the dependence
\begin{equation} \label{eq:muJprofileshort}
\mu_J = \mu_J(\mu_H, \mu_S(C),e_J)\,,
\end{equation}
where $e_J$ is a parameter that is varied in our theory scans to slightly modify the natural relation between the scales.  The exact functional form for $\mu_S$ and $\mu_J$ in \Eqs{eq:muSprofile2}{eq:muJprofileshort} is given in Appendix~\ref{ap:profile}.

To avoid large logarithms in the soft function subtractions $\delta$, the scale $R(C)$ is chosen to be the same as $\mu_S(C)$ in the resummation and fixed-order regions. In the nonperturbative region we need $R(C)<\mu_S(C)$ to obtain an ${\cal O}(\alpha_s)$ subtraction that stabilizes the soft function in this region (removing unphysical negative dips that appear in the $\overline{\rm MS}$ scheme). This introduces an additional parameter $R_0 \equiv R(C=0)$. Therefore we have
\begin{align}
R = R(\mu_S(C), R_0).
\end{align}
The exact functional form for $R$ is also given in Appendix~\ref{ap:profile}.

For the nonsingular scale $\mu_{\rm ns}$, we use the variations
\begin{equation} \label{eq:muNSprofile}
\mu_{\rm ns}(C) = \left\{\! \begin{array}{ll}
\frac{1}{2} \big[\,\mu_H(C) + \mu_J (C)\,\big]\,, &~ n_s \,= ~~\,1 \\
\mu_H\,,  & ~n_s \,= ~~\,0 \\
\frac{1}{2} \big[\,3\,\mu_H(C) - \mu_J (C)\,\big]\,, &~ n_s \,= -\,1
\end{array}
\right.\!\!.
\end{equation}
Here the three choices vary $\mu_{\rm ns}$ in a manner that allows it to have some independence from $\mu_H$ in the resummation and nonperturbative regions, while still being equal $\mu_H$ in the fixed-order region (where $\mu_J=\mu_H$). These variations of $\mu_{\rm ns}$ probe the higher-order fixed-order uncertainty in the nonsingular cross section contribution. In the fixed-order region the variation of $\mu_H$ alone precisely reproduces the standard fixed-order scale variation. 

\begin{table}[tbh!]
\begin{tabular}{ccc}
	parameter\ & \ default value\ & \ range of values \ \\
	\hline
	$\mu_0$ & $1.1$\,GeV & - \\
	%  $1$ to $1.3$\,GeV
	$R_0$ & $0.7$\,GeV &  - \\
	%  $0.6$ to $0.9$\,GeV
	$n_0$ & $12$ & $10$ to $16$\\
	$n_1$ & $25$ & $22$ to $28$\\
	$t_2$ & $0.67$ & $0.64$ to $0.7$\\
	$t_s$ & $0.83$ & $0.8$ to $0.86$\\
	$r_s$ & $2$ & $1.78$ to $2.26$\\
	%  $e_S$ & $0$ & $-4$ to $4$\\
	$e_J$ & $0$ & $-\,0.5$ to $0.5$\\
	$e_H$ & $1$ & $0.5$ to $2.0$\\
	$n_s$ & $0$ & $-\,1$, $0$, $1$\\
	\hline
	$\Gamma^{\rm cusp}_3$ & $1553.06$ & $-\,1553.06$ to $+\,4659.18$ \\
	$s_2^{\widetilde C}$ & $-\,43.2$ & $-\,44.2$ to $-\,42.2$ \\
	$j_3$ & $0$ & $-\,3000$ to $+\,3000$ \\
	$s_3^{\widetilde C}$ & $0$ & $-\,500$ to $+\,500$ \\
	\hline
	$\epsilon^\text{low}_{2}$ & $0$ & $-\,1$, $0$, $1$ \\
	$\epsilon^\text{high}_{2}$ & $0$ & $-\,1$, $0$, $1$ \\
	$\epsilon^\text{low}_{3}$ & $0$ & $-\,1$, $0$, $1$ \\
	$\epsilon^\text{high}_{3}$ & $0$ & $-\,1$, $0$, $1$ \\
\end{tabular}
\caption{C-parameter theory parameters relevant for estimating the theory uncertainty, their
default values and range of values used for the theory scan during the fit
procedure. The last four parameters control the statistical errors induced by fit functions used
in the non-singular terms at $\mathcal{O}(\alpha_s^2)$ ($\epsilon^\text{low}_{2}$ and $\epsilon^\text{high}_{2}$) and $\mathcal{O}(\alpha_s^3)$
($\epsilon^\text{low}_{3}$ and $\epsilon^\text{high}_{3}$) in the region below ($\epsilon^\text{low}_{2}$ and $\epsilon^\text{low}_{3}$) and above
($\epsilon^\text{high}_{2}$ and $\epsilon^\text{high}_{3}$) the shoulder; see Sec. V of Ref.~\cite{Hoang:2014wka}.}
\label{tab:theoryerr}
\end{table}
\begin{table}[tbh!]
\begin{tabular}{ccc}
	parameter\ & \ default value\ & \ range of values \ \\
	\hline
	$\mu_0$ & $1.1$\,GeV & -\\
	$R_0$ & $0.7$\,GeV & -\\
	$n_0$ & $2$ & $1.5$ to $2.5$ \\
	$n_1$ & $10$ & $8.5$ to $11.5$\\
	$t_2$ & $0.25$ & $0.225$ to $0.275$\\
	$t_s$ & $0.4$ & $0.375$ to $0.425$\\
	$r_s$ & $2$ & $1.77$ to $2.26$\\
	%  $e_S$ & $0$ & $-4$ to $4$\\
	$e_J$ & $0$ & $-\,1.5$ to $1.5$\\
	$e_H$ & $1$ & $0.5$ to $2.0$\\
	$n_s$ & $0$ & $-\,1$, $0$, $1$\\
	\hline
	$j_3$ & $0$ & $-\,3000$ to $+\,3000$ \\
	$s_3^{\tau}$ & $0$ & $-\,500$ to $+\,500$ \\
	\hline
	$\epsilon_{2}$ & $0$ & $-\,1$, $0$, $1$ \\
	$\epsilon_{3}$ & $0$ & $-\,1$, $0$, $1$ \\
\end{tabular}
\caption{\label{tab:theoryerrthrust}Thrust theory parameters relevant for estimating the theory uncertainty, their
default values and range of values used for the theory scan during the fit
procedure. The last two parameters control the statistical errors induced by fit functions used
in the non-singular terms at $\mathcal{O}(\alpha_s^2)$ ($\epsilon_2$) and $\mathcal{O}(\alpha_s^3)$
($\epsilon_3$); see Sec. E of Ref.~\cite{Abbate:2010xh}.}
\end{table}

\begin{figure}[t!]
	\begin{center}
		\includegraphics[width=0.95\columnwidth]{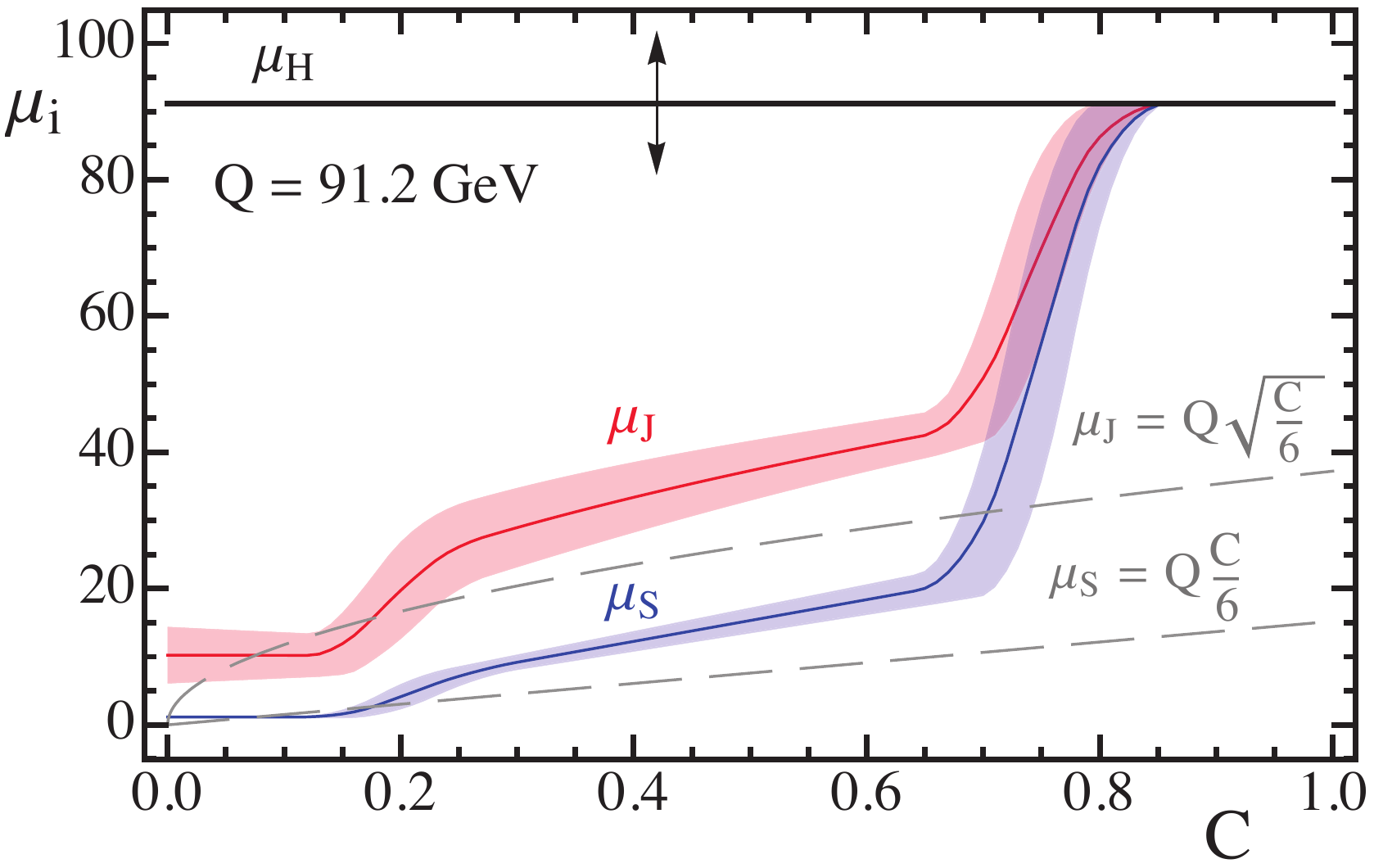}
		\caption{Bands for the profile functions for the renormalization scales
			$\mu_H$, $\mu_J(C)$, $\mu_S(C)$ when varying the profile parameters.}
		\label{fig:profile-variation}
	\end{center}
\end{figure}

The details of the variations of the profile function parameters used to assess uncertainties are given in Tab.~\ref{tab:theoryerr}. 
The plot in Fig.~\ref{fig:profile-variation} shows how the scales vary with the changes to our C-parameter profile parameters. The vertical arrow on the hard scale indicates the overall up/down variation, which causes a variation to all the scales.  Also shown (as gray dashed lines) are plots of the canonical soft scale $QC/6$ and canonical jet scale $Q\sqrt{C/6}$. In the resummation region, these correspond fairly well with the profile functions, indicating that in this region our analysis will avoid large logarithms. As discussed in detail in Ref.~\cite{Hoang:2014wka}, to improve the convergence of the C-parameter cross sections we take $r_s=2$ as our default slope parameter, explaining why our soft and jet scales are larger than the canonical values in Fig.~\ref{fig:profile-variation} (by a factor that does not induce further large logs).   In the analysis of Sec.~\ref{sec:results}, we will see how varying each of these profile parameters affects our final fit results.

For the numerical analyses carried out in this work we have created within our collaboration two completely independent codes. One code within Mathematica~\cite{mathematica} implements the theoretical expressions exactly as given in Ref.~\cite{Hoang:2014wka}, and another code is based on theoretical formulae in Fourier space and realized as a fast Fortran~\cite{gfortran} code suitable for parallelized runs on computer clusters. These two codes agree for the C-parameter distribution at the level of $10^{-6}$.

We will also repeat the thrust fits of Ref.~\cite{Abbate:2010xh}, implementing the same type of profile functions used here. These profiles have several advantages over those in Ref.~\cite{Abbate:2010xh}, including a free variable for the slope, a flat nonperturbative region, and parameters whose impact is much more confined to one of the three regions in \Eq{eq:profile-constraints}. For the thrust profiles we redefine \mbox{$r_s \to 6 \,r_s$}, which eliminates the factors of $6$ in \Eqs{eq:muSprofile}{eq:muRprofile}. This way, the canonical choice of slope is $r_s = 1$ for both C-parameter and thrust. We use $r_s=2$ as our default for thrust as well, again to improve the perturbative convergence, as discussed in  Ref.~\cite{Hoang:2014wka}. The profile parameters for thrust and their variations are summarized in \tab{theoryerrthrust}. These choices create profiles that are very similar to those used in  Ref.~\cite{Abbate:2010xh}. The new fit results for thrust are fully compatible with those of Ref.~\cite{Abbate:2010xh} in the resummation region used for the $\alpha_s$ fits. Additionally, they give a better description in the nonperturbative region which is outside of our fit range.

\subsection{Hadron-mass effects}\label{sec:hadron}
In Ref.~\cite{Mateu:2012nk} it was shown that hadron-mass effects, apart from breaking the universality properties of the leading power correction for various event shapes, also induce a nontrivial running. Since these are single logarithms, they start at NLL order. In Ref.~\cite{Mateu:2012nk} the corresponding leading anomalous dimension was determined, which yields the NLL resummation of larger logs between the scales $\mu_S$ and $\Lambda_{\rm QCD}$ for a large set of event shapes. The pieces necessary for a higher-order resummation have not yet been computed. One might be worried that accounting for only the NLL running for $\Omega_1$ in an expression as accurate as N$^3$LL in cross-section logs could be inadequate, or that it could leave significant perturbative uncertainties. However, one should recall that the hadronic parameter $\Omega_1$ itself is a correction, and hence it is valid to account for the related resummation with less precision.  In this section we show that the $\Omega_1$ evolution at NLL order suffices for the precision of our N$^3$LL$^\prime+{\cal O}(\alpha_s^3)$ analysis. Indeed, it turns out that the effect of the hadron mass running on the fit outcome is very small as compared to other uncertainties, and
therefore can be safely neglected. 

For our C-parameter analysis the implementation of hadron mass running effects has been explained at length in Ref.~\cite{Hoang:2014wka}, and we only summarize here the most relevant aspects needed to understand the fit results. 
In the $\overline{\rm MS}$ scheme the leading power correction can be written as an integral of a universal hadron function, $\overline\Omega_1(\mu,r)$, common to all event shapes
\begin{equation} \label{eq:omega1functionofg}
\overline\Omega_1^{\,e}(\mu) \,=\, c_e\! \int_0^1\! \df r \, g_e(r)\, \overline\Omega_1(\mu,r)\,,
\end{equation}
where $r$ is the transverse velocity, $e$ denotes a specific event shape, $c_e$ is a calculable normalization factor, and $g_e(r)$ is an event-shape-dependent function encoding the hadron-mass effects. The functions $g_e(r)$ are known analytically, and specific examples can be found in Ref.~\cite{Mateu:2012nk}. For C-parameter $c_C=3\pi$, while for thrust $c_\tau =2$. For the simple case of the $\overline{\rm MS}$ scheme the running between the initial reference scale $\mu_0$ where the universal hadron function is specified,
 and the soft scale $\mu_S$, is given at leading order by
\begin{equation} \label{eq:omega1-r-murunning}
\overline \Omega_1(\mu_S,r) = \overline \Omega_1(\mu_0,r)\!
\left[ \frac{\alpha_s(\mu_S)}{\alpha_s(\mu_0)} \right]^{\hat\gamma_1(r)},
\end{equation}
with $\hat\gamma_1(r) = \frac{2 C_{\!A}}{\beta_0} \ln (1-r^2)$.
The corresponding evolution formula for the Rgap scheme is considerably more complex, as shown by the form displayed in \Eq{eq:hadron-masses} above. Ensuring that the renormalon is not reintroduced by the renormalization group evolution requires an additional evolution in the scale $R$, so $\Delta^{\rm diff}(R_\Delta,R,\mu_\Delta,\mu,r)$ contains evolution in both the $\mu$ and $R$ scales. Also here we have two reference scales $\mu_\Delta$ and $R_\Delta$ to specify the initial parameter $\Omega_1(R_\Delta,\mu_\Delta,r)$. The full formula for $\Delta^{\rm diff}$ is given in Eq.~(67) of Ref.~\cite{Hoang:2014wka}. 

Note that the renormalization group evolution is a function of $r$ and takes place independently for each $r$, but the required result for C-parameter or thrust requires an integral over $r$. Due to this integration the functional form that we assume for the initial condition $\overline\Omega_1(r,\mu_0)$ or $\Omega_1(R_\Delta,\mu_\Delta,r)$ gets entangled with the perturbative resummation.

\begin{figure}[t!]
	\begin{center}
		\includegraphics[width=0.95\columnwidth]{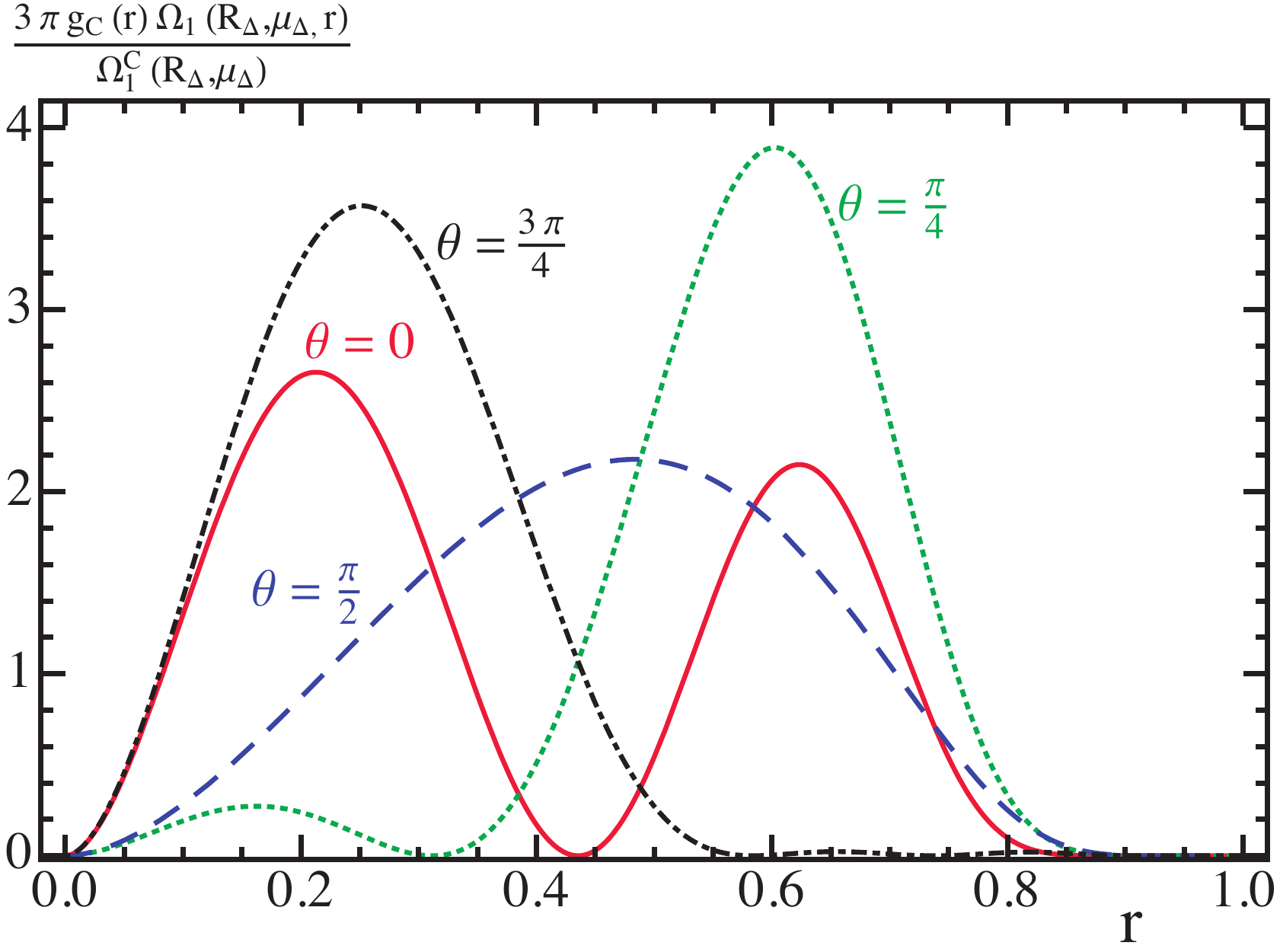}
		\caption{Plots of the $r$ dependence of $g_C(r)\, \Omega_1(R_\Delta,\mu_\Delta,r)$ for different values of $\theta(R_\Delta,\mu_\Delta)$. We normalize to $\Omega_1^C(R_\Delta,\mu_\Delta)$, since it is simply an overall multiplicative factor.}
		\label{fig:Omega1-different-thetas}
	\end{center}
\end{figure}

With current constraints on the $r$ dependence, and with the lack of even more precise experimental data to probe this issue, a model-independent formulation (like a complete functional basis for the $r$ dependence at the reference scales, $R_\Delta$ and $\mu_\Delta$) is not feasible. To implement this running we have therefore tested several models for the $r$ dependence in Ref.~\cite{Hoang:2014wka}, and found that generically the experimental data is sensitive to the normalization which specifies $\Omega_1$, but not to the detailed form used for the $r$ dependence as long as it satisfies several reasonable constraints. Therefore for the fits performed here we simply adopt 
the default form from Ref.~\cite{Hoang:2014wka},
\begin{align} \label{eq:omega1-ansatz}
\Omega_1(R_\Delta,\mu_\Delta,r) &= [\, a(R_\Delta,\mu_\Delta) f_a(r) + b(R_\Delta,\mu_\Delta) f_b(r) \,]^2, \nn \\
f_a(r) &= 3.510 \, e^{-\frac{r^2}{1-r^2}}, \\
f_b(r) &= 13.585 \, e^{-\frac{2\,r^2}{1-r^2}} - 21.687 \,\, e^{-\frac{4\,r^2}{1-r^2}}. \nn
\end{align}
This model ensures that $\Omega_1(R_\Delta,\mu_\Delta,r)$ is always positive definite and smoothly goes to zero
at the $r=1$ endpoint (where the ratio of the hadron mass to $p_T$ goes to zero). The functions $f_{a,b}$ form an orthonormal basis upon integration with $g_C(r)$, which yields
the following relation:
\begin{align} \label{eq:omegafunctionofaandb}
\Omega_1^C(R_\Delta,\mu_\Delta) & = 3\pi\, [\,a(R_\Delta,\mu_\Delta)^2 + b(R_\Delta,\mu_\Delta)^2\,]\,,
\end{align}
which determines the normalization. We also define
\begin{align} \label{eq:thetadef}
 \theta(R_\Delta,\mu_\Delta) & \equiv \arctan\left(\frac{b(R_\Delta,\mu_\Delta)}{a(R_\Delta,\mu_\Delta)}\right),
\end{align}
which was chosen to have an effect orthogonal to the more relevant parameter
$\Omega_1^C(R_\Delta,\mu_\Delta)$. By examining our ability to simultaneously measure $\Omega_1^C(R_\Delta,\mu_\Delta)$ and $\theta(R_\Delta,\mu_\Delta)$ we have a means to test for the impact that the initial $r$ dependence has on our fits. As we can see in \Fig{fig:Omega1-different-thetas}, our model captures different behavior for the $r$ dependence of $\Omega_1(R_\Delta,\mu_\Delta,r)$ by choosing different values of $\theta(R_\Delta,\mu_\Delta)$.  Over the interval $r\in [0,1]$, all the curves in this figure are normalized so that they integrate to $1$.

\section{Experimental Data}
\label{sec:data}
Data on the C-parameter cross section are given by several experiments for a 
range of $Q$ from $35$ to $207$ GeV.  We use data from ALEPH\,\footnote{The ALEPH dataset with $Q=91.2\,$GeV has  two systematic uncertainties for each bin. The second of
these uncertainties is treated as correlated while the first one is treated as an uncorrelated uncertainty
and simply added in quadrature to the statistical uncertainty.} with $Q = \{91.2$,
$133$, $161$, $172$, $183$, $189$, $200$, $206\}$ GeV \cite{Heister:2003aj},
DELPHI with $Q = \{45$, $66$, $76$, $89.5$, $91.2$, $93$, $133$, $161$, $172$, $183$, 
$189$, $192$, $196$, $200$, $202$, $205$, $207\}$ GeV \cite{Abdallah:2004xe, 
Abreu:1996mk, Abreu:1999rc, Wicke:1999zz}, JADE with $Q=\{35$, $44\}$ GeV 
\cite{Biebel:1999zt}, L3 with $Q=\{91.2$, $130.1$, $136.1$, $161.3$, $172.3$, $182.8$, 
$188.6$, $194.4$, $200.2$, $206.2\}$ GeV \cite{Achard:2004sv, Adeva:1992gv}, OPAL with 
$Q=\{91$, $133$, $177$, $197\}$ GeV \cite{Abbiendi:2004qz}, and SLD with $Q=91.2$ GeV \cite{Abe:1994mf}. As each of these datasets is given in binned form, our cross section in Eq.~(\ref{eq:singular-nonperturbative-gap}) is integrated over each bin before being compared to the data.
The default range on $C$ used in fitting the data is $25\,\text{ GeV}/Q \le C \le 0.7$. A lower limit of $25\text{ GeV}/Q$  eliminates the peak region where higher nonperturbative moments $\Omega_{n>1}^C$  become important. The upper limit is  chosen to be 0.7 in order to avoid the far-tail region as well as the Sudakov shoulder at  $C=0.75$. Any bin that contains one of the end points of our range ($C= 25\, \text{GeV}/Q$  or 0.7) is included if more than half of that bin lies within the range. Using the default range and datasets gives a total of $404$ bins. As a further test of the stability of our analysis, both this C-parameter range and the selection of datasets is varied in the numerical analysis  contained in Sec.~\ref{sec:results}. 

\begin{figure}[t!]
\begin{center}
\subfigure[{}]{
\includegraphics[width=0.95\columnwidth]{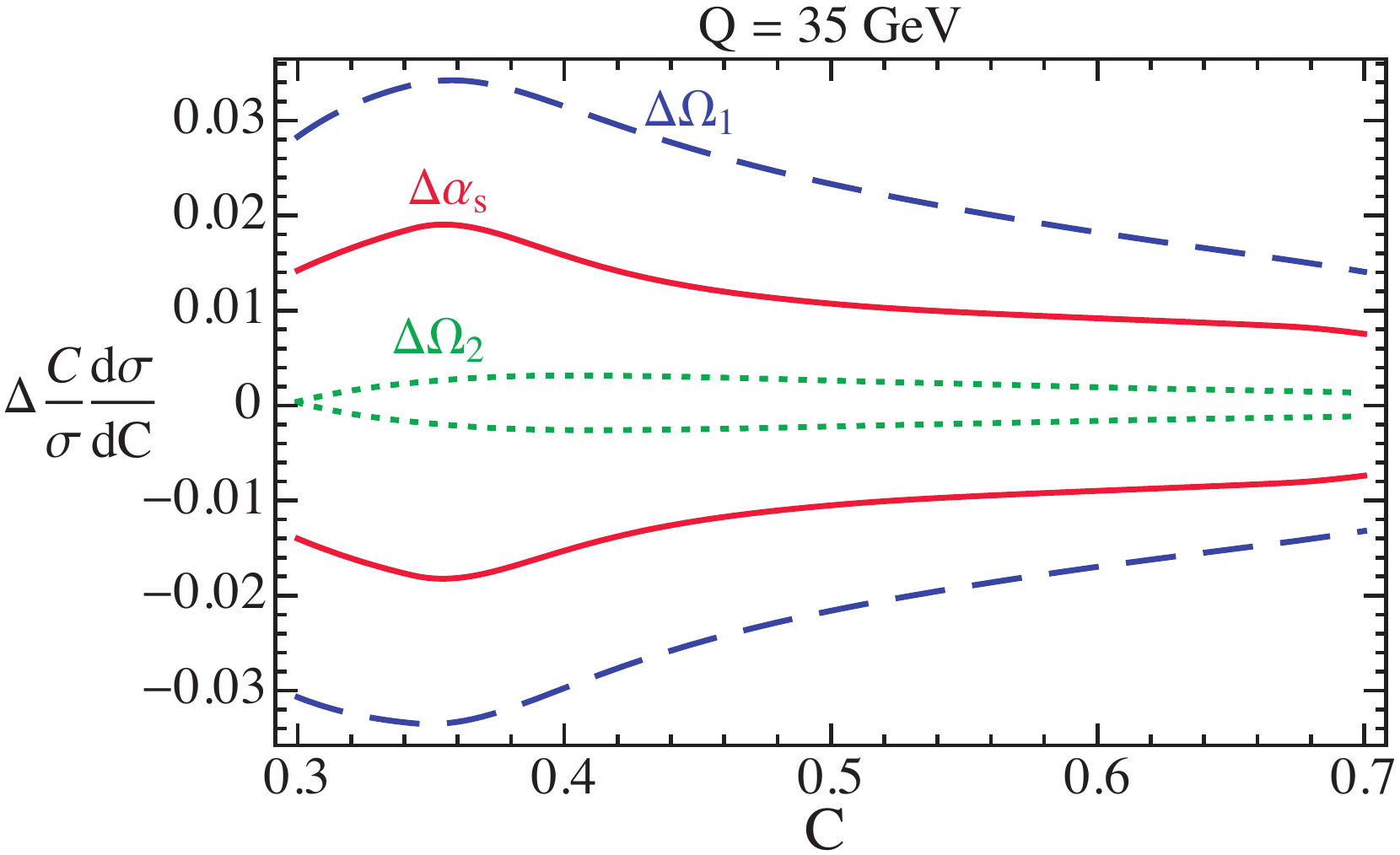}
}

\subfigure[{}]{
\includegraphics[width=0.95\columnwidth]{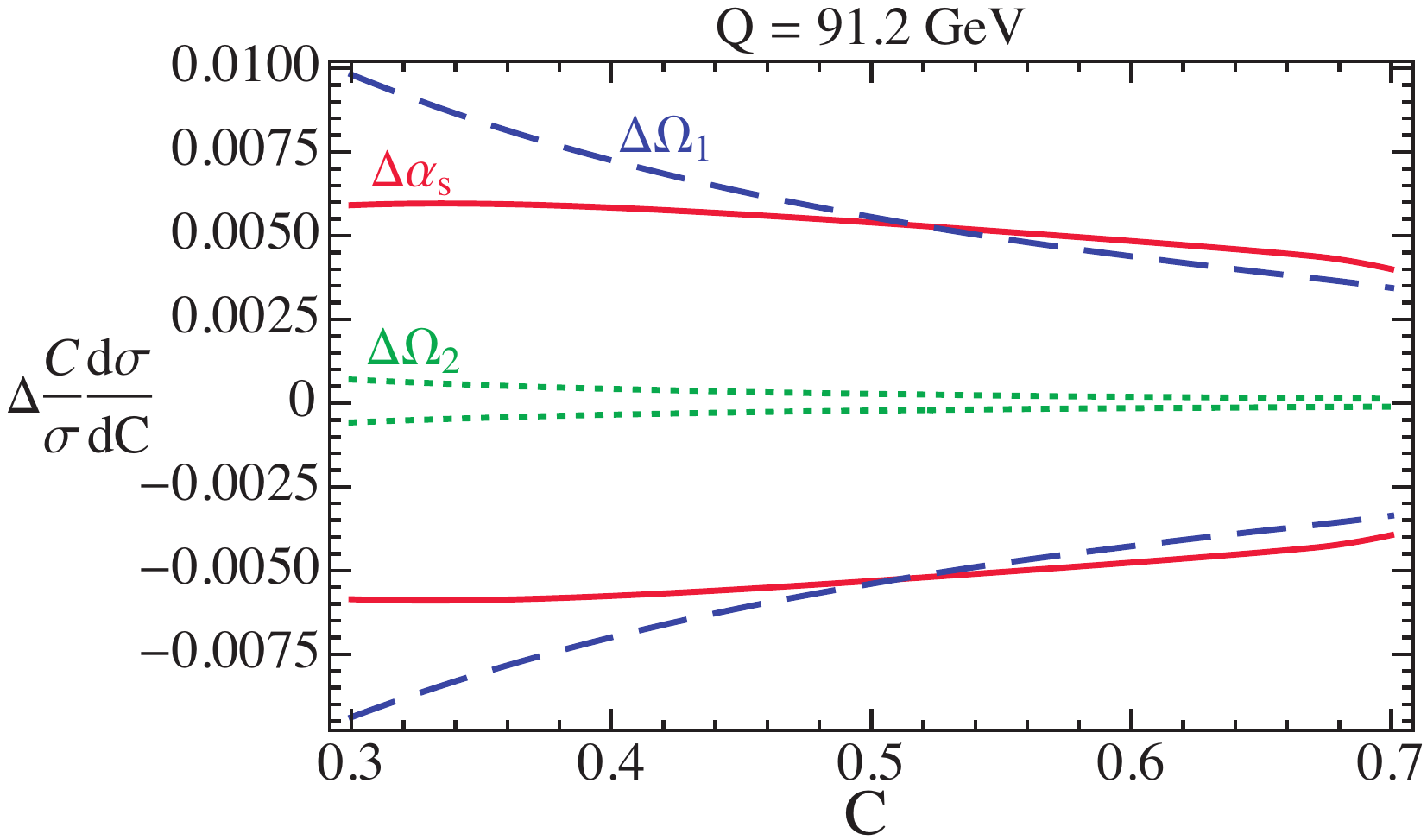}
}
\subfigure[{}]{
\includegraphics[width=0.95\columnwidth]{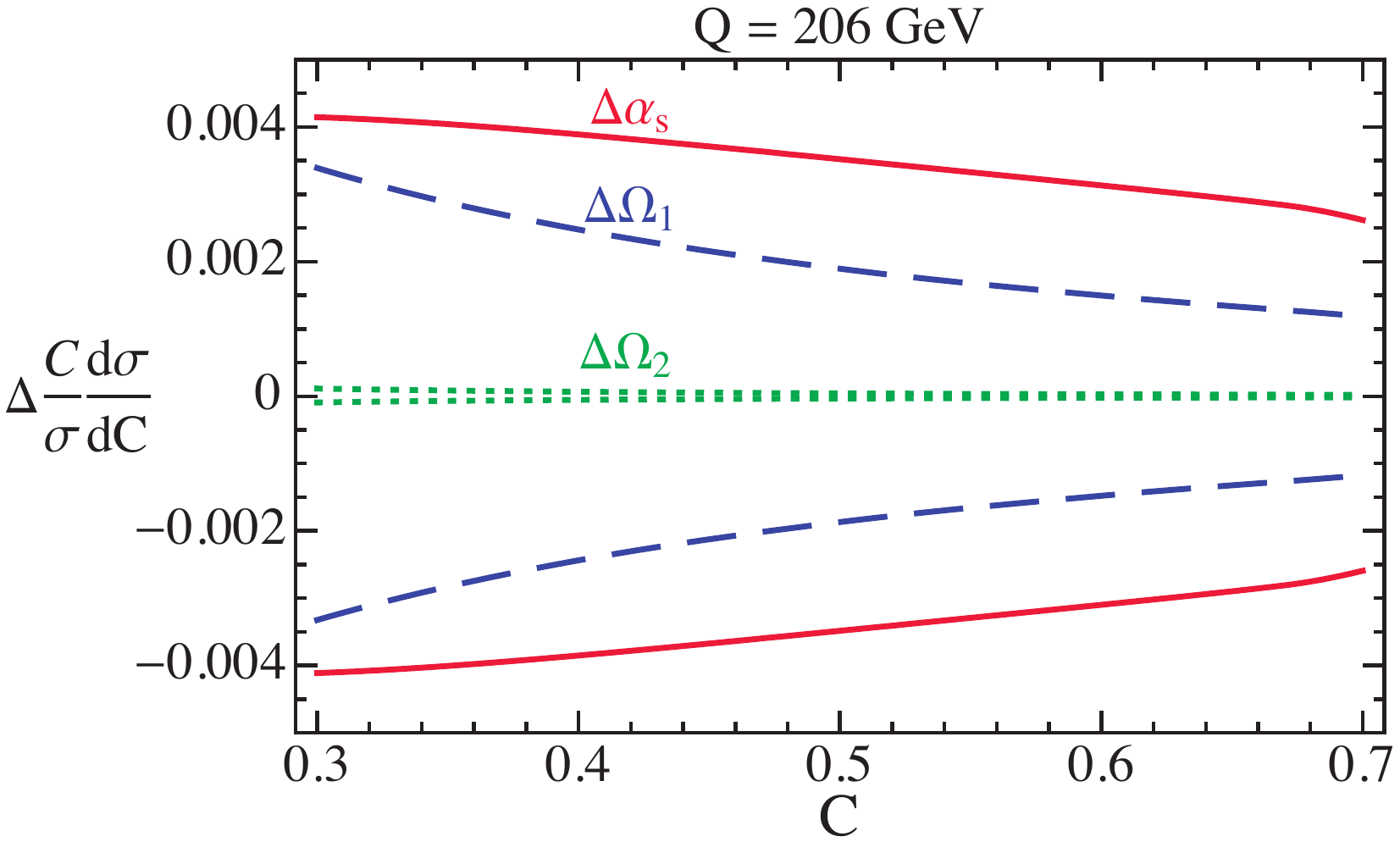}
}
\caption{Difference between the default cross section and the cross section varying only one parameter.
We vary $\alpha_s (m_Z)$ by $\pm\, 0.001$ (solid red), $2\,\Omega_1$ by $\pm \,0.1$ (dashed blue) and
$\Omega_2^C$ by $\pm\,0.5$ (dotted green). The three plots correspond to three different center
of mass energies: (a)~$Q=35$\,GeV, (b)~$Q=91.2$\,GeV, (c)~$Q=206$\,GeV.}
\label{fig:Degeneracy}
\end{center}
\end{figure}
In our fitting procedure, we consider both the statistical and systematic experimental uncertainties. The statistical uncertainties can be treated as independent between bins. The systematic experimental uncertainties come from various sources and full documentation of their correlations are not available, so dealing with them in our $\chi^2$ analysis is more complicated, and we have to use a correlation model. For this purpose we follow the LEP QCD working group~\cite{Heister:2003aj,Abbiendi:2004qz} and use the minimal overlap model. Within one C-parameter dataset, which consists of various C-parameter bins at one $Q$ value for one experiment, we take for the bin $i$, bin $j$ off-diagonal entry of the experimental covariance matrix $[{\rm min}(\Delta_i^{\rm sys},\Delta_j^{\rm sys})]^2$. Here $\Delta_{i,j}^{\rm sys}$ are the quoted systematic uncertainties of the bins $i$ and $j$. Within each dataset, this model implies a positive correlation of systematic uncertainties. In addition to this default model choice, we also do the fits assuming uncorrelated systematic uncertainties, in order to test whether the minimal overlap model introduces any bias. See Sec.~\ref{sec:random} for more details on the correlation matrix.

\section{Fit Procedure} \label{sec:fitprocedure}

In order to accurately determine both $\alpha_s(m_Z)$ and the leading power correction in the same fit, it is important to perform a global analysis, that is, simultaneously fitting $C$-spectra for a wide range of center-of-mass energies $Q$. For each $Q$, effects on the cross sections induced by changes in $\alpha_s(m_Z)$ can be partly compensated by changes in $\Omega_1$, resulting in a fairly strong degeneracy. This is resolved by the global fit, just as in the thrust analysis of Ref.~\cite{Abbate:2010xh}. Fig.~\ref{fig:Degeneracy} shows the difference between the theoretical prediction for the cross section at three different values of $Q$, when
$\alpha_s(m_Z)$ or $\Omega_1$ are varied by $\pm\,0.001$ and $\pm\,0.05$ GeV, respectively. It is clear that the potential degeneracy in these parameters is broken by having data at multiple $Q$ values. In Fig.~\ref{fig:Degeneracy}  we also vary the higher-order power correction parameter $\Omega^C_2$, which clearly has a much smaller effect than the dominant power correction parameter $\Omega_1$.

To carry out a fit to the experimental data we fix the profile and theory parameters to the values shown in \tab{theoryerr}.  The default values are used for our primary theory cross section.  We integrate the resulting theoretical distribution over the same C-parameter bins as those available experimentally, and construct a $\chi^2$ function with the uncorrelated statistical experimental uncertainties and correlated systematic uncertainties. This $\chi^2$ is a function of $\alpha_s(m_Z)$ and $\Omega_1$, and is very accurately described by a quadratic near its global minimum, which therefore determines the central values and experimental uncertainties. The value of $\Omega_1$ and its associated uncertainties encode the dominant hadronization effect as well as the dominant residual uncertainty from hadronization.  

To obtain the perturbative theoretical uncertainty we consider the range of values shown for the theory parameters in \tab{theoryerr}. Treating each of these as a flat distribution, we randomly generate values for each of these parameters and then repeat the fit described above with the new $\chi^2$ function.  This random sampling and fit is then repeated 500 times. We then construct the minimum ellipse that fully contains all 500 of the central-fit values by first creating the convex envelope that contains all of these points within it. Then, we find the equation for the ellipse that best fits the points on the envelope, with the additional restrictions that all values lie within the ellipse and its center is the average of the maximum and minimum values in each direction. This ellipse determines the perturbative theoretical uncertainty, which turns out to be the dominant uncertainty in our fit results. In our final results the perturbative and experimental uncertainties are added in quadrature.  This procedure is similar to that discussed in the Appendix of Ref.~\cite{Abbate:2012jh}.

\section{Results}
\label{sec:results}
In this section we discuss the results from our global analysis. We split the presentation into several subsections. In Sec.~\ref{sec:resumpower} we discuss the impact that resummation and the inclusion of power corrections have on the fit results. 
In Sec.~\ref{sec:random} we present the analysis which yields the perturbative uncertainty in detail, cross-checking our method by analyzing the order-by-order convergence.  We also analyze the impact of removing the renormalon.  In Sec.~\ref{sec:exptfit} we discuss the experimental uncertainties obtained from the fit.  Section~\ref{sec:up-down} discusses the impact that varying the theory parameters one by one has on the best-fit points, allowing us to determine which parameters dominate the theoretical uncertainty. The impact of hadron-mass resummation is discussed in detail in \Sec{sec:hadmassresum}. We examine the effects of changing the default dataset in \Sec{sec:dataset}. The final fit results are collected in Sec.~\ref{sec:final}. When indicating the perturbative precision, and whether or not the power correction $\Omega_1$ is included and at what level of precision, we use the following notation:
\begin{align}
& {\cal O}(\alpha_s^k) 
& \phantom{x} & \text{fixed order up to ${\cal O}(\alpha_s^k)$} 
\nn\\ 
& \text{N}^k\text{LL}^\prime \!+\! {\cal O}(\alpha_s^k) 
& \phantom{x} & \text{perturbative resummation} 
\nn\\ 
& \text{N}^k\text{LL}^\prime \!+\! {\cal O}(\alpha_s^k)  \!+\! {\overline \Omega}_1
& \phantom{x} & \text{$\overline{\rm MS}$ scheme for $\Omega_1$} 
\nn\\ 
& \text{N}^k\text{LL}^\prime \!+\! {\cal O}(\alpha_s^k) 
\!+\!  {\Omega}_1(R,\mu)
& \phantom{x} & \text{Rgap scheme for $\Omega_1$} 
\nn\\ 
& \text{N}^k\text{LL}^\prime \!+\! 
{\cal O}(\alpha_s^k) \!+\! {\Omega}_1(R,\mu,r)
& \phantom{x} & \text{Rgap scheme with }
\nn\\
& & \phantom{x} & \  \text{hadron masses for $\Omega_1$} 
\,. \nn
\end{align}

\begin{figure*}[t!]
	\begin{center}
		\includegraphics[width=1.5\columnwidth]{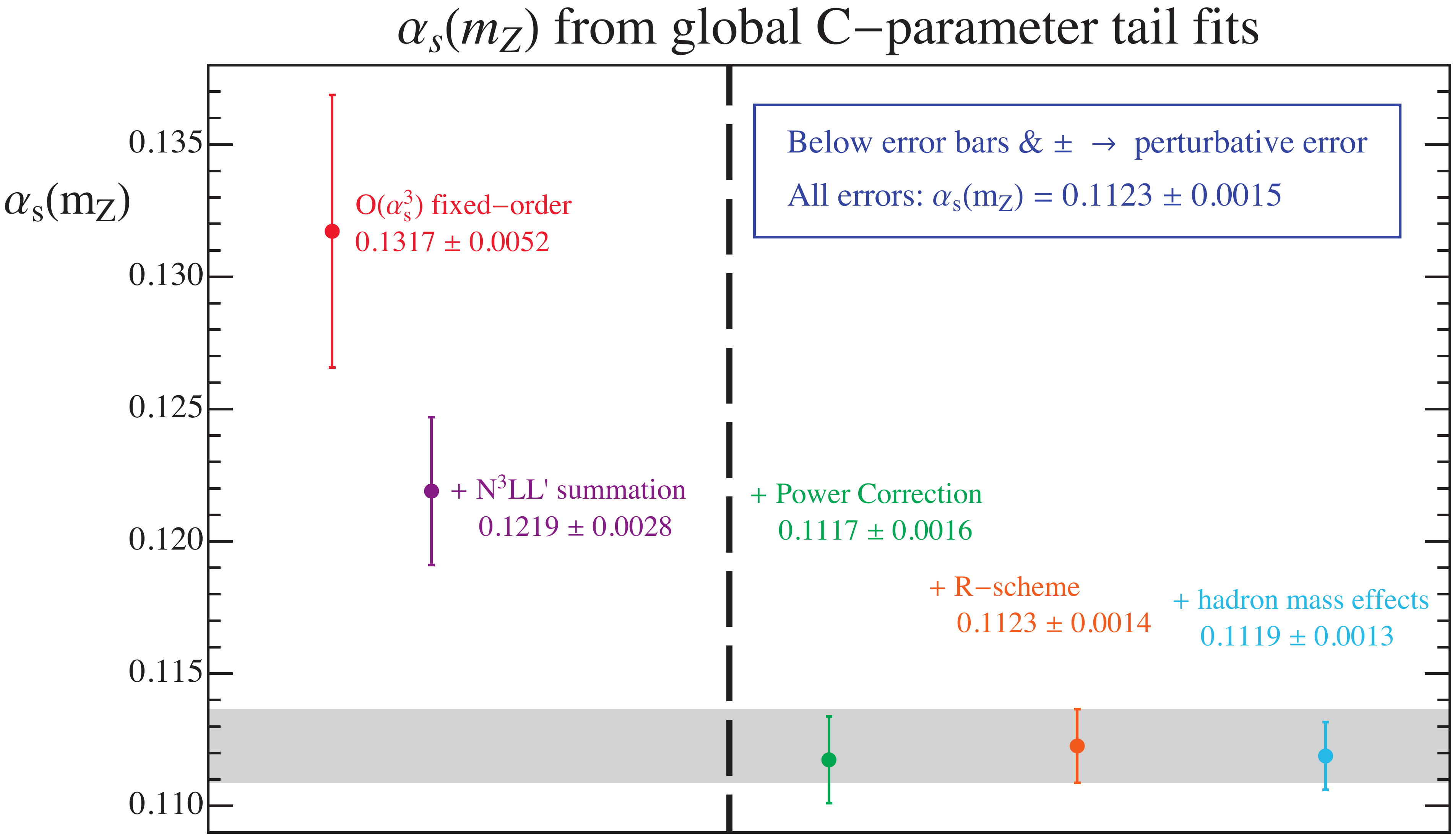}
		\caption{The evolution of the value of $\alpha_s(m_Z)$ adding components of the calculation. An additional $\sim 8\%$ uncertainty from not including power corrections is not included in the two left points. }
		\label{fig:alpha-evolution-C}
	\end{center}
\end{figure*}

\subsection{Impact of Resummation and Power Corrections} 
\label{sec:resumpower}
In Fig.~\ref{fig:alpha-evolution-C} we show $\alpha_s(m_Z)$ extracted
from fits to the tail of the C-parameter distribution including sequential improvements to the treatment of perturbative and nonperturbative components of our code, using the highest perturbative accuracy at each stage. The sequence from left to right shows the fit results using:
${\cal O}(\alpha_s^3)$ fixed-order results only, adding N$^3$LL resummation, adding the $\overline\Omega_1$ power correction, adding renormalon
subtractions and using the Rgap power correction parameter $\Omega_1(R_\Delta,\mu_\Delta)$, and adding hadron-mass effects. These same results together with the corresponding $\chi^2/{\rm dof}$ are also collected in Tab.~\ref{tab:lesser}.  
The fit with only fixed-order ${\cal O}(\alpha_s^3)$ results has a relatively large $\chi^2/{\rm dof}$ and also its central value has the largest value of $\alpha_s(m_Z)$.  Including the resummation of large logarithms decreases the central $\alpha_s(m_Z)$ by 8\% and also decreases the perturbative uncertainty by $\sim 50\%$. Due to this smaller perturbative uncertainty it becomes clear that the theoretical cross section has a different slope than the data, which can be seen, for example, at $Q=m_Z$ for $0.27 < C< 0.35$. This leads to the increase in the $\chi^2/{\rm dof}$ for the ``N$^3$LL$^\prime$ no power corr.''~fit, and makes it quite obvious that power corrections are needed. When the power correction parameter $\Omega_1$ is included in the fit, shown by the third entry in Tab.~\ref{tab:lesser} and the result just to the right of the vertical dashed line in Fig.~\ref{fig:alpha-evolution-C}, the $\chi^2/{\rm dof}$ becomes $1.004$ and this issue is resolved. Furthermore, a reduction by $\sim 50\%$ is achieved for the perturbative uncertainty in $\alpha_s(m_Z)$. This reduction makes sense since some of the perturbative uncertainty of the cross section is now absorbed in $\Omega_1$, and a much better fit is achieved for any of the variations associated to estimating higher-order corrections. The addition of $\Omega_1$ also caused the fit value of $\alpha_s(m_Z)$ to drop by another 8\%, consistent with our expectations for the impact of power corrections and the estimate made in Ref.~\cite{Hoang:2014wka}. 
Note that the error bars of the first two purely perturbative determinations, shown at the left-hand side of the vertical thick dashed line in Fig.~\ref{fig:alpha-evolution-C} and in the last two entries in \tab{lesser}, do not include the $\sim 8\%$ uncertainties associated with the lack of power corrections.

The remaining corrections we consider are the  use of the R-scheme for  $\Omega_1$ which includes the renormalon subtractions, and the inclusion of the log-resummation effects associated to the hadron-mass effects.  Both of these corrections have a fairly small impact on the determination of $\alpha_s(m_Z)$, shifting the central value by $+0.5\%$ and $-\,0.3\%$ respectively. Since adding the $-\,0.3\%$ shift from the hadron mass corrections in quadrature with the $\simeq 1.2\%$ perturbative uncertainty does not change the overall uncertainty we will use the R-scheme determination for our main result. This avoids the need to fully discuss the extra fit parameter $\theta(R_\Delta,\mu_\Delta)$ that appears when hadron masses are included. Further discussion of the experimental uncertainties and the perturbative uncertainty from the random scan are given below in \Secs{sec:random}{sec:up-down}, and a more detailed discussion of the impact of hadron-mass resummation is given below in \Sec{sec:hadmassresum}.

The values of $\Omega_1$ obtained from the fits discussed above can be directly compared to the $\Omega_1$ power correction obtained from the thrust distribution. Values for $\Omega_1$ from the C-parameter fits are given below in \Secs{sec:random}{sec:up-down} and the comparison with thrust is considered in \Sec{sec:universality}.

\begin{table}[t!]
	\begin{tabular}{l|c c}
		& $\alpha_s(m_Z)$ & $\chi^2/{\rm dof}$ \\
		\hline
		N$^3$LL$^\prime$ + hadron                          & ~~$0.1119(13)(06)$ & $0.991$\\
		N$^3$LL$^\prime$ with $\Omega_1(R,\mu)$        & ~~$0.1123(14)(06)$ & $0.988$\\
		N$^3$LL$^\prime$ with $\overline\Omega_1$ & ~~$0.1117(16)(06)$ & $1.004$\\
		N$^3$LL$^\prime$ no power corr.\,\,                & ~~$0.1219(28)(02)$& $2.091$\\
		\!\!\parbox{20ex}{${\cal O}(\alpha_s^3)$ fixed order\\[2pt]
			no power corr.} & ~~$0.1317(52)(03)$ & $1.486$
	\end{tabular}
	\caption{
		Comparison of C-parameter tail fit results for analyses when we add various components of the theoretical result (from the bottom to top). The first parentheses gives the theory uncertainty, and the second is the experimental and hadronic uncertainties added in quadrature for the first three rows, and experimental uncertainty for the last two rows.
		\label{tab:lesser}}
\end{table}

\subsection{Perturbative Uncertainty from the Scan}
\label{sec:random}
\begin{figure*}[t!]
\subfigure[]{
\includegraphics[width=0.485\textwidth]{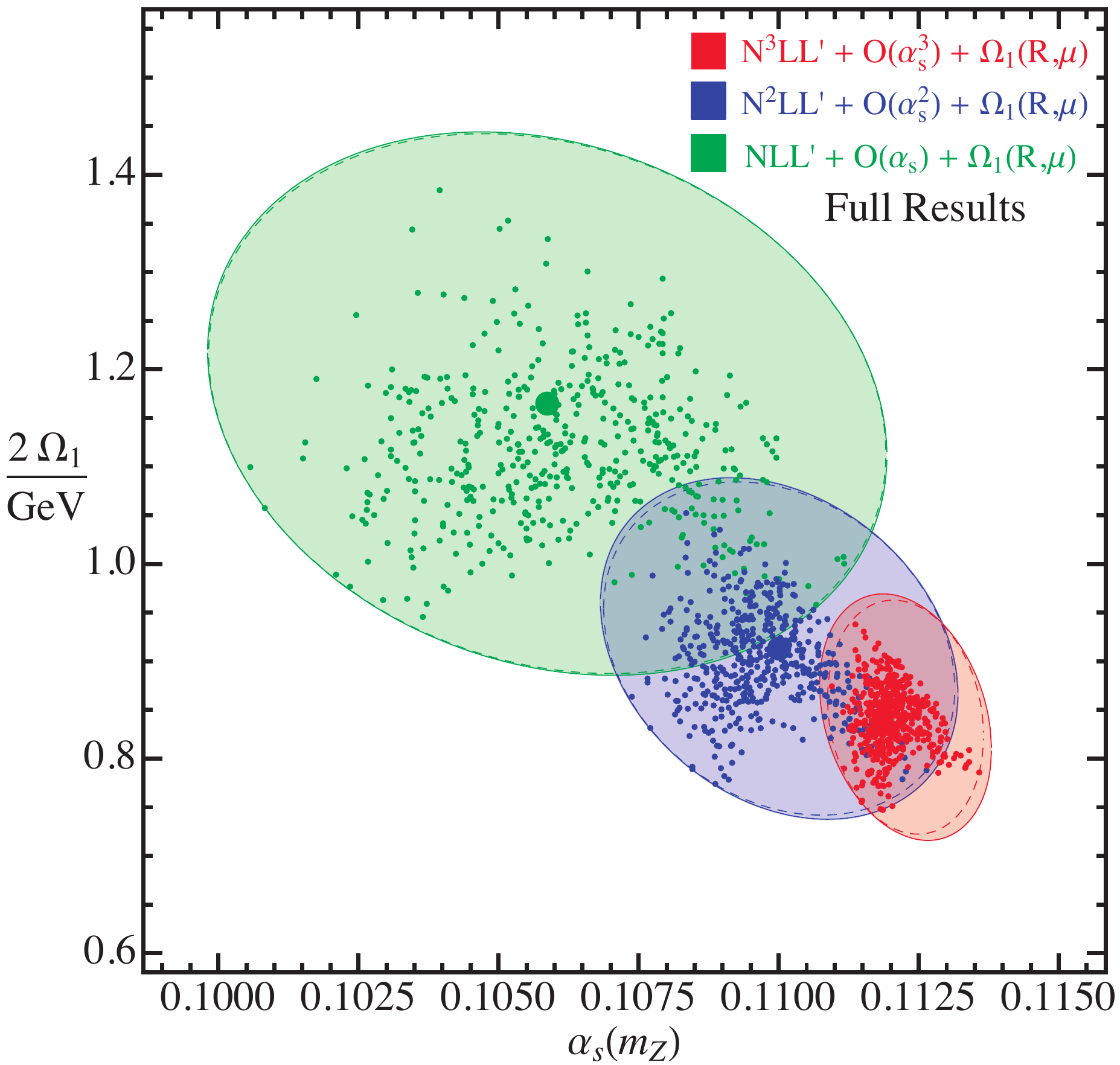}
\label{fig:alphagap}
}
\subfigure[]{
\includegraphics[width=0.485\textwidth]{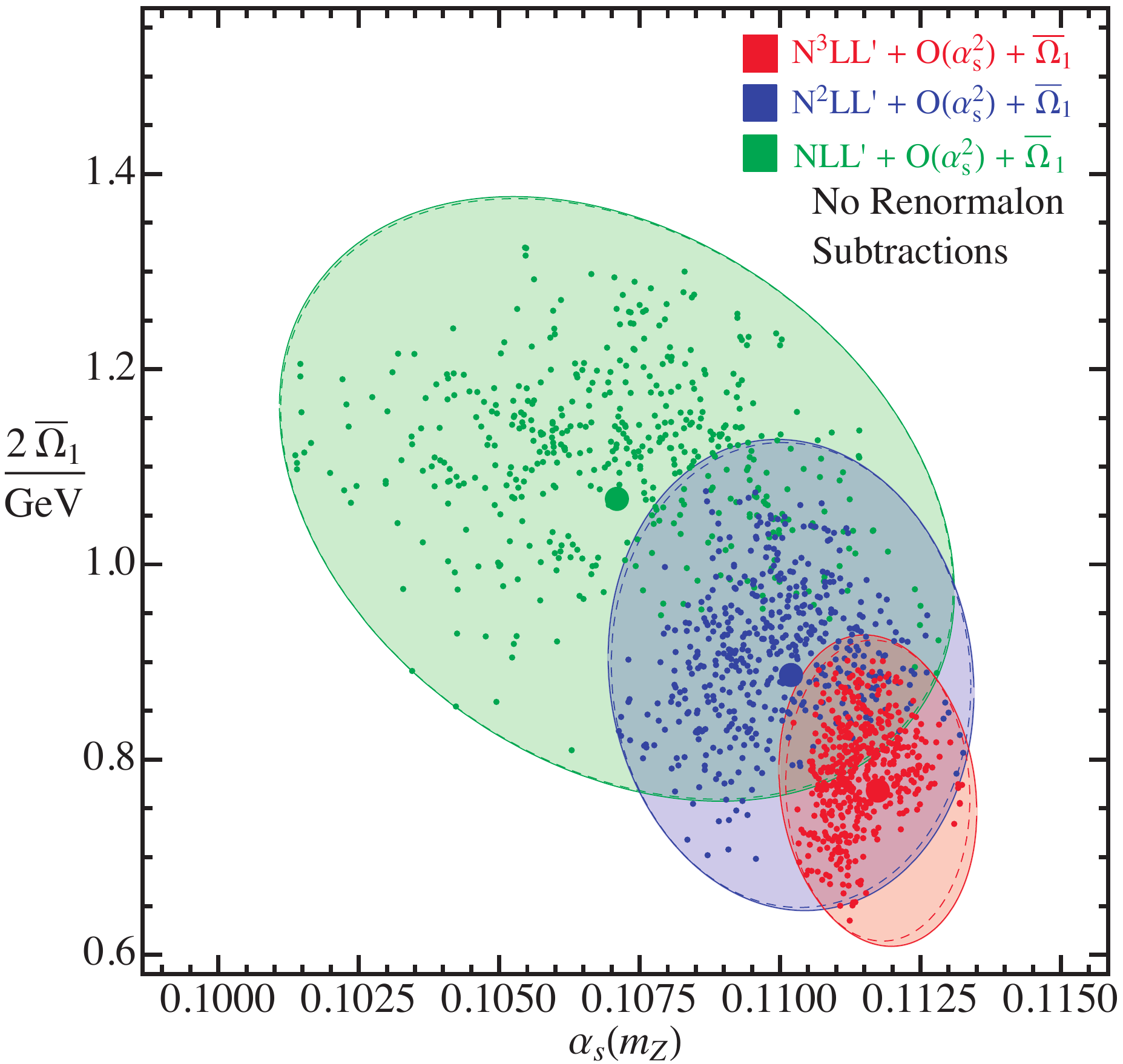}
\label{fig:alphanogap}
}
\subfigure[]{
\includegraphics[width=0.485\textwidth]{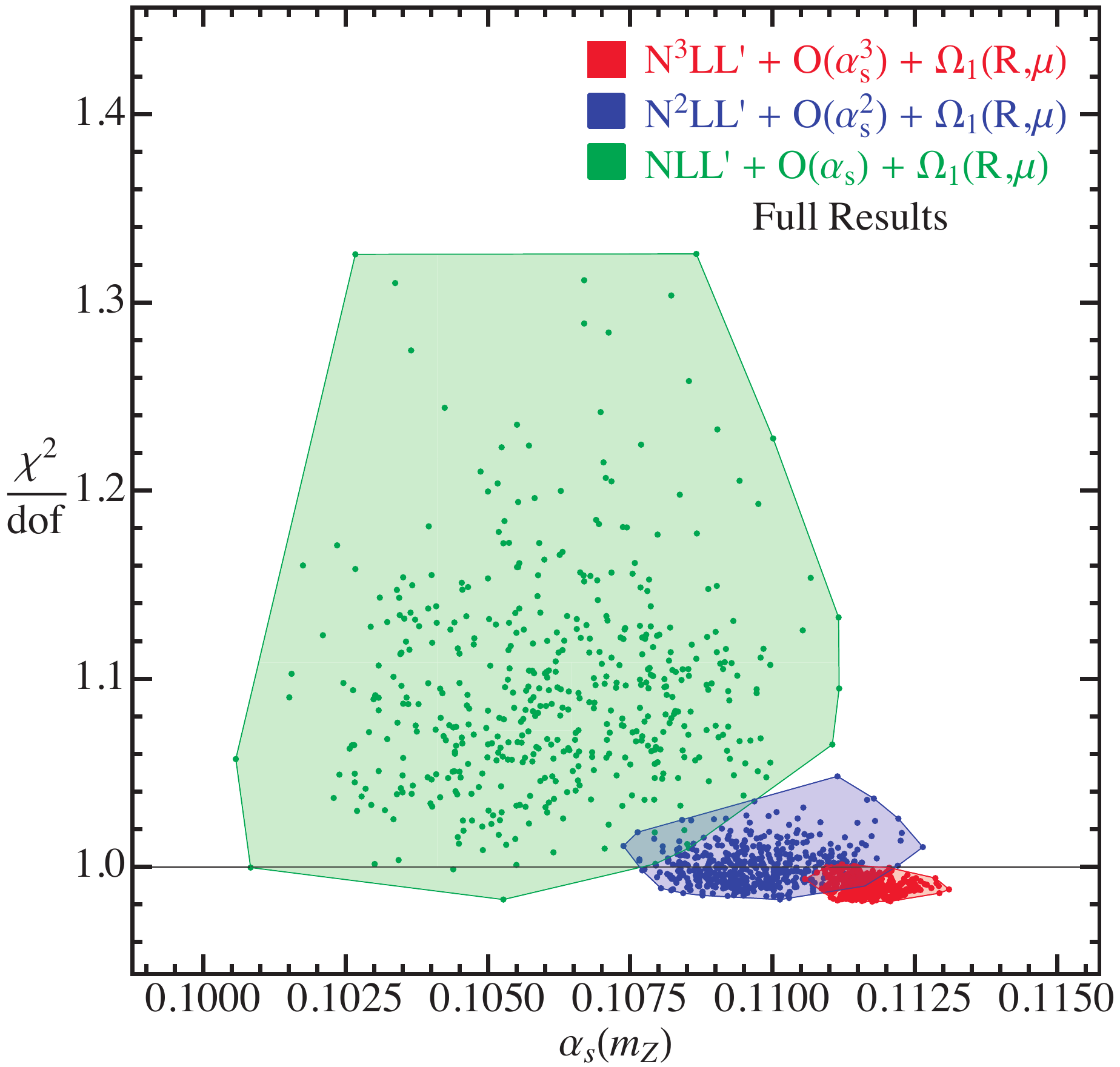}
\label{fig:chi2alphagap}
}
\subfigure[]{
\includegraphics[width=0.485\textwidth]{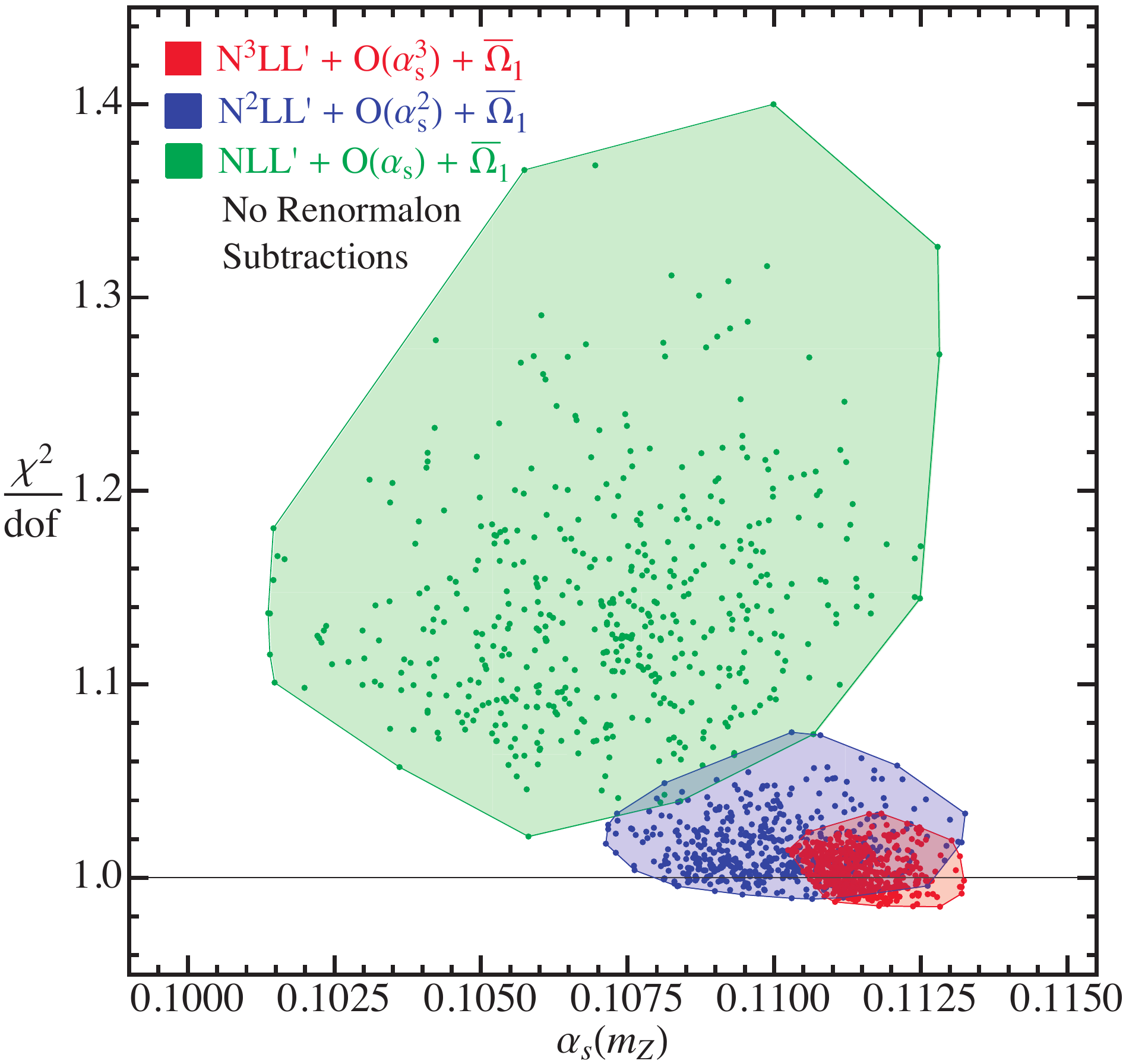}
\label{fig:chi2alphanogap}
}
\vspace{-0.2cm}
\caption{The first two panels show the distribution of best-fit points in the $\alpha_s(m_Z)$-$2\Omega_1$ and
$\alpha_s(m_Z)$-$2\overline\Omega_1$ planes. Panel~(a) shows results including perturbation theory, resummation of large
logs, the soft nonperturbative function and $\Omega_1$ defined in the Rgap scheme with renormalon subtractions.
Panel~(b) shows the results as in panel~(a), but with $\overline\Omega_1$ defined in the $\msbar$ scheme without renormalon
subtractions. In both panels the dashed lines corresponds to an ellipse fit to the contour of the best-fit points to
determine the theoretical uncertainty. The respective total (experimental\,+\,theoretical) 39\% CL standard uncertainty
ellipses are displayed (solid lines), which correspond to $1$-$\sigma$ (68\% CL) for either one-dimensional projection.
The big points represent the central values in the random scan for $\alpha_s(m_Z)$ and $2\,\Omega_1$.
Likewise, the two panels at the bottom show the distribution of best-fit points in the $\alpha_s(m_Z)$-$\chi^2/{\rm dof}$
plane. Panel~(c) shows the $\chi^2/{\rm dof}$ values of the points given in 
panel~(a), whereas panel~(b) shows the $\chi^2/{\rm dof}$ values of the points given in panel~(b).
\label{fig:alpha} }
\end{figure*}

To examine the robustness of our method of determining the perturbative uncertainty by the random scan, we consider the convergence and overlap of the results at different perturbative orders. Figure~\ref{fig:alpha} shows the spread of best-fit values at NLL$^\prime$, N$^2$LL$^\prime$ and N$^3$LL$^\prime$.
The upper left panel, Fig.~\ref{fig:alphagap}, shows results
from fits performed in the Rgap scheme, which implements a renormalon subtraction for $\Omega_1$,
and the upper right-panel, Fig.~\ref{fig:alphanogap}, shows results in the $\msbar$
scheme without renormalon subtractions. Each point in the plot represents the
outcome of a single fit, and different colors correspond to different orders in perturbation theory. Not unexpectedly, fits in the Rgap scheme show generally smaller theory uncertainties.

In order to estimate correlations induced by theoretical uncertainties, each ellipse in the $\alpha_s$-$2\Omega_1$ plane is constructed following the procedure discussed in \Sec{sec:fitprocedure}.
Each theory ellipse constructed in this manner is interpreted as an estimate for the \mbox{1-$\sigma$} theoretical uncertainty ellipse for each individual parameter (39\% confidence for the two parameters), and is represented by a dashed ellipse in Fig.~\ref{fig:alpha}.
The solid lines represent the combined (theoretical plus experimental) standard uncertainty ellipses at 39\% confidence for two parameters, obtained by adding the theoretical and experimental error matrices from the individual
ellipses, where the experimental ellipse corresponds to $\Delta \chi^2 = 1$.
Figure~\ref{fig:alpha} clearly shows a substantial reduction of the perturbative uncertainties when increasing the resummation accuracy, and given that they are 39\% confidence regions for two parameters, also show good overlap between the results at different orders. 

The results for $\alpha_s(m_Z)$ and $\Omega_1$ from the theory scan at different perturbative orders are collected in Tabs~\ref{tab:power-results} and \ref{tab:Omega1results}. Central values here are determined from the average of the maximal and minimal values of the theory scan, and are very close to the central values obtained when running with our default parameters.  The quoted perturbative uncertainties are one-parameter uncertainties.

\begin{table}[t!]
\begin{tabular}{ccc}
order &$\alpha_s(m_Z)$ (with $\overline\Omega_1$) & $\alpha_s(m_Z)$
(with $\Omega_1(R_\Delta,\mu_\Delta)$)\\
\hline
NLL$^\prime$             & $0.1071(60)(05)$ & $0.1059(62)(05)$  \\
N$^2$LL$^\prime$            & $0.1102(32)(06)$ & $0.1100(33)(06)$ \\
N$^3$LL$^\prime$ (full)  & $0.1117(16)(06)$ & $\mathbf{0.1123(14)(06)}$
\end{tabular}
\caption{Central values for $\alpha_s(m_Z)$ at various 
  orders with theory uncertainties from the parameter scan (first value in 
  parentheses), and experimental and hadronic uncertainty added in quadrature (second
  value in parentheses). The bold N$^3$LL$^\prime$ value is our final result.}
\label{tab:power-results}
\end{table}
\begin{table}[t!]
\begin{tabular}{ccc}
order & \hspace{5mm}$\overline\Omega_1$ [GeV]\hspace{4mm} 
 & \hspace{1mm}$\Omega_1(R_\Delta,\mu_\Delta)$ [GeV]\hspace{1mm} \\
\hline
NLL$^\prime$             & $0.533(154)(18)$ & $0.582(134)(16)$ \\
N$^2$LL$^\prime$            & $0.443(119)(19)$  & $0.457(83)(19)$ \\
N$^3$LL$^\prime$ (full)  & $0.384(91)(20)$  & $\mathbf{0.421(60)(20)}$ \\
\end{tabular}
\caption{Central values for  $\Omega_1$ at the 
  reference scales $R_\Delta=\mu_\Delta=2$\,GeV and for $\overline\Omega_1$ and at various
  orders. The parentheses show first the theory uncertainties from the parameter scan,
  and second the experimental plus the uncertainty due to the imprecise determination of $\alpha_s$ (added in quadrature). The bold N$^3$LL$^\prime$ value is our final result.}
\label{tab:Omega1results}
\end{table}

In Tab.~\ref{tab:lesser} above we also present $\alpha_s(m_Z)$ results with no power corrections and either using resummation or fixed-order perturbative results. Without power corrections there is no fit for $\Omega_1$, so we take the central value to be the average of the maximum and minimum value of $\alpha_s(m_Z)$ that comes from our parameter scan. Our estimate of the uncertainty is given by the difference between our result and the maximum fit value. For the fixed-order case, since there is only one renormalization scale, we know that the uncertainties from our parameter variation for $e_H$, $s_2^{\widetilde C}$, $\epsilon_2^{\rm low}$ and $\epsilon_3^{\rm low}$ are uncorrelated. So, we take the fit value for $\alpha_s(m_Z)$ with the default parameters as our result and add the uncertainties from variations of these parameter in quadrature to give the total uncertainty.

An additional attractive result of our fits is that the experimental data is better described when increasing the order of the resummation and fixed-order terms. This can be seen by looking at the minimal $\chi^2/$dof values for the best-fit points, which are shown in Fig.~\ref{fig:alpha}. In Figs.~\ref{fig:chi2alphagap} and \ref{fig:chi2alphanogap} we show the distribution of $\chi^2_{\rm min}/{\rm dof}$ values for the various $\alpha_s(m_Z)$ best-fit points. 
Figure~\ref{fig:chi2alphagap}
displays the results in the Rgap scheme, whereas Fig.~\ref{fig:chi2alphanogap} shows the results in the $\msbar$ scheme. In both cases we find that the $\chi^2_{\rm min}$ values
systematically decrease with increasing perturbative order. The highest-order analysis in the
$\msbar$ scheme leads to $\chi^2_{\rm min}/{\rm dof}$ values around unity and thus provides
an adequate description of the whole dataset, however one also observes that accounting for the
renormalon subtraction in the Rgap scheme leads to a substantially improved theoretical
description having $\chi^2_{\rm min}/{\rm dof}$ values below unity essentially for all points
in the random scan. Computing the average of the $\chi^2_{\rm min}$ values we find at  N$^3$LL$^\prime$ order for the Rgap and $\msbar$ schemes $0.988$ and $1.004$, respectively (where the spread of values is smaller in the Rgap scheme). Likewise for N$^2$LL$^\prime$ we find $1.00$ and $1.02$, and for NLL$^\prime$ we find $1.09$ and $1.14$.
These results show the excellent description of the experimental data for
various center-of-mass energies. They also validate the smaller theoretical uncertainties
obtained for $\alpha_s$ and $\Omega_1$ at N$^2$LL$^\prime$ and N$^3$LL$^\prime$ orders in the Rgap scheme.

\subsection{Experimental Fit Uncertainty} \label{sec:exptfit}
\begin{figure}[t]
\vspace{0pt}
\includegraphics[width=1\linewidth]{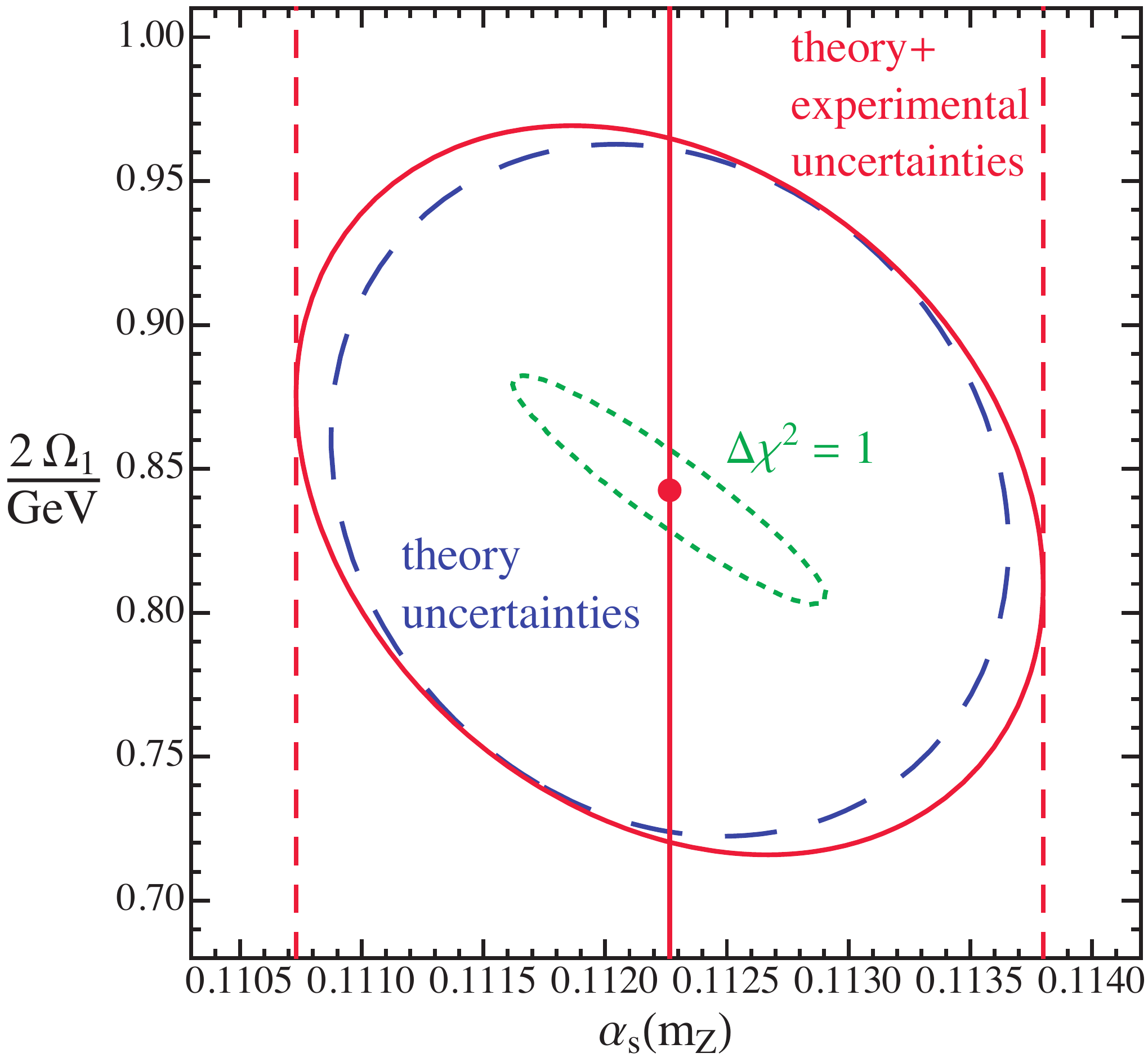}
\caption{Experimental $\Delta\chi^2=1$ standard uncertainty ellipse (dotted green) at
N${}^3$LL$^\prime$ accuracy with renormalon subtractions, in the
$\alpha_s$\,-\,$2\Omega_1$ plane.  The dashed blue ellipse represents the
theory uncertainty which is obtained by fitting an ellipse to the contour of the
distribution of the best-fit points. This ellipse should be interpreted as
the $1$-$\sigma$ theory uncertainty for one parameter (39\% confidence for
two parameters). The solid red ellipse represents the total (combined
experimental and perturbative) uncertainty ellipse.}
\label{fig:ellipses}
\end{figure}
Next we discuss in more detail the experimental uncertainty in $\alpha_s(m_Z)$ and the hadronization parameter $\Omega_1$ as well as the combination with the perturbative uncertainty done to obtain the total uncertainty. 

Results are depicted in Fig.~\ref{fig:ellipses} for our highest order fit including resummation, power corrections and renormalon subtractions. The inner green dotted ellipse, blue dashed ellipse, and solid red ellipse represent the  $\Delta \chi^2=1$ uncertainty ellipses for the experimental, theoretical, and combined theoretical and experimental uncertainties respectively. These ellipses correspond to the \mbox{one-dimensional} projection of the
uncertainties onto either $\alpha_s(m_Z)$ or $\Omega_1$ (39\% confidence ellipse for two parameters). The correlation matrix of the experimental,
theory, and total uncertainty ellipses are (for $i,j=\alpha_s, 2\,\Omega_1$),
\begin{align} \label{eq:Vijresult}
V_{ij}&  =\, 
\left( \begin{array}{cc}
\sigma_{\alpha_s}^2
   & \,\, 2 \sigma_{\alpha_s} \sigma_{\Omega_1}\rho_{\alpha\Omega}\\
2\sigma_{\alpha_s} \sigma_{\Omega_1}\rho_{\alpha\Omega} 
   & \,\,4 \sigma_{\Omega_1}^2
\end{array}\right) ,
 \\
V^{\rm exp}_{ij}& =
\left( \begin{array}{cr}
4.18(52)\cdot 10^{-7}  & \,\, -\,0.24(5)\cdot 10^{-4}\,\mbox{GeV}\\ 
-\,0.24(5)\cdot 10^{-4}\,\mbox{GeV} & \,\, 1.60(47)\cdot 10^{-3}\,\mbox{GeV}^2
\end{array}\right) \! , \nn\\
V^{\rm theo}_{ij}& =
\left( \begin{array}{cr}
1.93\cdot 10^{-6}  & \,\, -\,0.27\cdot 10^{-4}~\mbox{GeV}\\
-\,0.27\cdot 10^{-4}~\mbox{GeV} & \,\, 1.45\cdot 10^{-2}~\mbox{GeV}^2
\end{array}\right), \nn\\
V^{\rm tot}_{ij}& =
\left( \begin{array}{cr}
2.35(5)\cdot 10^{-6}  & \,\, -\,0.51(5)\cdot 10^{-4}\,\mbox{GeV}\\ 
-\,0.51(5)\cdot 10^{-4}\,\mbox{GeV} & \,\, 1.61(5)\cdot 10^{-2}\,\mbox{GeV}^2
\end{array}\right)\! . \nn
\end{align}
Note that the theoretical uncertainties dominate by a significant amount. 
The experimental correlation coefficient is significant and
reads
\begin{align} \label{eq:rhoaO}
  \rho^{\rm exp}_{\alpha\Omega}\,=\,-\,0.93(15) \,.
\end{align}
The theory correlation coefficient is small, $\rho^{\rm theo}_{\alpha\Omega}\,=\,-\,0.16$, and since these uncertainties dominate 
it reduces the correlation coefficient for the total uncertainty to
\begin{align}\label{eq:rhoaOtot}
\rho_{\alpha\Omega}^{\rm total} \,=\, -\,0.26(2)\,.
\end{align}
In both \eqs{rhoaO}{rhoaOtot} the numbers in parentheses indicate a $\pm$ range that captures all values obtained from the theory scan.  The correlation exhibited by the green dotted experimental uncertainty ellipse in
Fig.~\ref{fig:ellipses} is given by the line describing the semimajor axis
\begin{align}
  \frac{\Omega_1}{30.84\,{\rm GeV}} = 0.1257 - \alpha_s(m_Z) \,.
\end{align}
Note that extrapolating this correlation to the extreme case where we neglect
the nonperturbative corrections ($\Omega_1=0$) gives $\alpha_s(m_Z)\to 0.1257$ which is consistent with the $0.1219 \pm 0.0028$ result of our fit without power corrections in \tab{lesser}.

\begin{figure*}[t!]
\begin{center}
\includegraphics[width=0.48\textwidth]{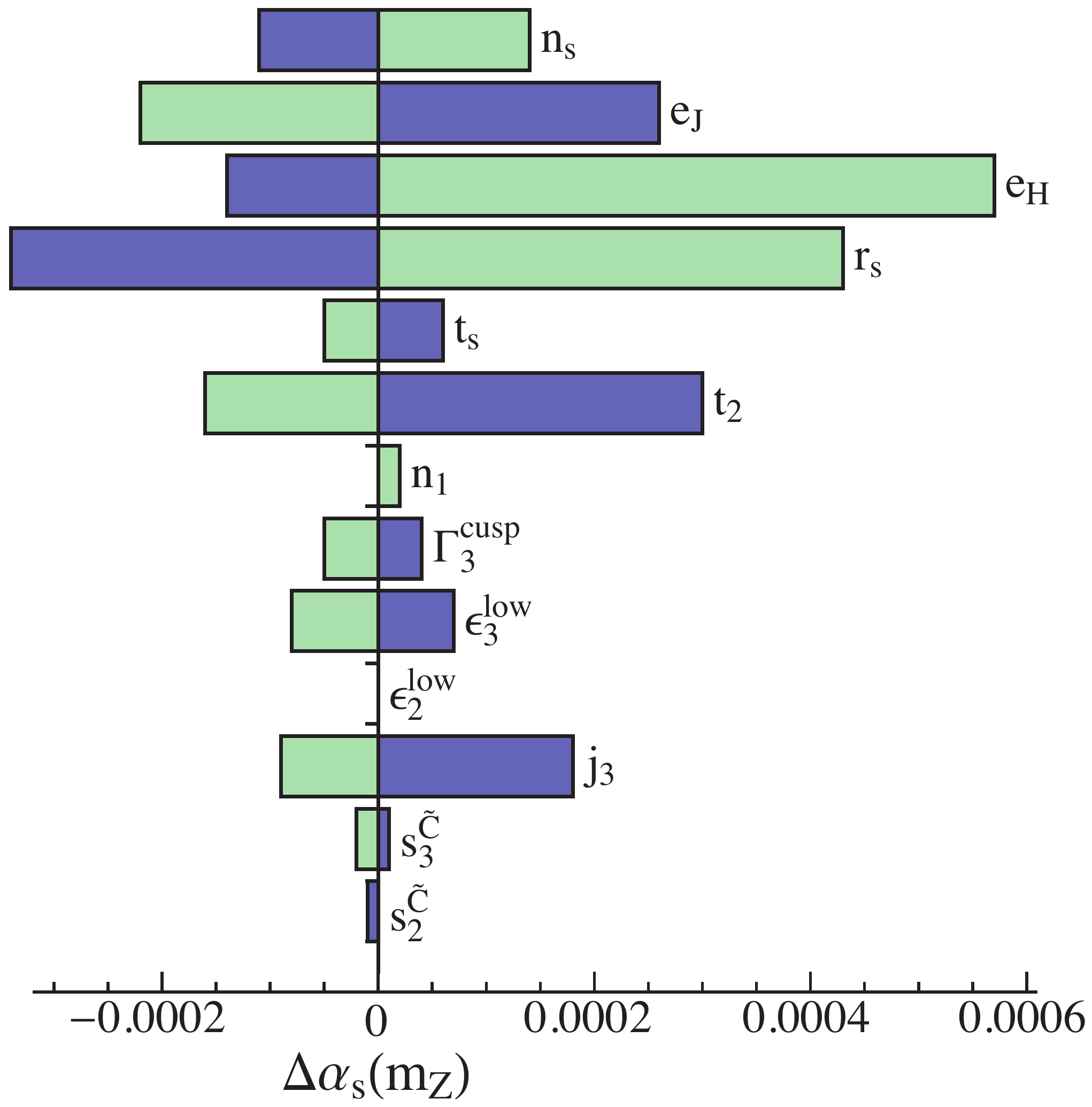}~~~
\includegraphics[width=0.48\textwidth]{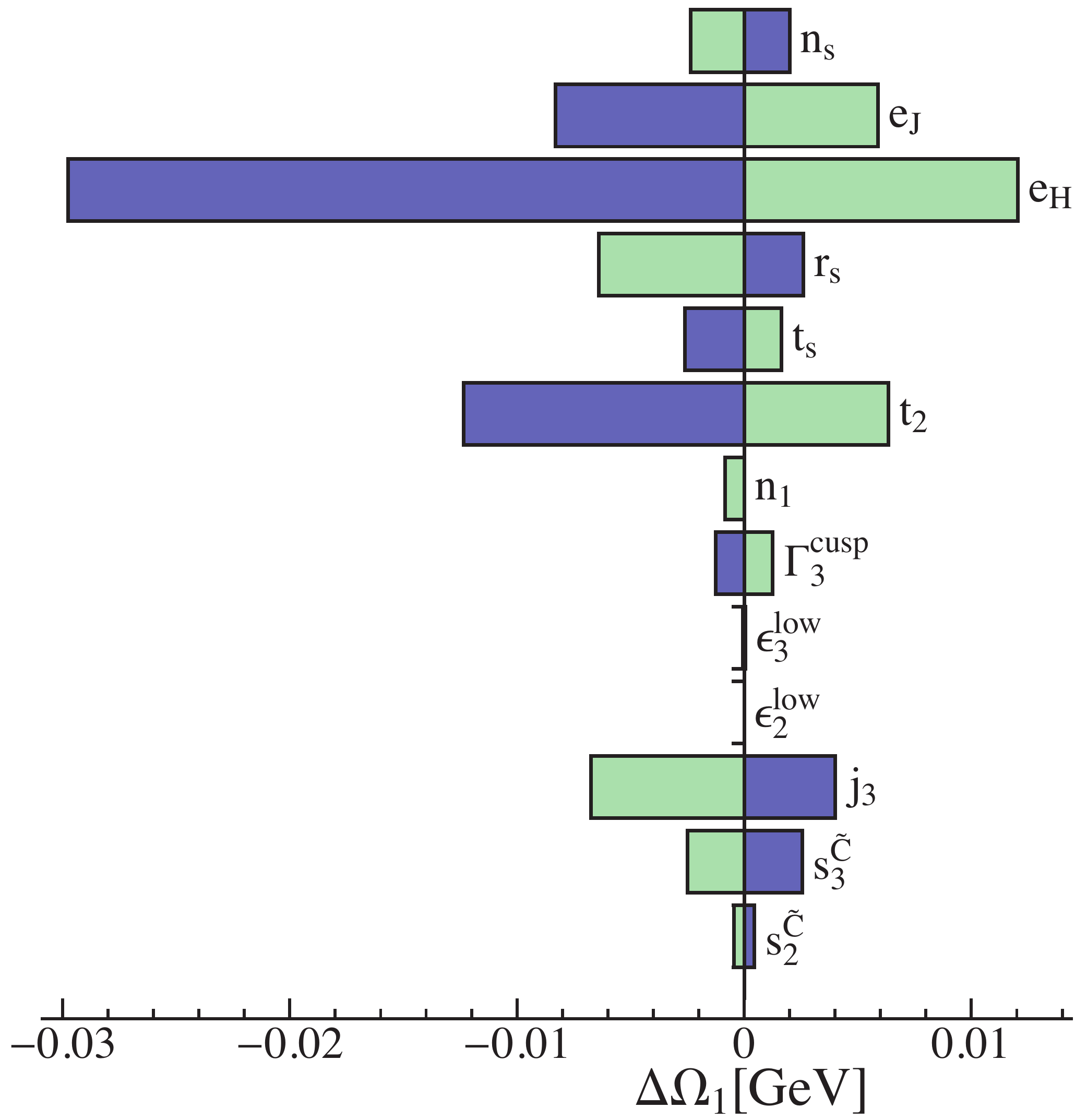}
\caption{Variations of the best-fit values for $\alpha_s(m_Z)$ and $\Omega_1$ from up (dark blue) and down (light green) 
variations for the theory parameters according to Tab.~\ref{tab:theoryerr}. We do not display those parameters 
which do not affect the fit region ($\epsilon_2^{\rm high}$, $\epsilon_3^{\rm high}$, $\mu_0$, $R_0$, $n_0$).}
\label{fig:UpDown}
\end{center}
\end{figure*}

From $V_{ij}^{\rm exp}$ in Eq.~(\ref{eq:Vijresult}) it is possible
to extract the experimental uncertainty for $\alpha_s(m_Z)$ and the uncertainty due to the imprecise determination of $\Omega_1$,
\begin{align}
\sigma_{\alpha_s}^{\rm exp} 
  & = \,\sigma_{\alpha_s}\sqrt{1-\rho^2_{\alpha\Omega}}
  =  \,0.0002 \,,
\nonumber\\
\sigma_{\alpha_s}^{\rm \Omega_1} 
  & = \,\sigma_{\alpha_s}\, |\rho_{\alpha\Omega}|\,
  =  \,0.0006 \,,
\end{align}
and to extract the experimental uncertainty for $\Omega_1$ and its uncertainty due to the imprecise determination of $\alpha_s(m_Z)$,
\begin{align}
\sigma_{\Omega_1}^{\rm exp} 
  & = \,\sigma_{\Omega_1}\sqrt{1-\rho^2_{\alpha\Omega}}
  =  \,0.014~\mbox{GeV} 
  \,, \nn \\
\sigma_{\Omega_1}^{\rm \alpha_s} 
  & = \,\sigma_{\Omega_1}\, |\rho_{\alpha\Omega}|\,
  =  \,0.037~\mbox{GeV}
\,.
\end{align}

The projections of the outer solid ellipse in Fig.~\ref{fig:ellipses}\,  show the total  uncertainty in our final one-parameter results obtained from $V_{ij}^{\rm tot}$, which are quoted below in
\eq{asfinal}.

\subsection{Individual Theory Scan Errors}
\label{sec:up-down}

To gain further insight into our theoretical precision and in order to estimate the dominant source for theory uncertainty from missing higher-order terms, we look at the size of the theory uncertainties caused by the individual variation of each one of the theory parameters included in our random scan. In Fig.~\ref{fig:UpDown}
two bar charts are shown with these results for $\alpha_s(m_Z)$ (left panel) and
$\Omega_1(R_\Delta,\mu_\Delta)$ (right panel) for fits corresponding to our best theoretical setup
(with N$^3$LL$^\prime$ accuracy and in the Rgap scheme). The dark blue bars correspond to the result of the fit with an upward variation of the given parameter from \tab{theoryerr}, while the light green bars correspond to the fit result from the downward variation in \tab{theoryerr}.
Here we vary a single parameter keeping the rest fixed at
their default values. 
We do not show parameters that have a negligibly small impact in the fit region, e.g.\ $\epsilon_2^{\rm high}$ and $\epsilon_3^{\rm high}$, which only
have an effect on the cross section to the right of the shoulder, or $n_0$, which only affects the cross section in the nonperturbative region.

We see that the dominant theory uncertainties are related to variations of the profile functions
($e_H, r_s, e_J, t_2$), where $e_H$ is the largest source of uncertainty, and is particularly dominant for $\Omega_1$.
The second most important uncertainty comes from $r_s$ for $\alpha_s$ and $t_2$ for $\Omega_1$, and $e_J$ also has a significant effect on both parameters. 

As expected, the parameters associated to the transitions on the sides of our fit region, $n_1$ and $t_s$, hardly matter. The
renormalization scale parameter $n_s$ for the nonsingular partonic distribution
${\rm d}\hat\sigma_{\rm ns}/{\rm d}C$ also causes a very small uncertainty since the nonsingular terms are always dominated by the singular terms in our fit region. The uncertainties
related to the numerical uncertainties of the perturbative constants ($s_2^{\widetilde C}$, $s_3^{\widetilde C}$, $j_3$) as well as the numerical uncertainties in the extraction of the nonsingular distribution for small $C$ values,
($\epsilon_2^{\rm low}$, $\epsilon_3^{\rm low}$) are -- with the possible exception of $j_3$
-- much smaller and do not play an important role. The uncertainty related to the unknown
$4$-loop contribution to the cusp anomalous dimension is always negligible. Adding quadratically
the symmetrized individual uncertainties shown in Fig.~\ref{fig:UpDown}, we find $0.0007$ for $\alpha_s$
and $0.05$ GeV for $\Omega_1$. This is about one half of the theoretical uncertainty
we have obtained by the theory parameter scan for $\alpha_s$ (or five sixths for $\Omega_1$), demonstrating that incorporating correlated variations through the theory parameter scan represents a more realistic method to estimate the theory uncertainty.

\subsection{Effects of $\mathbf{\Omega_1}$ hadron-mass resummation}
\label{sec:hadmassresum}

\begin{figure}[t]
\vspace{0pt}
\includegraphics[width=1\linewidth]{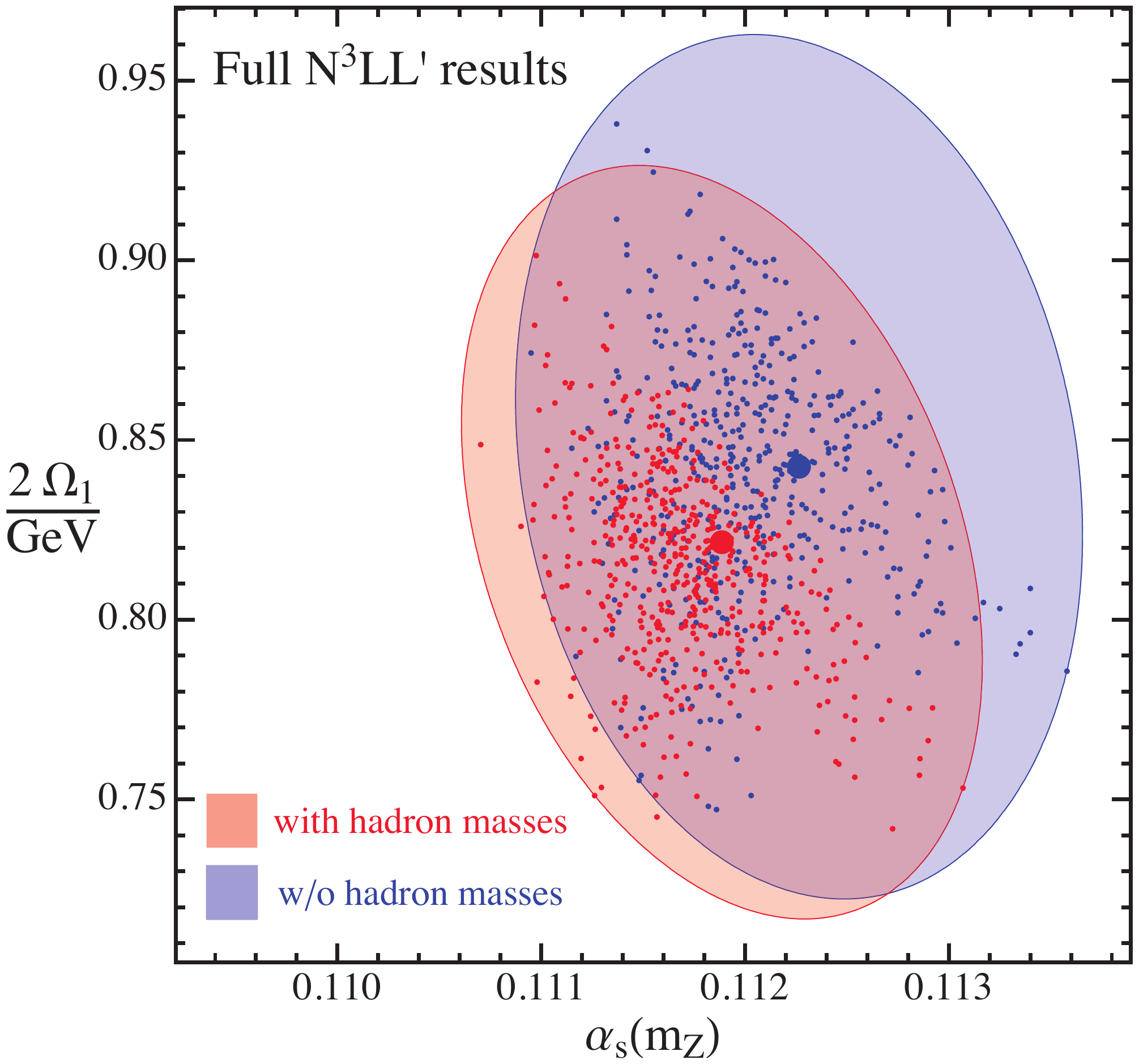}
\caption{Comparison of fits to the C-parameter tail distribution with theory prediction which include/ignore
hadron-mass effects (in red/blue). Although a direct comparison of $\alpha_s$ values is possible, one has
to keep in mind that $\Omega_1(\mu_\Delta, R_\Delta)$ has a different meaning once hadron mass running effects are included.}
\label{fig:hadron}
\end{figure}

The fit results presented in the previous two sections ignored the small hadron-mass effects. These effects are analyzed in greater detail in this section. We again perform $500$ fits for a theory setup which includes N$^3$LL$^\prime$ accuracy and a power correction in the Rgap scheme, but this time it also includes hadron-mass-induced running.

Since the impact of hadron-mass effects is small, one finds that the experimental data in the tail of the distribution is not accurate enough to fit for $\theta(R_\Delta,\mu_\Delta)$ in \Eq{eq:thetadef}, in addition to $\alpha_s(m_Z)$ and $\Omega_1(R_\Delta,\mu_\Delta)$. This is especially true because it enters as a small modification to the power correction, which by itself is not the dominant term. Indeed, fitting for $a(R_\Delta,\mu_\Delta)$ and $b(R_\Delta,\mu_\Delta)$ as defined in Eq.~(\ref{eq:omega1-ansatz}) gives a strongly correlated determination of these two parameters. The dominant hadronic parameter $\Omega_1^C(R_\Delta,\mu_\Delta)$, which governs the normalization, is still as accurately determined from data as the $\Omega_1$ in Tab.~\ref{tab:Omega1results}. However, the orthogonal parameter $\theta(R_\Delta,\mu_\Delta)$ is only determined with very large statistical uncertainties. As discussed in Ref.~\cite{Hoang:2014wka}, the specific value of $\theta(R_\Delta,\mu_\Delta)$ has a very small impact on the cross section, which is consistent with the inability to accurately fit for it.

The results of our fit including hadron-mass effects are
\begin{align}
&\alpha_s(m_Z) = 0.1119 \pm 0.0006_{\rm exp+had} \pm 0.0013_{\rm pert}\,,\\[1mm]
&\Omega_1(R_\Delta,\mu_\Delta) = 0.411 \pm 0.018_{\rm exp+\alpha_s} \pm 0.052_{\rm pert}\,{\rm GeV}\,.\nn
\end{align}
Note that the meaning of $\Omega_1(R_\Delta,\mu_\Delta)$ here is different from the case in which hadron-mass running effects are ignored because there are extra evolution effects needed to translate this value to that used in the cross section at a given value of $C$, compared to the no-hadron-mass case.

In Fig.~\ref{fig:hadron} we compare the outcome of the $500$ fits at N$^3$LL$^\prime$ in the Rgap scheme. Results with hadron-mass effects give the red ellipse on the left, and without hadron-mass effects give the blue ellipse on the right. (The latter ellipse is the same as the one discussed above in Sec.~\ref{sec:random}.) The effects of hadron masses on $\alpha_s(m_Z)$ are to decrease its central value by $0.3\%$ and reduce the percent perturbative uncertainty by $0.1\%$. Given that the total perturbative uncertainties are $1.2\%$, these effects are not statistically significant. When studying the effect on $\Omega_1$ one has to keep in mind that its meaning changes when hadron-mass effects are included. Ignoring this fact we observe that hadron masses shift the central value downwards by $2.4\%$, and reduce the percent theoretical uncertainty by $1.6\%$. Again, given that the perturbative uncertainty for $\Omega_1$ is $14\%$, this shift is not significant.

Since the theory uncertainties become slightly smaller when hadron-mass effects are incorporated, one could use this setup as our default. However we take a more conservative approach and consider the $0.3\%$ shift on the central value as an additional source of uncertainty, to be added in quadrature to the hadronization uncertainty already discussed in Sec.~\ref{sec:random}. This increases the value of the hadronization uncertainty from $0.0006$ to $0.0007$, and does not affect the total $\alpha_s$ uncertainty. The main reason we adopt this more conservative approach is that, while well motivated, the ansatz that we take in Eq.~(\ref{eq:omega1-ansatz}) is not model independent. We believe that this ansatz serves as a good estimate of what the numerical effect of hadron masses are, but should likely not be used for the central fit until further theoretical insight on the form of $\Omega_1(r)$ is gained. We do not add an additional uncertainty to $\Omega_1$ since hadron-mass effects change its meaning and uncertainties for $\Omega_1$ are large enough that these effects are negligible.

In \App{ap:subtractions} we also consider fits performed using the Rgap scheme with C-parameter gap subtractions, rather than our default Rgap scheme with thrust gap subtractions. The two results are fully compatible. As discussed in Ref.~\cite{Hoang:2014wka} the thrust gap subtractions give better perturbative convergence, and hence are used for our default cross section.

\subsection{Dataset dependence}
\label{sec:dataset}
\begin{figure}[t]
\vspace{0pt}
\includegraphics[width=1\linewidth]{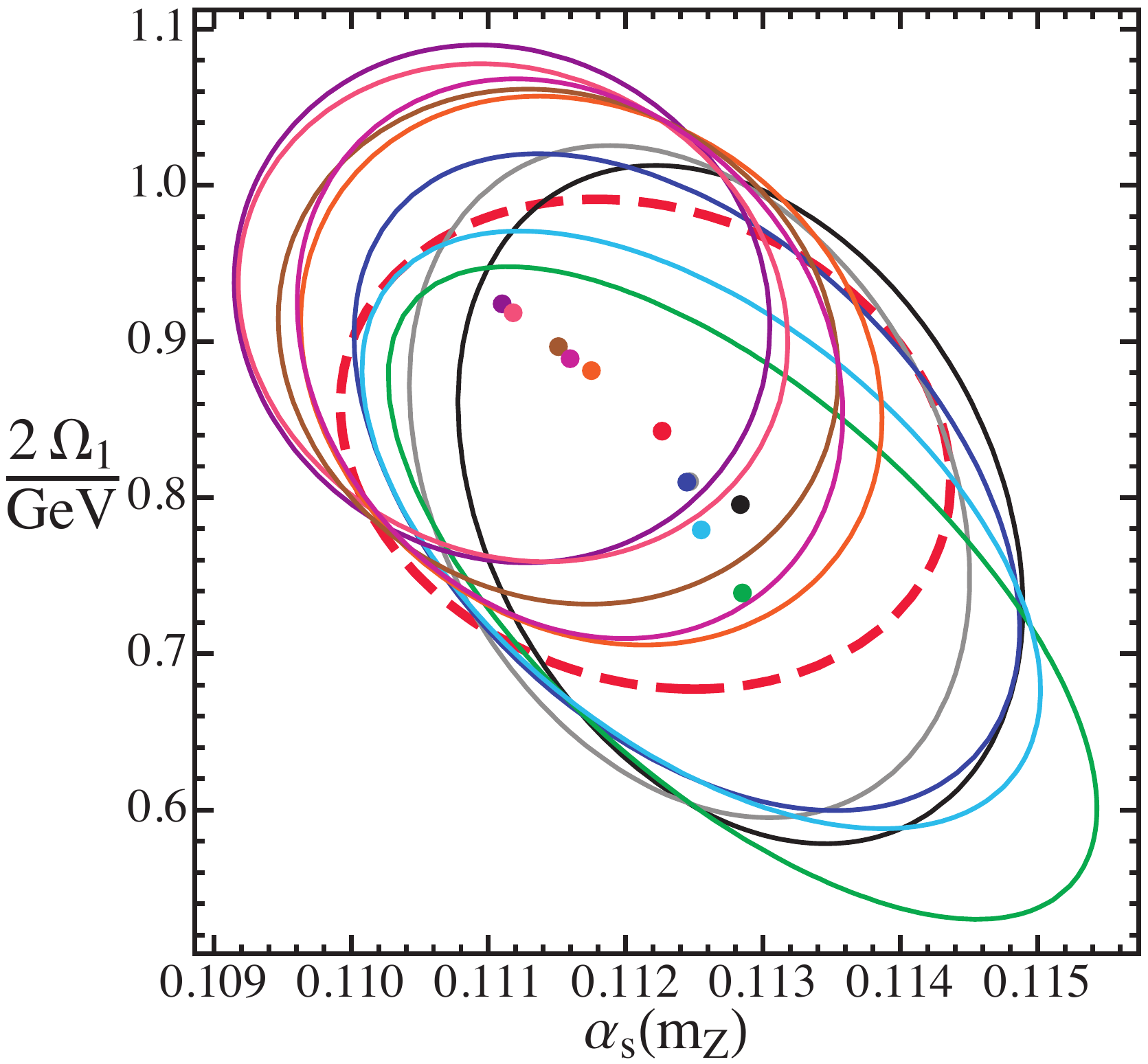}
\caption{Global fit results for different choices of dataset, using our best theory setup at N$^3$LL$^\prime$ with
power corrections in the Rgap scheme. Considering the central values from left to right, the datasets read
$[\,C_{\rm min}, C_{\rm max}\,]_{\rm \#~of~bins}$: $[\,29/Q, 0.7\,]_{371}$, $[\,22/Q, 0.75\,]_{453}$,
$[\,23/Q, 0.7\,]_{417}$, $[\,0.24, 0.75\,]_{403}$, $[\,24/Q, 0.7\,]_{409}$, $[\,25/Q, 0.7\,]_{404}$ (default),
$[\,25/Q, 0.6\,]_{322}$, $[\,25/Q, 0.75\,]_{430}$,
$[\,27/Q, 0.7\,]_{386}$, $[\,25/Q, 0.65\,]_{349}$, $[\,22/Q, 0.7\,]_{427}$. We accept bins which are at least $50\%$ inside these
fit regions. The ellipses correspond to total $1$-$\sigma$ uncertainties (experimental + theory) for two variables ($\alpha_s$ and
$\Omega_1$), which are suitable for a direct comparison of the outcome of two-parameter fits. The center of the ellipses are
also shown.}
\label{fig:dataset}
\end{figure}
In this section we discuss how much our results depend on the dataset choice. Our default global dataset accounts for all experimental bins for $Q\ge 35\,$GeV in the intervals $[\,C_{\rm  min}, C_{\rm max}\,]=[\,25/Q,0.7\,]$, (more details are given in Sec.~\ref{sec:data}). The upper limit in this range is motivated by the fact that we do not want to include data too close to the shoulder, since we do not anticipate having the optimal theoretical description of this region. The lower limit avoids including data too close to the nonperturbative region, which is near the cross section peak for $Q=m_Z$,  since we by default only include the leading power correction $\Omega_1$ in the OPE of the shape function. To consider the impact of this dataset choice we can vary the upper and lower limits used to select the data.

In Fig.~\ref{fig:dataset} the best fits and the respective total experimental + theory $68\%$ CL uncertainty ellipses (for two parameters) are shown for global datasets based on different choices of data ranges. The result for our default global dataset is given in red, with a thicker, dashed ellipse. In the caption of Fig.~\ref{fig:dataset} the data ranges and the number of bins are specified for each one of the plotted ellipses.

Interestingly all uncertainty ellipses have very similar correlation and are lined up approximately along the line
\begin{align}
\frac{\Omega_1}{41.26 \,{\rm GeV}} = 0.1221 - \alpha_s(m_Z) \,.
\end{align}
As expected, the results of our fits depend only weakly on the $C$ range and the size of the global
datasets, as shown in Fig.~\ref{fig:dataset}. The size and tilt of the total uncertainty ellipses is very similar
for all datasets (with the exception of $[\,22/Q, 0.7\,]$, which clearly includes too much peak data).
Since the centers and the sizes of the uncertainty
ellipses are fully statistically compatible at the $1$-$\sigma$ level, this indicates that our theory uncertainty estimate at
N$^3$LL$^\prime$ really reflects the accuracy at which we are capable of describing the different regions of the spectrum.
Therefore a possible additional uncertainty that one could consider due to the arbitrariness of the dataset choice is actually already represented in our final uncertainty estimates.
\vspace*{-0.3cm}
\subsection{Final Results}
\label{sec:final}
\vspace*{-0.3cm}
As our final result for $\alpha_s(m_Z)$ and $\Omega_1$, obtained at N$^3$LL$^\prime$ order
in the Rgap scheme for $\Omega_1(R_\Delta,\mu_\Delta)$, we get
\begin{align} \label{eq:asfinal}
\alpha_s(m_Z) & \, = \,
 0.1123 \,\pm\, 0.0002_{\rm exp}
\\[1mm] & \,\pm\, 0.0007_{\rm hadr} \,\pm \, 0.0014_{\rm pert},
\nonumber\\[2mm]
\Omega_1(R_\Delta,\mu_\Delta) & \, = \,
 0.421 \,\pm\, 0.007_{\rm exp} 
\nonumber\\[1mm] &        \,\pm\, 0.019_{\rm \alpha_s(m_Z)} 
\,\pm \, 0.060_{\rm pert}\,\mbox{GeV},
\nonumber
\end{align}
where $R_\Delta=\mu_\Delta=2$~GeV and we quote individual \mbox{$1$-$\sigma$} uncertainties for each parameter.
Here $\chi^2/\rm{dof}=0.99$. Equation~(\ref{eq:asfinal}) is the main result of this work.

Equation~(\ref{eq:asfinal}) accounts for the effect of hadron mass running through an additional (essentially negligible) uncertainty. Also, it neglects QED and finite bottom-mass corrections, which were found to be small effects in the corresponding thrust analysis in Ref.~\cite{Abbate:2010xh}.

\begin{figure}[t!]
\begin{center}
\includegraphics[width=0.95\columnwidth]{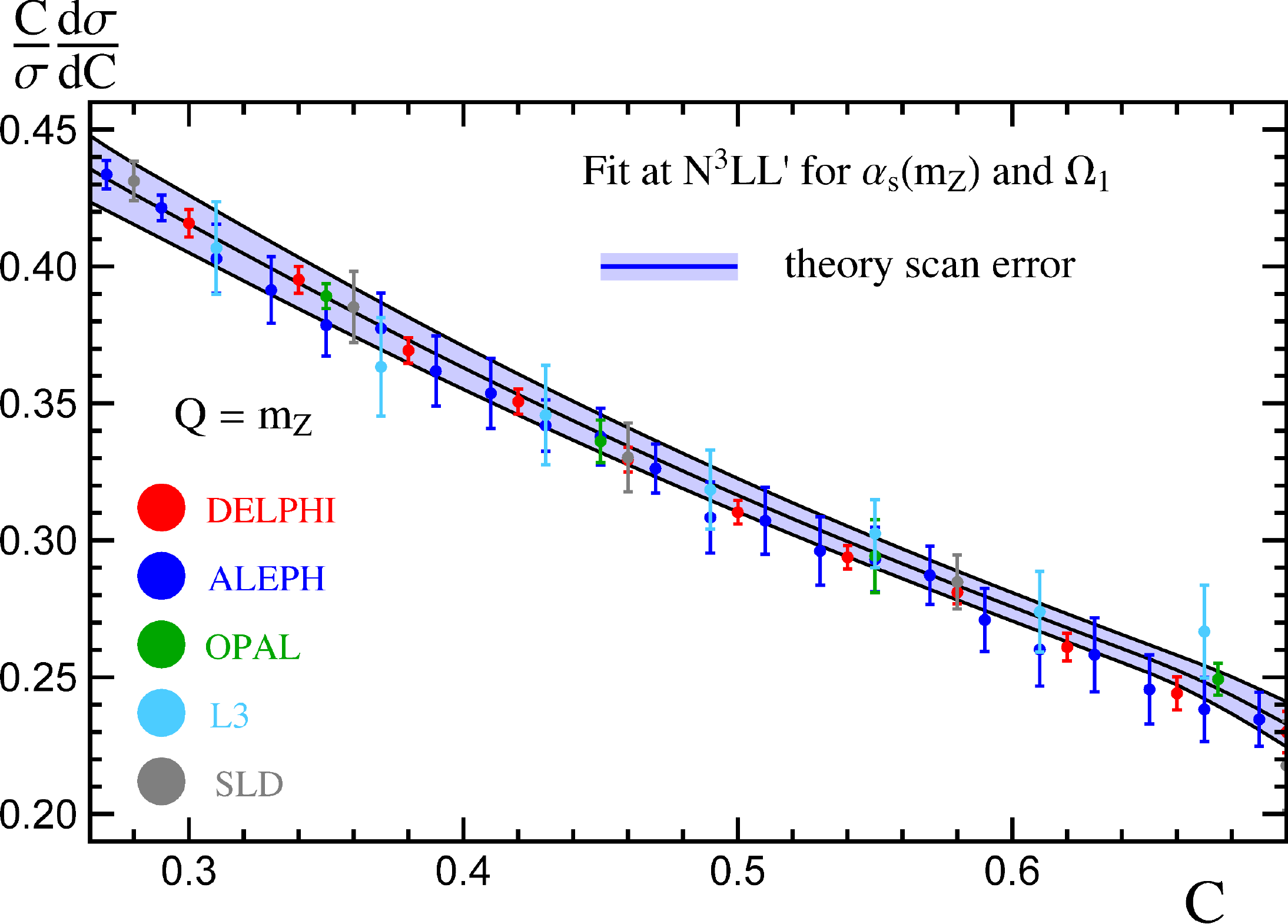}
\caption{C-parameter distribution at N${}^3$LL$^\prime$ order for $Q=m_Z$ showing the fit result for the values for $\alpha_s(m_Z)$ and
$\Omega_1$. The blue band corresponds to the theory uncertainty as described in Sec.~\ref{sec:random}.
Experimental data is also shown.}
\label{fig:Theory-data-comparison}
\end{center}
\end{figure}

Given that we treat the correlation of the systematic experimental uncertainties
in the minimal overlap model, it is useful to examine the
results obtained when assuming that all systematic experimental uncertainties are uncorrelated. At N$^3$LL$^\prime$ order in the Rgap scheme the results that are
analogous to Eq.~(\ref{eq:asfinal}) read
$\alpha_s(m_Z) = 0.1123 \pm 0.0002_{\rm exp} \pm 0.0007_{\rm hadr} \pm
0.0012_{\rm pert}$ and $\Omega_1(R_\Delta,\mu_\Delta) = 0.412 \,\pm\, 0.007_{\rm exp}
\pm 0.022_{\alpha_s} \pm 0.061_{\rm pert}$\,GeV
with a combined correlation coefficient of $\rho_{\alpha\Omega}^{\rm total}=-\,0.091$.
The results are compatible with Eq.~(\ref{eq:asfinal}), indicating that the ignorance of the precise
correlation of the systematic experimental uncertainties barely affects the outcome of the fit.

In Fig.~\ref{fig:Theory-data-comparison} we show the theoretical fit for the \mbox{C-parameter} distribution
in the tail region, at a center-of-mass energy corresponding to the $Z$-pole. We use the best-fit values given in 
Eq.~(\ref{eq:asfinal}). The band corresponds to the perturbative
uncertainty as determined by the scan. The fit result is shown in comparison with experimental
data from DELPHI, ALEPH, OPAL, L3 and SLD. Good agreement is observed for this spectrum, as well as for spectra at other center of mass values.

\begin{figure}
\begin{center}
\includegraphics[width=0.95\columnwidth]{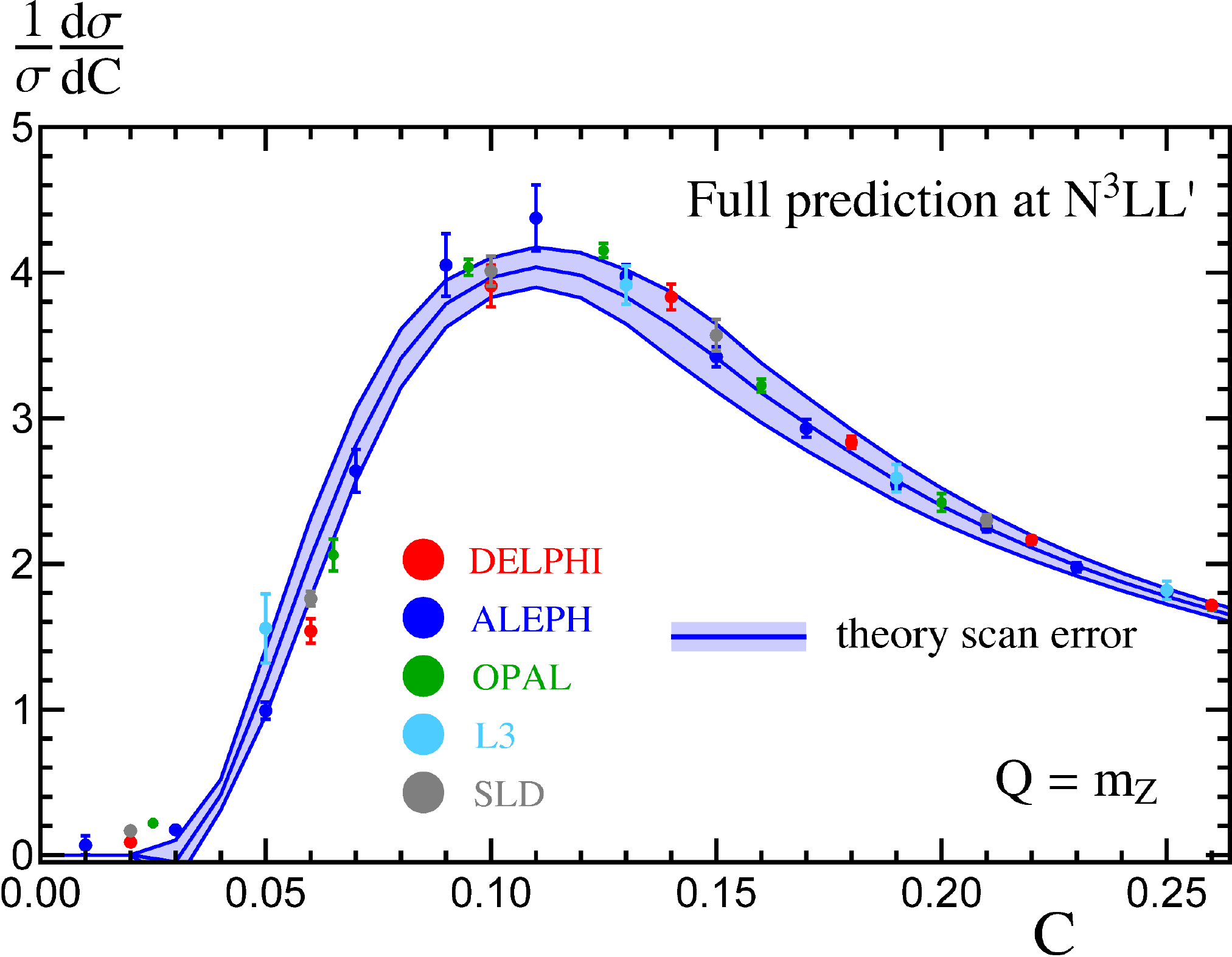}
\caption{C-parameter distribution below the fit region, shown at N${}^3$LL$^\prime$ order for $Q=m_Z$ using the best-fit values for $\alpha_s(m_Z)$ and
$\Omega_1$. Again the blue band corresponds to the theory uncertainty and error bars are used for experimental data.}
\label{fig:Peak-plot}
\end{center}
\end{figure}
\begin{figure}[t!]
\begin{center}
\includegraphics[width=0.95\columnwidth]{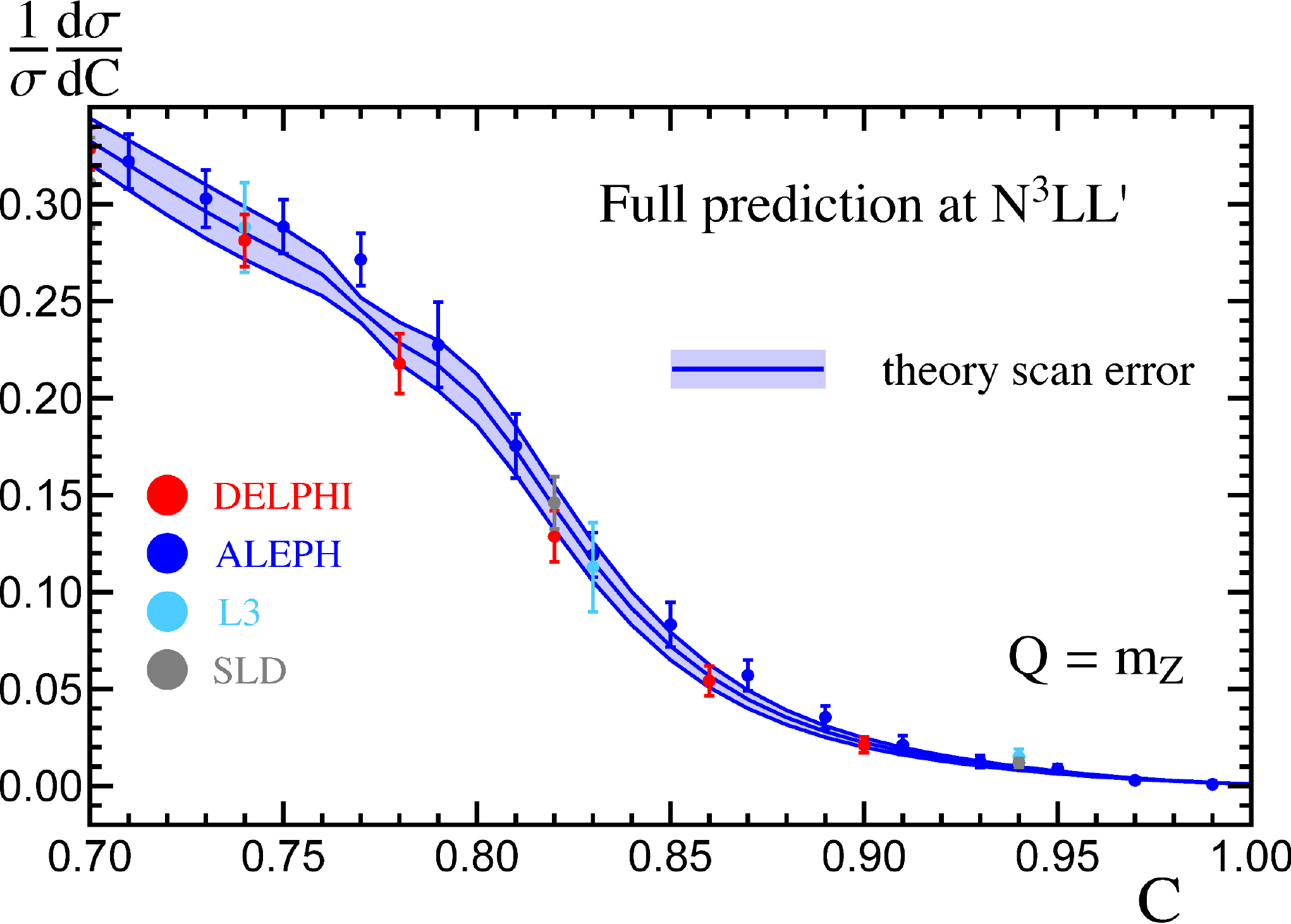}
\caption{C-parameter distribution above the fit range, shown at N${}^3$LL$^\prime$ order for $Q=m_Z$ using the best-fit values for $\alpha_s(m_Z)$ and
$\Omega_1$. Again the blue band corresponds to the theory uncertainty and the error bars are used for experimental data. }
\label{fig:Far-tail-plot}
\end{center}
\end{figure}

\section{Peak and Far Tail Predictions}
\label{sec:peak-tail}

Even though our fits were performed in the resummation region which is dominated by tail data, our theoretical results also apply for the peak and far-tail regions. As an additional
validation for the results of our global analysis in the tail region, we use the best-fit values obtained for
$\alpha_s$ and $\Omega_1$ to make predictions in the peak and the far-tail regions where the corresponding data was not included in the fit.

Predictions from our full N$^3$LL$^\prime$ code in the Rgap scheme for the C-parameter cross section at the $Z$-pole in the peak
region are shown in Fig.~\ref{fig:Peak-plot}. The nice agreement within theoretical uncertainties (blue band) with the
precise data from DELPHI, ALEPH, OPAL, L3, and SLD indicates that the value of $\Omega_1$ obtained from the fit to the
tail region is the dominant nonperturbative effect in the peak. The small deviations between the theory band and the
experimental data can be explained due to the fact that the peak is also sensitive to higher-order power corrections
$\Omega_{k\ge 2}^C$, which have not been tuned to reproduce the peak data in our analysis.

In Fig.~\ref{fig:Far-tail-plot} we compare predictions from our full N$^3$LL$^\prime$ code in the Rgap scheme to the accurate
DELPHI, ALEPH, L3, and SLD data at $Q=m_Z$ in the far-tail region.\footnote{The OPAL data was excluded from the plot because its bins are
rather coarse in this region, making it a bad approximation of the differential cross section.} We find excellent agreement with the data within the
theoretical uncertainties (blue band). The key feature of our theoretical prediction that matters most in the far-tail
region is the merging of the renormalization scales toward $\mu_S=\mu_J=\mu_H$ at $C\sim 0.75$ in the profile functions. This is a necessary condition for the cancellations between singular and nonsingular terms in the cross section to occur above the shoulder region.\footnote{It is worth mentioning that
in the far-tail region we employ the $\overline{\rm MS}$ scheme for $\Omega_1$, since the subtractions implemented in the
Rgap scheme clash with the partonic shoulder singularity, resulting in an unnatural behavior of the cross section around
$C=0.75$. The transition between the Rgap and $\overline{\rm MS}$ schemes is performed smoothly, by means of a hybrid
scheme which interpolates between the two in a continuous way. This hybrid scheme has been discussed at length in
Ref.~\cite{Hoang:2014wka}.} At $Q=m_Z$ the theoretical cross section presented here 
obtains accurate predictions in the region both below and above the shoulder that agree with the data. Our analysis does not include the full ${\cal O}(\alpha_s^k \Lambda_{\rm QCD}/Q)$ power corrections (for $k < 4$), since they are not part of our master formula. Nevertheless, and in analogy with what was found in the case of thrust, agreement with the experimental data seems to indicate that these missing power corrections may be smaller than naively expected.

\section{Universality and Comparison to Thrust}
\label{sec:universality}

\begin{figure}[t!]
\begin{center}
\includegraphics[width=0.95\columnwidth]{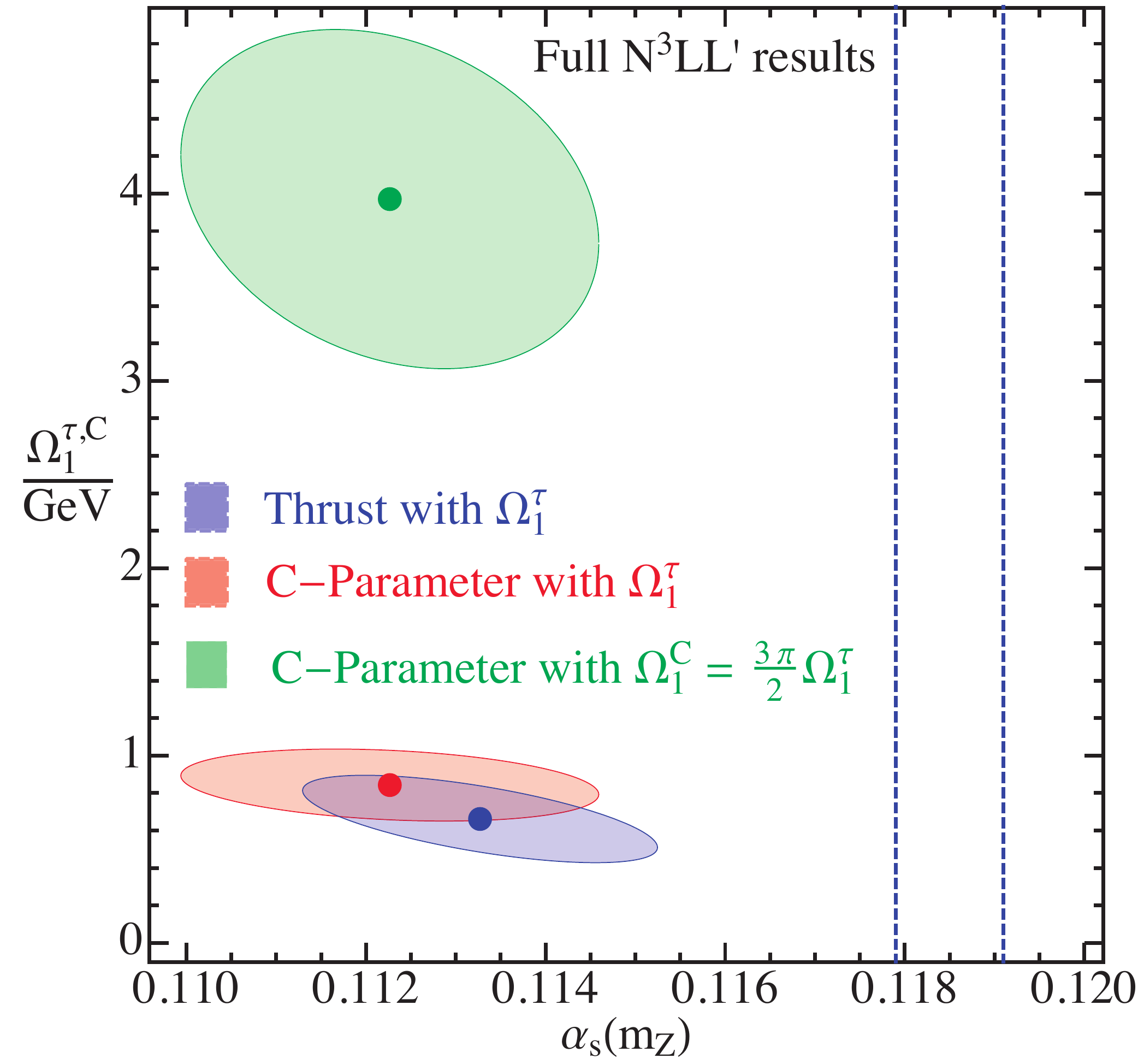}
\caption{Comparison of determinations of $\alpha_s(m_Z)$ and $\Omega_1$ with the corresponding total 1-$\sigma$
uncertainty ellipses. As an illustration we display the determination of $\Omega_1^C$ obtained from fits
to the C-parameter distribution (green), which is clearly different from $\Omega_1^\tau$ obtained from thrust fits (blue),
and the determination of $\Omega_1^\tau$ as obtained from \mbox{C-parameter} distribution fits (red). All fits have been performed with N$^3$LL$^\prime$ theoretical predictions with power corrections and in the Rgap scheme. The dashed vertical lines indicate the PDG 2014~\cite{Agashe:2014kda} determination of $\alpha_s(m_Z)$. }
\label{fig:Universality}
\end{center}
\end{figure}

\begin{figure}[t!]
\begin{center}
\includegraphics[width=0.95\columnwidth]{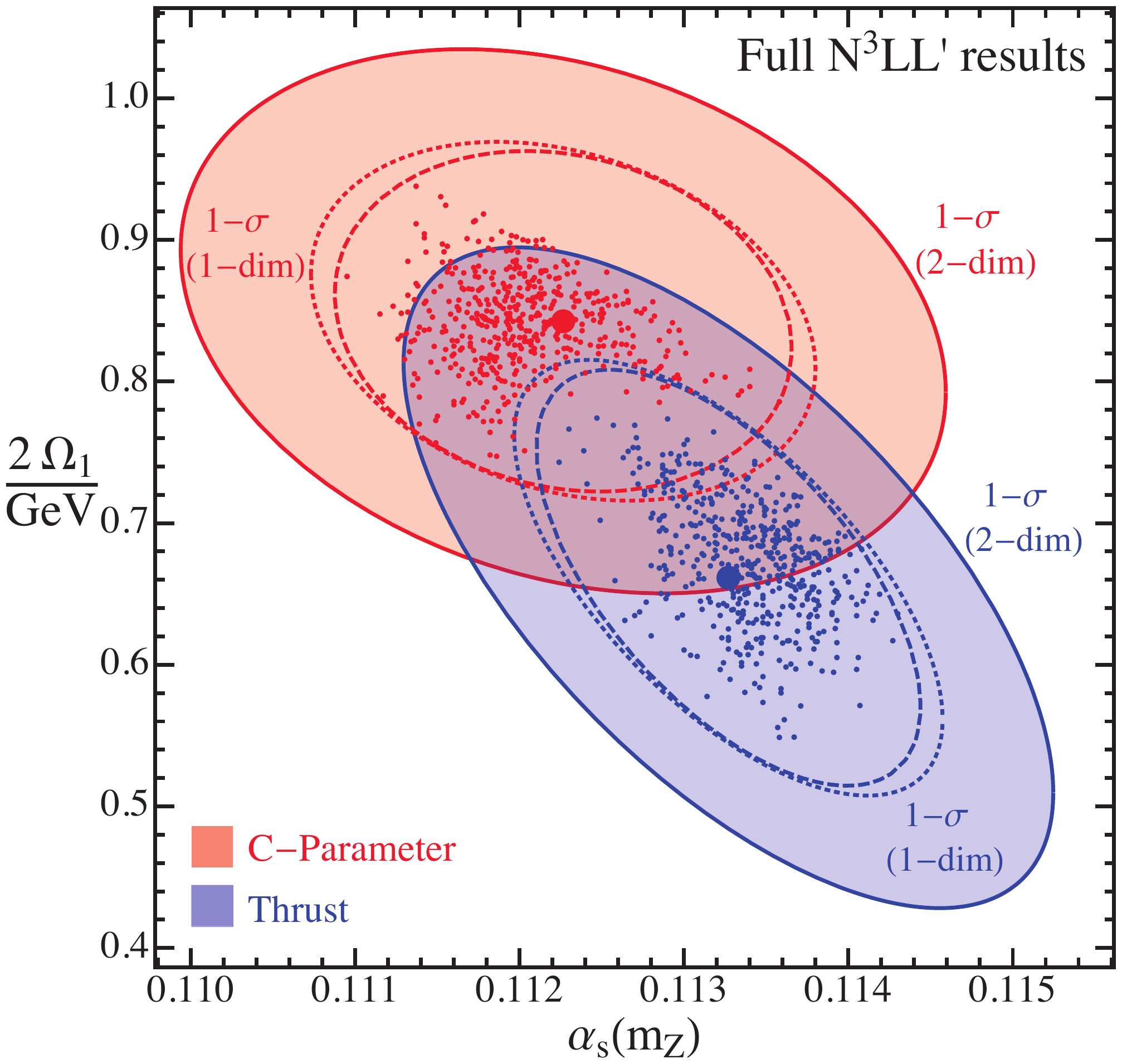}
\caption{Distribution of best-fit points in the \mbox{$\alpha_s(m_Z)$-$2\Omega_1$} plane for both thrust (blue) and
C-parameter (red) at N${}^3$LL$^\prime+\mathcal{O}(\alpha_s^3) + \Omega_1(R,\mu)$. The outer solid ellipses show the $\Delta \chi^2 = 2.3$ variations, representing 1-$\sigma$ uncertainties for two variables. The inner dashed ellipses correspond to the 1-$\sigma$ theory uncertainties for each one of the fit parameters. The dotted ellipses correspond to $\Delta \chi^2 = 1$ variations of the total uncertainties. All fits have been performed with N$^3$LL$^\prime$ theoretical predictions with power corrections and in the Rgap scheme. This plot zooms in on the bottom two ellipses of Fig.~\ref{fig:Universality}.}
\label{fig:Thrust-Cparam-comparison}
\end{center}
\end{figure}

\begin{figure}[t!]
\begin{center}
\includegraphics[width=0.95\columnwidth]{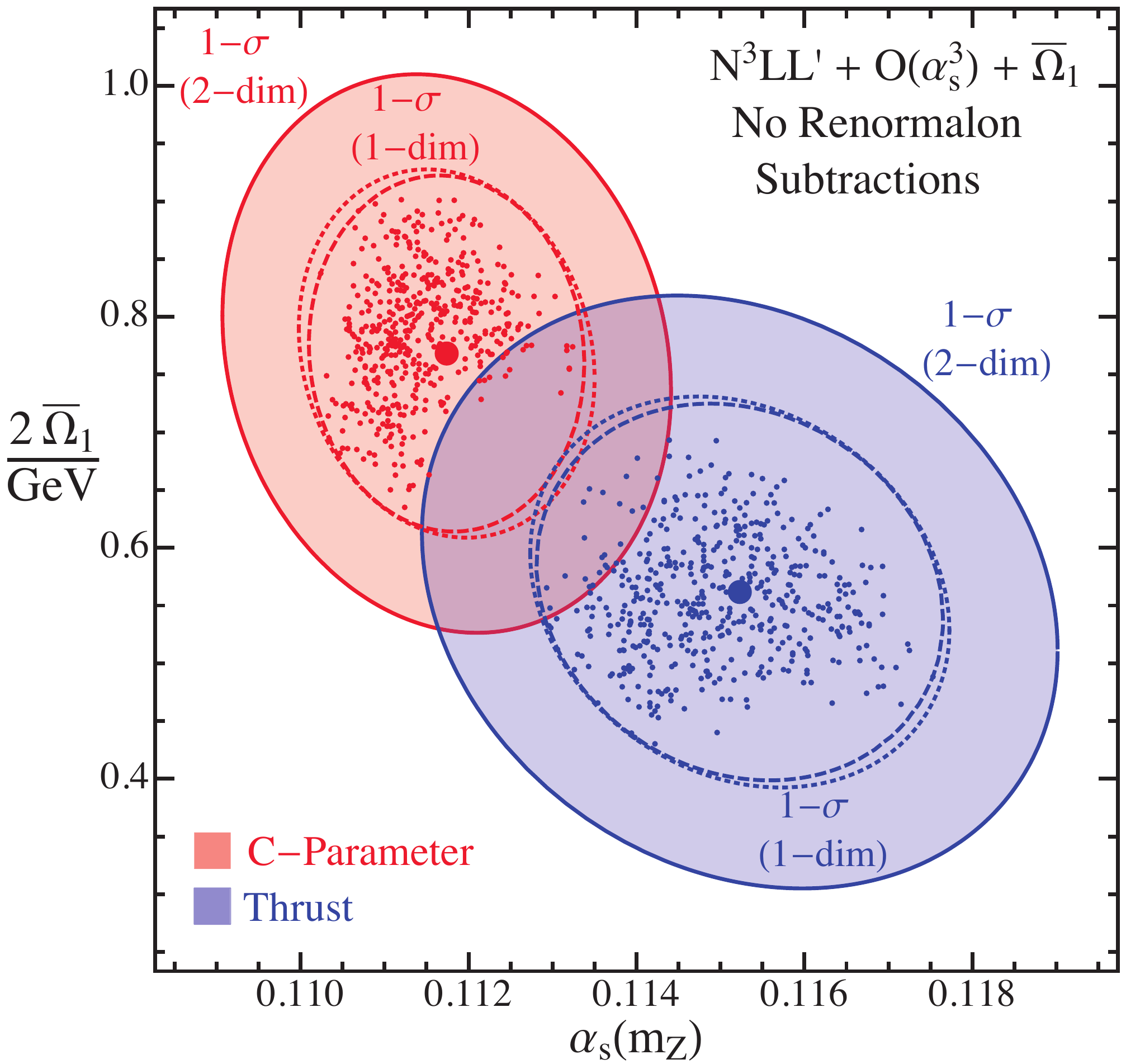}
\caption{Distribution of best-fit points in the \mbox{$\alpha_s(m_Z)$-$2\overline\Omega_1$} plane for both thrust (blue) and
C-parameter (red) at N${}^3$LL$^\prime+\mathcal{O}(\alpha_s^3) + \overline\Omega_1$. The meaning of the different
ellipses is the same as in Fig.~\ref{fig:Thrust-Cparam-comparison}.}
\label{fig:Thrust-Cparam-NoGap}
\end{center}
\end{figure}

An additional prediction of our theoretical formalism is the universality of $\Omega_1$ between the thrust and \mbox{C-parameter} event shapes. Therefore, a nontrivial test of our formalism can be made by comparing our result for $\Omega_1$ with the determination from the earlier fits of the thrust tail distributions in Ref.~\cite{Abbate:2010xh} and the first moment of the thrust distribution in Ref.~\cite{Abbate:2012jh}.

Since we have updated our profiles for thrust, it is expected that the outcome of the $\alpha_s$ and $\Omega_1$ determination is slightly (within theoretical uncertainties) different from that of Ref.~\cite{Abbate:2010xh}. We also have updated our code to match that of Ref.~\cite{Abbate:2012jh} (higher statistics for the two-loop nonsingular cross sections and using the exact result for the two-loop soft function non-logarithmic constant). In addition we have corrected the systematic uncertainty for the ALEPH data, $Q = 91.2$\,GeV of Ref.~\cite{Heister:2003aj}.\footnote{In Ref.~\cite{Abbate:2010xh} we assumed that two quoted uncertainties where asymmetric uncertainties, but it turns out they are two sources of systematic uncertainties that need to be added in quadrature. This has no significant effect on the results of  Ref.~\cite{Abbate:2010xh}.} When we compare thrust and C-parameter we neglect bottom-mass and QED effects in both event shapes. In this setup, we find an updated result for thrust:
\begin{align} \label{eq:thrustQCD}
\alpha_s(m_Z) & \, =
 0.1134 \pm 0.0002_{\rm exp}
\\[1mm] & \pm 0.0005_{\rm hadr} \pm  0.0011_{\rm pert},
\nonumber\\[2mm]
\Omega_1(R_\Delta,\mu_\Delta) & \, = 
 0.329 \pm 0.009_{\rm exp}
\nonumber\\[1mm] &        \pm 0.021_{\rm \alpha_s(m_Z)} 
\pm  0.060_{\rm pert}\,\mbox{GeV}.
\nonumber
\end{align}
For completeness we also quote an updated thrust result when both QED and bottom-mass effects are taken into account:
\begin{align} \label{eq:thrustQED}
\alpha_s(m_Z) & \, =
 0.1128 \pm 0.0002_{\rm exp} 
\\[1mm] & \pm 0.0005_{\rm hadr} \pm 0.0011_{\rm pert},
\nonumber\\[2mm]
\Omega_1(R_\Delta,\mu_\Delta) & \, =
 0.322 \pm 0.009_{\rm exp} 
\nonumber\\[1mm] &       \pm 0.021_{\rm \alpha_s(m_Z)} 
\pm  0.064_{\rm pert}\,\mbox{GeV}.
\nonumber
\end{align}
Both the results in Eqs.~(\ref{eq:thrustQCD}) and (\ref{eq:thrustQED}) are fully compatible at 1-$\sigma$ with those in Ref.~\cite{Abbate:2010xh}, as discussed in more detail in \App{ap:thrustresults}.

When testing for the universality of $\Omega_1$ between thrust and C-parameter, there is an important calculable numerical factor of $3\pi/2=4.7$ between $\Omega_1^\tau$ and $\Omega_1^C$ that must be accounted for; see  \Eq{eq:O1univ}. If we instead make a direct comparison of $\Omega_1^\tau$ and $\Omega_1^C$, as shown in Fig.~\ref{fig:Universality} (lowest blue ellipse vs uppermost green ellipse, respectively) then the results are $4.5$-$\sigma$ away from each other. Accounting for the $3\pi/2$ factor to convert from $\Omega_1^C$ to $\Omega_1^\tau$ the upper green ellipse becomes the centermost red ellipse, and the thrust and C-parameter determinations agree with one another within uncertainties. Due to our high-precision control of perturbative effects, the $\Omega_1$ parameters have only $\sim 15\%$ uncertainty, yielding a test of this universality at a higher level of precision than what has been previously achieved. 

A zoomed-in version of this universality plot is shown in Fig.~\ref{fig:Thrust-Cparam-comparison}. The upper red ellipse again shows the result from fits to the C-parameter distribution, while the lower blue 
ellipse shows the result from thrust tail fits. For both we show the theory uncertainty
(dashed lines) and combined theoretical and experimental (dotted lines) 39\% CL uncertainty ellipses,
as well as the solid ellipses which correspond to \mbox{$\Delta\chi^2=2.3$} which is the standard 1-$\sigma$ uncertainty for a two-parameter fit
(68\% CL). We see that
the two analyses are completely compatible at the 1-$\sigma$ level. An important  
ingredient to improve the overlap is the fact that we define the power corrections in the
renormalon-free Rgap scheme. This is shown by contrasting the Rgap result in Fig.~\ref{fig:Thrust-Cparam-comparison} with the overlap obtained when using the ${\overline {\rm MS}}$ scheme for $\Omega_1$, as shown in Fig.~\ref{fig:Thrust-Cparam-NoGap}.

\section{Conclusions and Comparison to Other $\mathbf{\alpha_s}$ Determinations}
\label{sec:conclusions}
In this paper an accurate determination of $\alpha_s$ from fits to the C-parameter distribution in the resummation region was presented. We fit to the tail of the distribution defined by \mbox{$3\pi\Lambda_{\rm QCD}/Q \ll C \lesssim 3/4$}, where the dominant hadronization effects are encoded in the first moment of the shape function $\Omega_1$, which is a power correction to the cross section. By fitting to data at multiple $Q$'s, the strong coupling $\alpha_s(m_Z)$ and $\Omega_1$ can be simultaneously determined. The key points to our precise theoretical prediction are: a) higher-order resummation accuracy (N$^3$LL$^\prime$), achieved through an SCET factorization theorem, b) $\mathcal{O}(\alpha_s^3)$ matrix elements and fixed-order kinematic power corrections, c) field-theoretical treatment of nonperturbative power corrections, and d) switching to a short-distance Rgap scheme, in which the sensitivity to infrared physics is reduced.

As our final result from the \mbox{\it C-parameter global fit} we obtain,
\begin{align} \label{eq:allerror}
\alpha_s(m_Z) &  = \, 0.1123 \pm 0.0015\,,\\
\Omega_1(R_\Delta,\mu_\Delta) &  = \, 0.421 \pm 0.063\,\mbox{GeV},\nn
\end{align}
where $\alpha_s$ is defined in the $\msbar$ scheme, and $\Omega_1$ in the Rgap scheme (without hadron-mass effects) at the reference scales $R_\Delta=\mu_\Delta=2$\,GeV. Here the respective total $1$-$\sigma$ uncertainties are shown. The results with
individual $1$-$\sigma$ uncertainties quoted separately for the different sources of uncertainties are given in Eq.~(\ref{eq:asfinal}). Neglecting the nonperturbative effects incorporated by $\Omega_1$, the fit yields $\alpha_s(m_Z)=0.1219$ which exceeds the result in Eq.~(\ref{eq:allerror}) by $8\%$. This is consistent with a simple scaling argument one can derive from experimental data, presented in Ref.~\cite{Hoang:2014wka}. We have also presented an updated thrust result, using our improved profiles for thrust and including bottom-mass and QED effects (but neglecting hadron-mass effects). This {\it global fit for thrust} gives
\begin{align} \label{eq:allerrortau}
\alpha_s(m_Z) & \, = \, 0.1128 \,\pm\, 0.0012\,,\\
\Omega_1(R_\Delta,\mu_\Delta) & \, = \, 0.322 \,\pm\, 0.068\,\mbox{GeV}.\nn
\end{align}

Our theoretical prediction is the most complete treatment of C-parameter at this time, and, to the best of our knowledge, all sources of uncertainties have been included in our final uncertainty. Possible improvements which are expected to be negligible relative to our final uncertainty include finite bottom-mass effects, QED effects, and axial-singlet contributions. From our results there are a number of theoretical avenues that lead to small effects but which would be interesting to investigate further in the future.  These are common to almost every event-shape analysis in the literature and include (i) resummation of logarithms for the nonsingular partonic cross section; (ii) the structure of nonperturbative power corrections for the nonsingular contributions (the last two points can be clarified with subleading SCET factorization theorems); (iii) analytic perturbative computations of the ${\cal O}(\alpha_s^3)$ nonlogarithmic coefficients in the partonic soft function and the jet function, as well as the four-loop QCD cusp anomalous dimension (and to a lesser extent, the numerically determined $s_2^{\widetilde C}$ constant of the two-loop partonic soft function); (iv) a better understanding of hadron-mass effects, and in particular their resummation beyond NLL; (v) a better theoretical description of the region around and above the shoulder. Concerning (i), and following the common lore, we have incorporated in our analysis the nonsingular contributions in fixed-order perturbation theory. However we have estimated the uncertainty related to the higher-order logarithms through the usual renormalization scale variation. Concerning (ii) we observe that the effect of these neglected power corrections is much smaller than naively expected, as can be seen from a comparison of our theoretical prediction and LEP data in the far-tail region. A first step towards clarifying (i) and (ii) has been taken in Refs.~\cite{Freedman:2013vya,Freedman:2014uta}, for the case of thrust. The computation of missing perturbative terms (iii) is a priori feasible with current computational knowledge but they do not dominate our perturbative uncertainties. Concerning (iv) we have shown that hadron-mass effects have a very small impact on the determination of $\alpha_s$, and hence unless the rest of the sources of uncertainty become substantially smaller, our lack of knowledge does not constitute a problem. As for (v), our fits do not include data above the shoulder, so this problem has no impact on our fit. Nevertheless an analysis of these subleading effects would be interesting.

The same theoretical program carried out for thrust and C-parameter can be applied to other event-shapes, and
the most prominent one is Heavy-Jet-Mass. This has partially worked out already in Ref.~\cite{Chien:2010kc}
at the purely perturbative level using fully canonical profiles. Their determination of $\alpha_s$ is discussed below. For recoil-sensitive observables such as Jet Broadening~\cite{Catani:1992jc, Dokshitzer:1998kz,Chiu:2011qc,Becher:2011pf,Chiu:2012ir}, one needs to deal with rapidity singularities,
which imply that additional logs need to be resummed, and more complicated nonperturbative power corrections. The former
has been pushed to the N$^2$LL order in Ref.~\cite{Becher:2012qc}, and the latter has been studied in Ref.~\cite{Becher:2013iya}.
Recoil-insensitive versions of Broadening have also been derived~\cite{Larkoski:2014uqa}, but not yet studied experimentally.
Finally, it is very straightforward to generalize our theoretical treatment to the case of oriented event
shapes~\cite{Mateu:2013gya}, in which one additionally measures the angle between the beam and the thrust axes.

\begin{figure}[t!]
\begin{center}
\includegraphics[width=0.95\columnwidth]{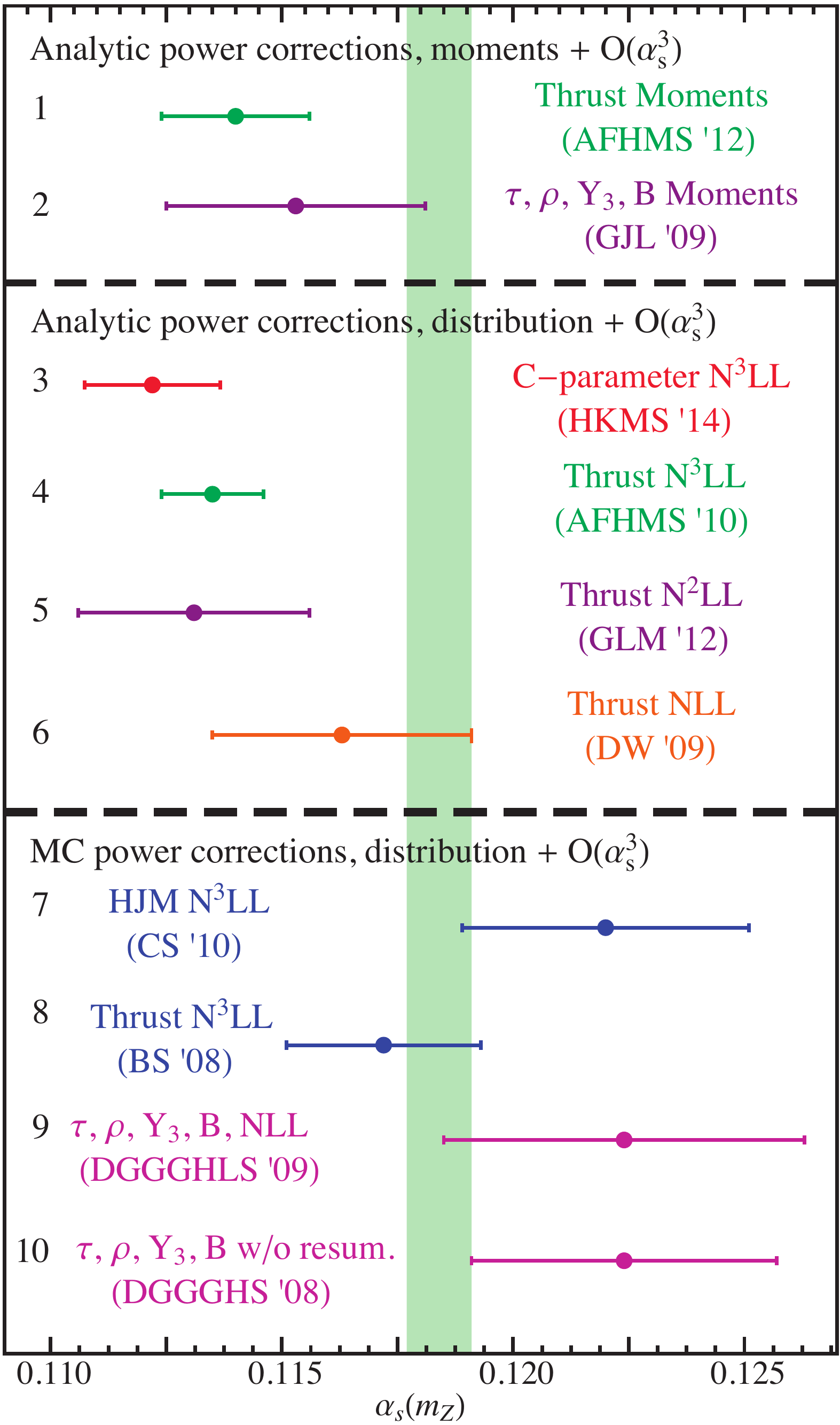}
\caption{Comparison of our determination of $\alpha_s(m_Z)$ (red) with similar analyses from 
thrust~\cite{Abbate:2010xh,Abbate:2012jh} (green) and other determinations from fits to event-shape distributions
using $\mathcal{O}(\alpha_s^3)$ theory predictions and different levels of resummation. Results shown below the lower
dashed line include power corrections as predicted by MC generators, and results above this line treat power corrections
either from a shape function (red and green) or from the dispersive model (orange~\cite{Davison:2008vx} and 
purple~\cite{Gehrmann:2012sc}). Determinations above the upper dashed line correspond to fits to moments of the 
distributions, and those below to fits to the tail of the differential distribution. The translucent green band corresponds to the world average from Ref.~\cite{Agashe:2014kda}. }
\label{fig:alpha-ES-comparison}
\end{center}
\end{figure}

At this point we compare our result for $\alpha_s$ with other determinations from event shapes at $\mathcal{O}(\alpha_s^3)$.
To the best of our knowledge, the only analyses which fit to the tail of the C-parameter distribution using three-loop input
are Ref.~\cite{Dissertori:2007xa} (using purely fix-order perturbation theory) and Ref.~\cite{Dissertori:2009ik} (including NLL resummation).
Both analyses use Monte Carlo (MC) event generators to estimate hadronization effects, and fit $\alpha_s$ for different $Q$ values,
finding values 
$\alpha_s(m_Z)=0.1288\,\pm\, 0.0043$ and $0.1252\,\pm\, 0.0053$ respectively for a fit to the \mbox{$Q = 91.2\,$GeV} data. These larger $\alpha_s(m_Z)$ values are consistent with our fits which neglect power corrections, and following Ref.~\cite{Abbate:2010xh} we can conclude from this that MCs does not provide a reasonable estimate of the power corrections when including the higher-order perturbative contributions.
In Ref.~\cite{Gehrmann:2009eh} two-parameter global fits to the first five moments of the C-parameter
distribution were performed. Hadronization effects are included via the frozen coupling model, and
the value obtained, $\alpha_s(m_Z) = 0.1181\pm 0.0048$, is fully consistent with our result in \Eq{eq:allerror} at 1-$\sigma$.

A graphical comparison with other event-shape determinations is shown in Fig.~\ref{fig:alpha-ES-comparison}. The figure includes determinations where power corrections are estimated with MC generators, labeled by 7-10. Analyses 1-6 correspond to those in which power corrections were incorporated with an analytic method (either a shape function or the dispersive model). In the analyses 1-6 global fits are performed, whereas in the 7-10 analyses $\alpha_s$ was determined at multiple $Q$ values and the final result is an average of those. Only analyses 1~\cite{Abbate:2012jh}, 3 (this work), and 4~\cite{Abbate:2010xh} used a completely field-theoretical approach for the power corrections. We also show both results from fits to the event-shape distributions (3-10) and from fits to moments of the event shape distribution (1 and 2). Although all analyses included $\mathcal{O}(\alpha_s^3)$ matrix elements, different levels of resummation have been achieved. Analyses 2~\cite{Gehrmann:2009eh} and 10~\cite{Dissertori:2007xa} did not included resummation; 6~\cite{Davison:2008vx} and 9~\cite{Dissertori:2009ik} included NLL resummation; 5~\cite{Gehrmann:2012sc} include N$^2$LL resummation; and analyses 3,4,7, and 8 included N$^3$LL resummation. Analyses 2, 9, and 10 simultaneously fit to many event shapes, whereas the others focused on a single observable: thrust (1, 4-6 and 8~\cite{Becher:2008cf}), Heavy-Jet-Mass (7~\cite{Chien:2010kc}), and \mbox{C-parameter} (3, which is this work). The analyses 1, 3, 4, 7 and 8 used SCET to perform Sudakov log resummation.  All results that used an analytic treatment of power corrections have smaller values of $\alpha_s$. This is consistent with a simple dimensional analysis argument (see Refs.~\cite{Abbate:2010xh,Hoang:2014wka}).  Higher order resummation results in a convergent perturbation series and smaller uncertainties, and the Rgap scheme also reduces uncertainties. Accounting for the fact that results relying on MC for the treatment of power corrections should likely have larger hadronization uncertainties,  all results are compatible among one another. The most precise results are however clearly in disagreement with the world average, which is dominated by lattice QCD results (see below) and shown as a translucent green band.

\begin{figure}[t!]
\begin{center}
\includegraphics[width=0.95\columnwidth]{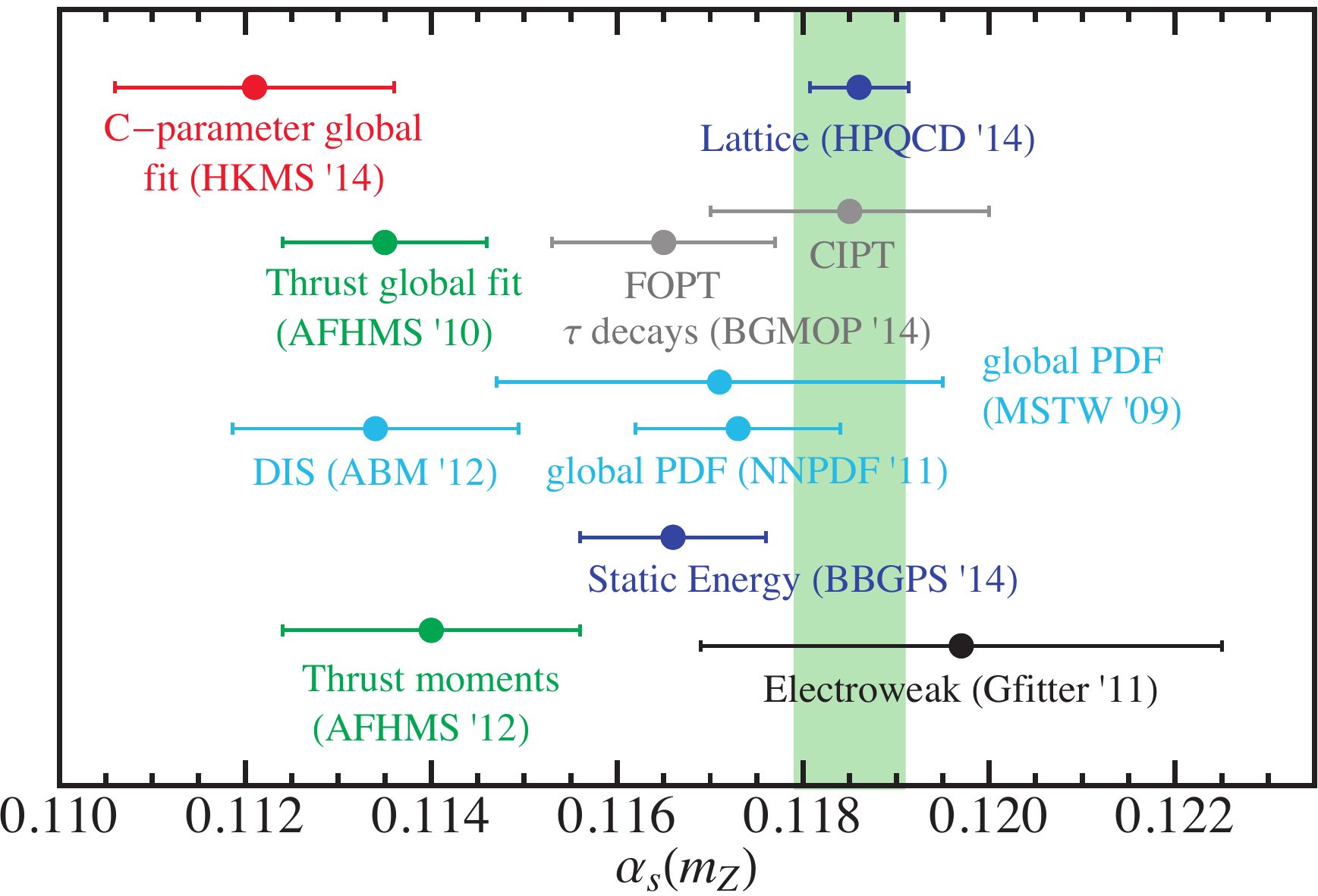}
\caption{Comparison of our determination of $\alpha_s(m_Z)$ (red) with similar previous analyses at N${}^3$LL$^\prime$ for thrust (green)~\cite{Abbate:2010xh,Abbate:2012jh} and other selected determinations: lattice~\cite{Chakraborty:2014aca} and static energy potential \cite{Bazavov:2014soa} (both use lattice input, in blue), Electroweak precision observables fits~\cite{Flacher:2008zq} (black), Deep Inelastic Scattering~\cite{Alekhin:2012ig} and global PDF fits \cite{Ball:2011us,Martin:2009bu}, and hadron $\tau$ decays~\cite{Boito:2014sta} (Fixed Order Perturbation Theory lower, and Contour Improved Perturbation Theory, both in gray). The current world average~\cite{Agashe:2014kda} is shown as a translucent green band.}
\label{fig:alpha-various-comparison}
\end{center}
\end{figure}

We conclude this work by comparing our result for $\alpha_s(m_Z)$ with the results of a selection of recent
analyses using other techniques and observables, as shown in Fig.~\ref{fig:alpha-various-comparison}. We include
a N$^3$LO analysis of data from deep inelastic scattering from the ABM group~\cite{Alekhin:2012ig}, the global PDF fits of the MSTW group~\cite{Martin:2009bu} and the NNPDF collaboration~\cite{Ball:2011us}; the most recent
(and accurate) determination from the HPQCD lattice collaboration~\cite{Chakraborty:2014aca}, from the analysis of Wilson
loops and pseudoscalar correlators; a determination analyzing the lattice prediction for the QCD static potential
\cite{Bazavov:2014soa}; a reanalysis of electroweak precision observables by the Gfitter collaboration~\cite{Flacher:2008zq};
the most recent analysis of tau decays in which the recently released ALEPH data was used together with the OPAL data; the
previous determinations from fits to the thrust distribution~\cite{Abbate:2010xh} and moments of the
thrust distribution~\cite{Abbate:2012jh}; and of course the current world average~\cite{Agashe:2014kda} (shown as the green band). The ABM (DIS) and thrust results are compatible with our determination, while in contrast the disagreement with either
lattice QCD or the world average is $4$-$\sigma$.  Many other determinations lie between these two values. The source of this disagreement is an important open question.

\begin{acknowledgments}
We thank the Erwin-Schr{\"o}dinger Institute (ESI) for partial support in the framework
of the ESI program ``Jets and Quantum Fields for LHC and Future Colliders''.
This work was supported by the offices of Nuclear and Particle Physics of the
U.S. Department of Energy (DOE) under Contract DE-SC0011090, and the
European Community's Marie-Curie Research Networks under contract
PITN-GA-2010-264564 (LHCphenOnet). IS was also supported in part by the Simons
Foundation Investigator grant 327942. VM was supported by a Marie
Curie Fellowship under contract PIOF-GA-2009-251174 while part of this work was
completed. AH, VM, and IS are also supported in part by MISTI global seed funds.
VM thanks H.~Stenzel for explaining how to interpret the systematic uncertainties in
the ALEPH dataset and Thomas Hahn for computer support.
\end{acknowledgments}

\appendix

\section{Profile Formulae}
\label{ap:profile}
In this appendix, we give the details for the profile functions that control the renormalization scales as laid out in \Sec{subsec:profiles}. For the soft profile function, we use the form,
\begin{equation} \label{eq:muSprofile}
\!\!\!\mu_{S} = \left\{ \begin{tabular}{p{.5\columnwidth} l}
$\mu_0$                                             & $0 \le C < t_0$ \\
$\zeta(\mu_0,\,0,\,0,\,\frac{r_s\,\mu_H}{6},\,t_0,\,t_1,\,C)$  & $t_0 \le C < t_1$ \\
$r_s \,\mu_H\, \frac{C}{6}$         & $t_1 \le C < t_2$ \\
$\zeta(0,\,\frac{r_s\,\mu_H}{6},\,\mu_H,0,\,t_2,\,t_s,\,C)$ & $t_2 \le C < t_s$ \\
$\mu_H$     &           $t_s \le C < 1$
\end{tabular}
\right.\!,
\end{equation}
where the physical meaning of the parameters is explained in Sec.~\ref{subsec:profiles}. The function $\zeta(a_1,b_1,a_2,b_2,t_1,t_2,\,t)$ (with $t_1 < t_2$), which
smoothly connects two straight lines of the form $l_1(t) = a_1 \,+\, b_1\,t$ for $t < t_1$
and $l_2(t) = a_2 \,+\, b_2\,t$ for $t > t_2$ is given by
\begin{align}
\zeta(t) &= \left\{ \!\begin{tabular}{p{.5\columnwidth} l}
$\hat a_1 + b_1(t - t_1) + e_1(t - t_1)^2$  &  $~t_1 \le t \le t_m$  \\
$\hat a_2 + b_2(t - t_2) + e_2(t - t_2)^2$  &  $~t_m \le t \le t_2$
\end{tabular}\nonumber
\right.\!,\\[0.1cm]
\hat a_1 & = a_1 + b_1\,t_1\,,\qquad \hat a_2 = a_2 + b_2\,t_2\,,\\
e_1 &=\frac{4\,(\hat a_2-\hat a_1)-(3\,b_1 + b_2)\,(t_2-t_1)}{2\,(t_2-t_1)^2}\,,\nn\\[0.1cm]
e_2 &=\frac{4\,(\hat a_1-\hat a_2)+(3\,b_2 + b_1)\,(t_2-t_1)}{2\,(t_2-t_1)^2}\,.\nn
\end{align}
For the jet scale, we use the form
\begin{equation} \label{eq:muJprofile}
\!\!\!\!\!\mu_J(C) = \left\{\!\! \begin{array}{lr}
\big[\,1 + e_J (C-t_s)^2\,\big] \sqrt{ \mu_H\, \mu_{S} (C)} & C \le t_s\\
\,\mu_H & C > t_s
\end{array}
\right.\!,
\end{equation}
which allows a slight modification of the natural relation between the scales $\mu_J = \sqrt{\mu_H \mu_S}$ in order to account for theoretical uncertainties.

For the subtraction scale, we have
\begin{equation} \label{eq:muRprofile}
\!\!\!\!\!R(C) = \left\{\!\! \begin{array}{ll}
R_0                             & 0 \le C < t_0 \\
\zeta(R_0,\,0,\,0,\,\frac{r_s\,\mu_H}{6},\,t_0,\,t_1,\,C) & t_0 \le C < t_1 \\
\mu_S(C)                        & t_1 \le C \le 1
\end{array}
\right.\!\!.
\end{equation}

\begin{figure}[t!]
	\begin{center}
		\includegraphics[width=0.95\columnwidth]{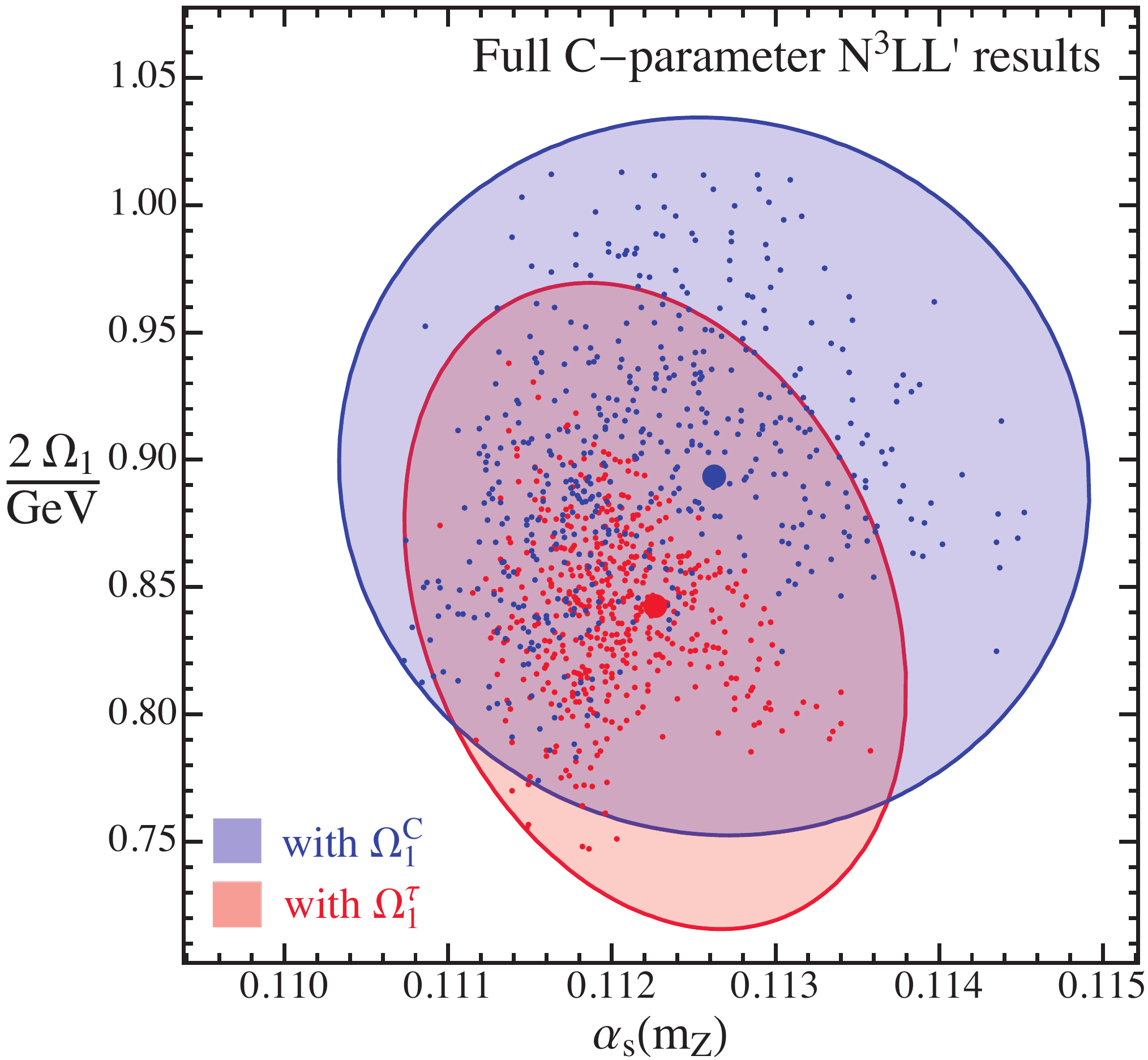}
		\caption{Comparison of $\alpha_s$ determinations from \mbox{C-parameter} tail fits in the thrust Rgap scheme (lower red ellipse) and the
			\mbox{C-parameter} Rgap scheme (upper blue ellipse). The leading power correction $\Omega_1^C$ in the \mbox{C-parameter} Rgap scheme is converted to $\Omega_1$ in the thrust Rgap scheme in order to have a meaningful comparison. Theoretical uncertainty ellipses are shown which are suitable for projection onto one dimension to obtain the $1$-$\sigma$ uncertainty, without experimental uncertainties.}
		\label{fig:Cgap}
		% total (theoretical + experimental)
	\end{center}
\end{figure}

As explained earlier, we take $R=\mu_S$ in the resummation region to avoid large logs and $R \neq \mu_S$ in the
nonperturbative region to remove the renormalon. The $\zeta$ function here interpolates smoothly between these
two regions.

It is necessary to vary the profile parameters to estimate the theory uncertainty. We hold the difference between
the parameters associated with the purely nonperturbative region constant: \mbox{$\mu_0 - R_0 = 0.4\,$GeV}, and
we set as default values \mbox{$\mu_0 = 1.1$ {\rm GeV}}, $R_0 = 0.7$ {\rm GeV}. We are
then left with nine profile parameters to vary during the theory
scan, whose central values and variation ranges used in our analysis are: %$\Delta\mu_0 = 0.2\,(2^{\pm1} - 1)$,
\mbox{$r_s = 2\times 1.13^{\pm1}$}, \mbox{$n_0=12\,\pm\,2$}, $n_1=25\,\pm\,3$, $t_2 = 0.67\,\pm\,0.03$,
\mbox{$t_s=0.83\,\pm\,0.03$}, $e_J = 0\,\pm\, 0.5$,  $e_H = 2^{\pm1}$ and $n_s = 0\pm1$. These variations are
shown in Tab.~\ref{tab:theoryerr}.

\begin{figure}[t!]
	\begin{center}
		\includegraphics[width=0.98\columnwidth]{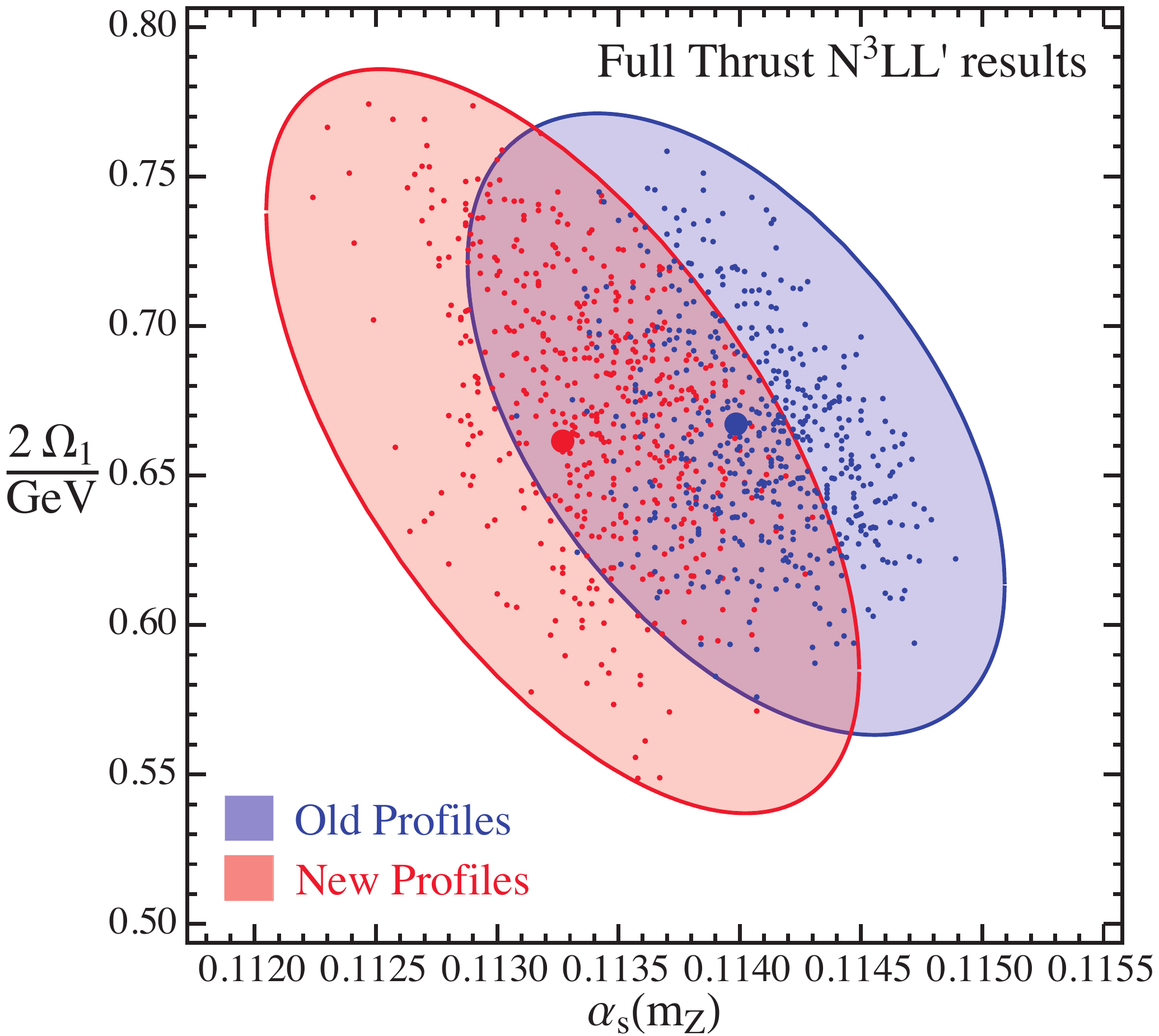}
		\caption{Comparison of thrust $\alpha_s$ determinations using our new profiles (left red ellipse) and the profiles of Ref.~\cite{Abbate:2010xh} (right blue ellipse). Theoretical uncertainty ellipses are shown which are suitable for projection onto one dimension to obtain the $1$-$\sigma$ uncertainty, without experimental uncertainties.}
		\label{fig:new-old-profiles}
	\end{center}
\end{figure}

\section{Comparison of thrust and C-parameter subtractions}
\label{ap:subtractions}

In Fig.~\ref{fig:Cgap} we compare fits performed in the Rgap scheme with C-parameter gap subtractions as the upper red ellipse, and for our default fits in the Rgap scheme with thrust gap subtractions as the lower blue ellipse. At N$^3$LL$^\prime$ order with C-parameter subtractions the results are $\alpha_s(m_Z) = 0.1126 \pm 0.0002_{\rm exp} \pm 0.0007_{\rm hadr} \pm
0.0022_{\rm pert}$ and $\Omega_1(R_\Delta,\mu_\Delta) = 0.447 \pm 0.007_{\rm exp}
\pm 0.018_{\alpha_s} \pm 0.065_{\rm pert}$~GeV, with $\chi^2_{\rm min}/{\rm dof} = 0.988$.
One can see that, even though both extractions are fully compatible, the thrust subtractions lead to smaller perturbative uncertainties. This is consistent with the better perturbative behavior observed for the cross section with thrust subtractions in Ref.~\cite{Hoang:2014wka}. 

\section{Comparison of thrust results with Ref.~\cite{Abbate:2010xh}}
\label{ap:thrustresults}
In Fig.~\ref{fig:new-old-profiles} we compare global fits for the thrust distribution using the profiles of Ref.~\cite{Abbate:2010xh} (shown by the right ellipse in blue) and the profiles used here (shown by the left ellipse in red). As mentioned earlier, the profiles used here have several advantages over those of Ref.~\cite{Abbate:2010xh} in terms of their ability to independently impact the different regions of the thrust distribution, and in particular do a better job in the nonperturbative region (which is outside our fit region). The two versions of the profiles are consistent within their variations, and the fit results shown for 39\% CL for two dimensions in Fig.~\ref{fig:new-old-profiles} (which is 68\% CL for each one-dimensional projection) are fully compatible.

\bibliography{../thrust3}

\end{document}